\documentclass[english,aps, usenatbib,onecolumn,times]{mn2e}
\usepackage[T1]{fontenc}
\usepackage[latin9]{inputenc}

\usepackage{hyperref}

\usepackage{amsmath, amssymb}
\usepackage{esint}
\usepackage{graphicx}
\usepackage{subfigure}

\usepackage[english,english]{babel}

\usepackage{array}
\usepackage{supertabular}
\usepackage{hhline}
\usepackage{hyperref}
\usepackage[usenames]{color}

\definecolor{grey}{rgb}{0.75,0.75,0.75}
\definecolor{Orange}{rgb}{1.0,0.5,0.15}
\definecolor{brown}{rgb}{0.7,0.25,0.0}
\definecolor{pink}{rgb}{1.0,0.5,0.5}
\definecolor{darkerred}{rgb}{0.8,0,0}
\definecolor{darkerblue}{rgb}{0,0,0.8}
\definecolor{Blue}{rgb}{0,0.08,0.65}
\definecolor{Red}{rgb}{0.65,0.08,0.05}
\definecolor{Green}{rgb}{0.15,0.45,0.25}

\hypersetup{dvips,colorlinks=true,breaklinks=true,bookmarks=true,hypertexnames=false,bookmarksnumbered=true,backref=page,pagebackref=true,citecolor= Blue,linkcolor= darkerblue,bookmarksopen=false,bookmarks=true,bookmarksnumbered=true}

\global\long\def\vx{{\bf x}}
 \global\long\def\vk{{\bf k}}
 
\def\gtrsim{\lower.5ex\hbox{$\; \buildrel > \over \sim \;$}}
\def\lwsim{\lower.5ex\hbox{$\; \buildrel < \over \sim \;$}}

\begin{document}


\title[Non-Gaussian statistics of critical sets]{Non-Gaussian statistics of critical sets in 2 and 3D:
\\ Peaks, voids, saddles, genus and skeleton}

\author[C. Gay, C. Pichon  and D. Pogosyan]{ C Gay$^{1,2}$ , C.
Pichon$^{1,3,4}$ and D. Pogosyan$^{2}$ \\
 $^{1}$Institut d'astrophysique
de Paris \& UPMC (UMR 7095), 98, bis boulevard Arago , 75 014, Paris,
France.\\
$^{2}$  Department of Physics, University of Alberta, 11322-89 Avenue,
Edmonton, Alberta, T6G 2G7, Canada.\\
$^{3}$ Astrophysics, University of Oxford, Keble Road, Oxford OX1 3RH, United
Kingdom.\\
$^{4}$ Institut de Physique Th\'eorique, CEA-CNRS, L'orme des meurisiers,
91470, Gif sur Yvette, France.\\
$^4$ Observatoire de Lyon (UMR 5574), 9 avenue Charles Andr\'e, F-69561 Saint Genis Laval, France. 
}
\maketitle

\begin{abstract}
 The formalism   to compute the geometrical and topological one-point statistics of mildly non-Gaussian 2D and 3D cosmological fields is developed.
 Leveraging the isotropy of the target statistics, the Gram-Charlier expansion is reformulated with  rotation invariant variables. This formulation allows  
  to track the geometrical statistics of the cosmic field to all orders. It then allows us
to connect the one point statistics of the critical sets  to the growth factor through perturbation theory, which predicts the redshift evolution of  higher order cumulants. In particular, the cosmic non-linear evolution of the skeleton's length, together with the statistics of extrema and Euler characteristic are investigated in turn. In 2D, the corresponding differential densities are analytic as a function of the excursion set threshold and the shape parameter.  In 3D, the Euler characteristics and the field isosurface area are also analytic to all orders in the expansion. Numerical integrations are performed and simple fits are provided whenever  closed
form expressions are not available. These statistics are compared  to  estimates from  N-body simulations and are shown to match well the cosmic evolution up to root mean square of the  density field of $\sim 0.2$. In 3D, gravitational perturbation theory is implemented to predict the cosmic evolution of all the relevant Gram-Charlier coefficients for universes with scale invariant matter distribution.  The one point statistics of critical sets could be used to constrain primordial non-Gaussianities and the dark energy equation of state on upcoming cosmic surveys; this is illustrated on idealized experiments.
\end{abstract}

\section{Introduction}
Random fields are ubiquitous in physics. 
In cosmology, the large scale distribution of matter (LSS) and the sky-maps of the polarized Cosmic Microwave Background (CMB) radiation are described as, respectively, 3D and 2D random fields.  In primordial  theories where initial seeds for cosmic structures have quantum origin, the fields of initial density fluctuations are close to Gaussian. Subsequent evolution  develop further non-Gaussian signatures via   non-linear gravitational dynamics. Hence these signatures  might be used to constrain the dark energy equation of state via 3D galactic surveys, or shed light on the physics of the early Universe. 
The non-Gaussianities can be accessed through the higher order moments of the field, which are generally difficult to estimate directly in real-life observations, due to their sensitivity to very rare events. The geometrical analysis of the critical sets of the field provides more robust measures of non-Gaussianity and has become an active field of investigation \citep{Gott2007,Park}. Another active area of research is centered around the use
of the bi- or trispectrum for evaluation of the non-Gaussian effects \citep{bispectrum,Scoccimarro2004,Sefusatti2007,Liguori2010, Sefusatti,Nishimichi2010}.
This paper  presents 
statistical tools that may help to detect unique signatures of the fundamental processes in our Universe.

Several statistics have been used to characterize the morphology of the density field of the large-scale structures. Peaks are one of the first components to have been studied in detail \citep{BBKS}. Maxima of the density fields are indeed considered as the location where gravitational collapse occurs to form the halos where galaxies form. Their distribution is thus a key ingredient in semi-analytical models and for predictions of the so-called halo model \citep{MoWhite,Cooray2002}. In 2D, the statistics of peaks of the cosmic microwave background has been used to constrain its Gaussianity \citep{Bond1987,Larson2004,Hou2009,PPG2011}.

The Euler characteristic and the genus, which only differ by their normalization, have also been thoroughly studied, since they describe in part the topology of the field. 
They provide a different approach to the usual geometrical probes such as the power-spectrum.
The genus has been used in  redshift surveys  \citep{Gott86, Vogeley1994, Canavezes1998, Hikage2003, Park}. 
It has also been applied to the CMB  to constrain the Gaussianity of primordial fluctuations, with results from COBE \citep{Smoot1994, Colley1996, Kogut1996} and WMAP \citep{Colley2003, Gott2007}. The genus can be defined in a probabilistic framework where its theory is  more easily derived. Gaussian predictions have been known for a long time \citep{Doroshkevich, Adler} and have been extended by Matsubara to non-Gaussian fields arising from gravitational non-linear dynamics \citep{Matsubara0} and to proper accounting of redshift distortions \citep{Matsubara1995, Matsubara1996}. More generally, the Euler characteristic is one of the Minkowski functionals (a complete set of morphological descriptors) \citep{Mecke} and have been combined to describe the shape of  structures \citep{Sahni1998}.
This paper  is an extension of this approach to other characteristics and higher orders. 

The skeleton has been introduced more recently \citep{Novikov2006}. It traces the filamentary structure of a field, extracting the ``ridges'' linking the maxima through saddle points. It has been applied in 2D for the CMB \citep{Eriksen2004,Hou2010} and in 3D for the study of the large-scale structure \citep{skellet,Rien1,Rien2}. Most of the efforts have focused up to now on improving the algorithm \citep{Sousbie2007, Sousbie2008,persistent} 
 and our theoretical knowledge:
\cite{pogoskel} gives a complete theoretical description of the properties of the skeleton for Gaussian fields. One of the motivations of the present paper is to extend this description of the skeleton beyond this specific Gaussian  case.
Indeed, mildly non-Gaussian random fields play an important role in cosmology: they can be present in the primordial fluctuations and open a window to the physics during the inflation phase \citep{bartolo}, and they also arise e.g. from the non-linear dynamics of gravitational clustering \citep{Bernardeau}. 

In this paper we develop the formalism for computing the geometrical and topological one-point
statistics for mildly non-Gaussian 2D and 3D cosmological fields.
Our formalism is based on 
the Gram-Charlier expansion of the Joint Probability Distribution  of the
field and its  derivatives using  rotation
invariant variables which allows to analyse the isotropic  one-point statistics to all orders in
cumulant expansion.
In the following we will  predict the evolution of quantities such as the Euler characteristic, the density of extrema and the differential length of the skeleton as a function of  root mean square of the cosmic density field, $\sigma$.  The evolution of these geometrical critical sets can thus be linked to the cosmic evolution of the underlying field.
The next generation of redshift surveys, such as Big-BOSS or EUCLID and WFIRST,\footnote{\texttt{http://bigboss.lbl.gov/}, \texttt{http://sci.esa.int/euclid/}, \texttt{http://wfirst.gsfc.nasa.gov/}} 
 aim at constraining the equation of state of dark energy by mapping millions of galaxies.  For example, \cite{Zunckel2010} showed how using the genus in Big-BOSS could give a precision of 5\% on a constant equation of state. 
 Similarly, \cite{Kratochvil} argue that a  joint powerspectrum  and  extrema counts probe  has approximately twice the cosmological sensitivity of powerspectrum estimation.
 As an alternative route, we will show below, for instance, that the length of the skeleton per unit volume within some range of some density threshold can be written as $L(z)=L^{(0)} + \sigma(z) \delta L^{(1)}$ where $L^{(0)}$ is the Gaussian prediction and $\delta L^{(1)}$ is the non-Gaussian prediction at first-order in $\sigma$. By knowing these two functions and  measuring $L(z)$, $\sigma(z)$ can be extracted and the skeleton could thus be used to measure the growth factor of the fluctuations in a manner which might be less sensitive to biases. Indeed, a common feature of the critical sets that we will  build below is that they are formally independent of any monotonic re-mapping of the underlying field.

The structure of the paper is the following.
In Section~\ref{sec:PDF}, we introduce the Gram-Charlier expansion and the rotational invariants used to describe an isotropic field. We obtain the expression of the expansion in two and three dimensions. In Section~\ref{sec:application}, these expressions are used to compute  a sequence of  topological invariants  in 2 and 3D: the Euler characteristic, the distribution of extrema and the length of the skeleton. The probability distribution of the field and the statistics of its iso-contours are also given.
Section~\ref{sec:applications} presents applications to CMB non-Gaussianities and to the cosmic evolution of the large scale structures. In particular,
 Section~\ref{sec:PT3} shows how to predict the cumulants that appear in these expansion  using perturbation theory when cosmic gravitational clustering is responsible for their growth, Section~\ref{sec:measurements} compares the results to measurements on numerical realizations and  Section~\ref{sec:growth} describes an idealized dark energy probe experiment.  
Appendix~\ref{sec:Edgeworth} details the link between the order in the Gram-Charlier expansion and that corresponding to gravitational perturbation theory, introducing the so-scalled Edgeworth expansion. It is then applied to the computation of the $\sigma^2$-order correction to the Euler characteristic in Appendix~\ref{sec:sigma2}. The  results from perturbation theory to scale invariant power spectra  are given in Appendix~\ref{sec:PTappendix}. Appendix~\ref{sec:euleralgorithm} presents the algorithm implemented to measure the Euler characteristic.

\section{Asymptotic expansion of the joint PDF of the field and its derivatives}
\label{sec:PDF}
The expectation of topological and geometrical statistics like the Euler characteristic, extrema counts,  the differential length of skeleton 
requires the knowledge of the  joint probability distribution (JPDF) of the field and its derivatives up to second order \citep{Adler}. 
In this paper we advance the formalism that considers these statistics in non-Gaussian regime treated as the correction to the Gaussian limit. We start with 
developing the relevant version of non-Gaussian JPDF of the field and its derivatives, following closely the exposition in  \cite{PGP2009}.
Starting with some notations, for a given field $\rho$, we define the moments
$\sigma^2 = \langle \rho^2 \rangle$, $\sigma_1^2 = \langle \left( \nabla \rho \right)^2 \rangle$, $\sigma_2^2 = \langle (\Delta \rho)^2 \rangle$.
Combining these moments, we can build two characteristic lengths,
$ R_0 ={\sigma}/{\sigma_1}$ and  $R_* = {\sigma_1}/{\sigma_2}$,
and the spectral parameter
$\gamma={\sigma_1^2}/({\sigma \sigma_2})$.
We choose to normalize the field, $\rho$  and its derivatives to have unit variances:
$ x={\rho}/{\sigma} $,  $x_i={1}/{\sigma_1} \nabla_i \rho$, and $ x_{ij} ={1}/{\sigma_2} \nabla_i \nabla_j \rho$.

\subsection{The Gram-Charlier expansion of the JPDF}

A statistically homogeneous N dimensional  random field $x$
is fully characterized by the
joint one-point distribution function (JPDF) of its value and its derivatives $P(\vx)$, $\vx\equiv (x,x_i,x_{ij},x_{ijk},\ldots)$.
The simplest case, frequently arising because of the central limit theorem (\emph{e.g.} simple inflation models), is the Gaussian JPDF:
\[G(\vx)\equiv(2\pi)^{-N/2}|{\bf C}|^{-1/2}\exp(-\frac{\scriptstyle 1}{\scriptstyle 2} \vx \cdot {\bf C}^{-1} \cdot \vx) \,,\] where ${\bf C}=\langle \vx^{\rm T} \cdot \vx\rangle$ is the covariance matrix of the $\vx$ variables and we have considered that all the variables are defined as having zero mean and unit variance (${\bf C}_{ii}=1$ for each $i$).
For a general non-Gaussian field, the JPDF can be expanded around a Gaussian
arranged to match the mean and the covariance of the 
field variables $\vx$, using a multivariate Gram-Charlier expansion
\citep{Chambers,Juszkiewicz,Amendola,Blinnikov}:
\begin{equation}
P(\vx)=G(\vx) \left[1+\sum_{n=3}^\infty \frac{1}{n!}\;
\mathrm{Tr}\left[ \langle {\mathbf x}^n\rangle_{\scriptscriptstyle{\mathrm{GC}}} \cdot {\mathbf h}_n(\vx)\right]
\right].
\label{eq:defGP}
\end{equation}
The correction to the Gaussian approximation is the series in Hermite tensors
\[ {\bf h}_n(\vx)=(-1)^{n}G^{-1}(\vx)\partial^n G(\vx) /\partial \vx^n \,,\]
of rank $n$. These tensors are the orthogonal polynomials related to the Gaussian kernel $G(\vx)$, which allows to construct the so called Gram Charlier coefficients from the moments:
\begin{equation} \langle {\vx}^n\rangle_{\scriptscriptstyle{\mathrm{GC}}}\equiv\langle \mathbf{h}_n(\vx) \rangle.
\label{eq:defGCfirst} \end{equation}
The statistics of the Euler characteristic (or equivalently the genus) and extremal points  \citep{Adler}, and
the basic description of the critical lines of the field in
the stiff approximation
\citep{pogoskel}, require to know the JPDF only up to
the second derivatives. In the Gaussian limit, in this case, the only 
non-trivial covariance parameter is the cross-correlation between the field and
the trace of the Hessian $\gamma=- \left\langle x \mathrm{Tr}(x_{ij}) \right\rangle$ 
\citep{BBKS,pogoskel}.

\subsection{Rotational invariants of the field and its derivatives}

The JPDF in the form of equation~(\ref{eq:defGP}) is not ideal to
study critical sets statistics since the coordinate representation 
does not take advantage of the isotropic nature of the statistical
descriptors. The pioneering works of \cite{Matsubara0,Matsubara},
where the first correction to Euler characteristic was computed, 
demonstrate the arising complexities.
Instead we developed the equivalent of the Gram-Charlier expansion
for the JPDF of the field variables that are invariant
under coordinate rotation \citep{PGP2009,PPG2011}. Such distribution can be computed via 
explicit integration of the series~(\ref{eq:defGP})
over rotations. However we obtain it here directly from general principles:
{\it the moment expansion of the non-Gaussian JPDF corresponds
to the expansion in the set of polynomials which are orthogonal with respect
to the weight provided by the JPDF in
the Gaussian limit}. Thus, the problem is reduced to finding such polynomials
for a suitable set of invariant variables.

The rotational invariants that are present in the problem 
are: the field value $x$ itself, the square modulus of its gradient, 
$q^2=\sum_{i} x_i^2$ and the invariants of the Hessian, the matrix
of the second derivatives $x_{ij}$.
A rank $N$ symmetric matrix has $N$ invariants with respect to rotations.
The eigenvalues $\lambda_i$ provide one such
representation of invariants, however
they are complex algebraic functions of the matrix components. An alternative 
representation is a set of invariants that are polynomial in $x_{ij}$,
with one independent invariant polynomial per order, from one to $N$. 
A familiar example is the set of coefficients, $I_s$, of the characteristic equation for the eigenvalues, where the linear invariant 
is the trace, $I_1=\sum_{i} \lambda_i$, 
the quadratic one is $I_2=\sum_{i<j} \lambda_i \lambda_j$
and the $N$-th order invariant is the determinant of the matrix.
$I_N=\prod_{i} \lambda_i $.
Aiming at simplifying
the JPDF in the Gaussian limit\footnote{
By Gaussian JPDF we mean
the limit of the Gaussian field $x$. Note that even in this limit, the
rotation invariant variables themselves are not Gaussian-distributed, if they
are non-linear in field variables $\vx$.
} 
\citep[e.g.][]{Doroshkevich,PB,pogoskel} we use their linear 
combinations, $J_s$,
\begin{equation}
J_1 = I_1 ~,\quad 
J_{s\ge 2}  = I_1^s - \sum_{p=2}^s \frac{(-N)^p C_{s}^p}{(s-1)C_N^p} I_1^{s-p} I_{p}\,,
\label{eq:Js}
\end{equation}
where $J_{s\ge2}$ are (renormalized) coefficients of the characteristic
equation of the {\it traceless part} of the Hessian and are independent
in the Gaussian limit on the trace $J_1$.

Our choice of normalization of the field and its
derivatives gives simple relations
for the linear and quadratic in the field moments of the invariant variables:
\begin{equation}
 \langle \zeta \rangle = 0, \qquad \langle J_1 \rangle = 0, \qquad \langle \zeta
J_1 \rangle =0, \qquad \langle \zeta^2 \rangle =1, \qquad \langle {J_1}^2
\rangle =1, \qquad
\langle q^2 \rangle = 1, \qquad \langle J_2 \rangle = 1 ~,
\label{eq:varnorm}
\end{equation}
i.e., the linear invariant variables have zero means and unit variances and the
quadratic ones have unit means.  Since we expand JPDF around the Gaussian form
that is arranged to match all these terms, the relations~(\ref{eq:varnorm}) are
retained in non-Gaussian case.

\subsection{The Gram Charlier non-Gaussian joint probability distribution in two dimensions }
The rotation invariant formalism is straightforwardly adapted to  2D random 
fields,  which are of interest for e.g. CMB and weak lensing studies.
There are just two invariants that describes the Hessian in 2D
\begin{equation}
I_1 \equiv \mathrm{Tr}(x_{ij}) = x_{11}+x_{22} 
\,, \quad{\rm and} \quad
I_2 \equiv \det\left| x_{ij} \right| = x_{11} x_{22} - x_{12}^2 
\,.
\end{equation}
which in the Hessian eigenframe have the form
\begin{equation}
I_1 \equiv \mathrm{Tr}(x_{ij}) =
\lambda_1+\lambda_2 
\,, \quad{\rm and} \quad
I_2 \equiv \det\left| x_{ij} \right| = \lambda_1 \lambda_2
\,.
\end{equation}
We shall also use the orthogonal set
\begin{equation}
J_1  =  I_1 ,  \quad{\rm and} \quad
J_2  =  I_1^2 - 4 I_2 = \left( x_{11}-x_{22} \right )^2 + 4 x_{12}^2 
.
\end{equation}
Introducing $\zeta=(x + \gamma J_1)/\sqrt{1-\gamma^2}$ in place of the
field value $x$ we can build the 2D Gaussian JPDF 
$G_{\rm 2D}(\zeta ,q^2,J_1,J_2)$,
normalized over ${\rm d}\zeta {\rm d}q^2 {\rm d}J_1 {\rm d}J_2$, which has a fully factorized form:
\begin{equation}
G_{\rm 2D}(\zeta,q^2 ,J_1, J_2)
d \zeta d q^2 d J_1 d J_2
= \frac{1}{2 \pi} 
\exp\left[-\frac{1}{2} \zeta^2 - q^2 - \frac{1}{2} J_1^2 - J_2 \right]
d \zeta d q^2 d J_1 d J_2.
\end{equation}
From this Gaussian kernel, we can build the basis of orthogonal polynomials associated with it. Since the variables are independent, the polynomials will not be coupled and we can consider each variable separately. For variables with a normal distribution ($\zeta$ and $J_1$), the orthogonal polynomials are the ``probabilists'' Hermite polynomials, which follow the orthogonality relation:
\begin{equation}
\int_{-\infty}^{+\infty} \frac{1}{\sqrt{2\pi}} \exp \left({-\frac{x^2}{2}} \right) H_n(x) H_m(x) {\rm d}x = n! \; \delta_{n}^{m}.
\end{equation}
For the quadratic variables ($q^2$ and $J_2$), the corresponding polynomials are the Laguerre polynomials:
\begin{equation}
 \int_{0}^{\infty}  \exp({-x})  L_n(x) L_m(x)  {\rm d}x  = \delta_{n}^{m}.
\end{equation}
The non-Gaussian rotation invariant JPDF can then be written in the form of the direct series in the products of
these polynomials:\footnote{The orthogonality expressions and the choice of normalization for the Gram-Charlier coefficient (equation~(\ref{eq:normalisation})) explain the quadruple factorial at the denominator together with the $(-1)^{j+l}$ factor.}
\begin{equation}
P_{\rm 2D}(\zeta, q^2, J_1, J_2) =  G_{\rm 2D} \left[ 1 + 
\sum_{n=3}^\infty \sum_{i,j,k,l}^{i+2 j+k+2 l=n} 
\frac{(-1)^{j+l}}{i!\;j!\; k!\; l!} 
\left\langle \zeta^i {q^2}^j {J_1}^k {J_2}^l \right\rangle_\mathrm{{\scriptscriptstyle GC}}
H_i\left(\zeta\right) L_j\left(q^2\right)
H_k\left(J_1\right) L_l\left(J_2\right)
\right]\,,
\label{eq:2DP_general}
\end{equation}
where $\sum_{i,j,k,l}^{i+2 j+k+2 l=n} $
stands for summation over all combinations
of non-negative $i,j,k,l$ such that $i+2j+k+2l$ adds 
to the order of the expansion $n$, defined as the number of
fields. This arrangement sorts the series in the order of increasing
power of the field variables.\footnote{ This ordering should not be confused
with the order given by perturbation theory. See Appendix \ref{sec:Edgeworth}
for further explanations.}
When counting the order of a term, $j$ and $l$ enter multiplied by two,
since each of $q^2$ and $J_2$ contributes a quadratic in the field
quantity.

The orthogonality of the polynomials \citep{Abramovitz} allows us to extract  the coefficients by projection:
\begin{equation}
\begin{split}
 \left\langle \vphantom{ \zeta^i {q^2}^j J_1^k J_2^l } \!
H_i\left(\zeta\right) L_j\left(q^2\right)
H_k\left(J_1\right) L_l\left(J_2\right) \!\right\rangle &\equiv \int_{-\infty}^{+\infty} {\rm d}\zeta \int_0^{+\infty} {\rm d}q^2 \int_{-\infty}^{+\infty} {\rm d}J_1 \int_0^{+\infty} {\rm d} J_2 P_{\rm 2D}(\zeta, q^2, J_1 ,J_2) H_i\left(\zeta\right) L_j\left(q^2\right)
H_k\left(J_1\right) L_l\left(J_2\right) \,,\\
&= \frac{(-1)^{j+l}}{ j! \; l!} \left\langle \zeta^i q^{2j} {J_1}^k {J_2}^l \right\rangle_{\scriptscriptstyle{\mathrm{GC}}}. \label{eq:normalisation}
\end{split}
\end{equation}
We can thus  express the Gram-Charlier coefficients as combinations of moments:
\begin{equation}
\left\langle \zeta^i {q^2}^j J_1^k J_2^l \right\rangle_{\scriptscriptstyle{\mathrm{GC}}} \!\! =
\frac{j! \; l!}{(-1)^{j+l}} 
\left\langle \vphantom{ \zeta^i {q^2}^j J_1^k J_2^l } \!
H_i\left(\zeta\right) L_j\left(q^2\right)
H_k\left(J_1\right) L_l\left(J_2\right) \!\right\rangle.
\label{eq:GCtomoments2D}
\end{equation} 
The chosen convention for the prefactor ensures that when the polynomial
expressions are expanded, the highest power
term is given directly by the corresponding moment:
\begin{equation}
 \left\langle \zeta^i {q^2}^j J_1^k J_2^l
\right\rangle_{\scriptscriptstyle{\mathrm{GC}}} 
\!\! = \left\langle \zeta^i {q^2}^j J_1^k J_2^l \right\rangle + \dots \quad .
\end{equation}
For example,
\begin{displaymath}
\begin{array}{rcl}
 \left\langle \zeta^3 \right\rangle_\mathrm{{\scriptscriptstyle GC}} &\equiv& 
\left\langle H_3(\zeta) \right\rangle = \left\langle \zeta^3  -3 \zeta
\right\rangle =
\left\langle \zeta^3 \right\rangle - 3\left\langle \zeta \right\rangle =
\left\langle \zeta^3 \right\rangle , \nonumber  \\
  \left\langle q^2 J_1 \right\rangle_\mathrm{{\scriptscriptstyle GC}} &\equiv&    \left\langle  L_1(q^2) H_1(J_1) \right\rangle =
   - \left\langle (1-q^2 ) J_1 \right\rangle = \left\langle q^2 J_1\right\rangle
- \left\langle J_1\right\rangle = \left\langle q^2 J_1\right\rangle , \\
  \left\langle J_1^2 J_2 \right\rangle_\mathrm{{\scriptscriptstyle GC}} &\equiv&
  \left\langle  H_2(J_1) L_1(J_2) \right\rangle =
   - \left\langle (J_1^2-1) (1 - J_2 ) \right\rangle = \left\langle 
J_1^2 J_2\right\rangle - \left\langle J_1^2 \right\rangle  - \left\langle J_2
\right\rangle + 1 = \left\langle J_1^2 J_2\right\rangle - 1 .  
\end{array}
\end{displaymath}
where we used  our normalization convention equation~(\ref{eq:varnorm}).
Consequently, the $n=3$ Gram-Charlier coefficients simply coincide with the
third-order moments, but the higher order ones are given by the
combination of moments.

The combinations of moments that constitute the Gram-Charlier coefficients vanish 
in the Gaussian limit. Thus they can be expressed via the cumulants of the field 
that have the same property. For the variables that are linear in the field, e.g. $\zeta, J_1$, the ordinary cumulants arise.
However, the variables $q^2$ and $J_2$ are not Gaussian distributed in the limit
of the Gaussian field
 due to their quadratic nature, and their own cumulants do not vanish in this
limit,
e.g.,
$\langle J_2^2 \rangle_{cum} = \langle J_2^2 \rangle - \langle J_2 \rangle^2 \to
1 $.
To take this effect into account we introduce the notion of ``field cumulants''. A ``field cumulant'' of a
polynomial variable is the cumulant of the expression in field variables, $x,x_i,x_{ij}$, that
constitutes it. By definition it vanishes in the limit of
the Gaussian field. For example, $\langle J_2^2 \rangle_{\rm fc} = \langle
\left[(x_{11} - x_{22})^2 + 4 x_{12}^2 \right]^2 \rangle_c = \langle J_2^2
\rangle - 2 \langle J_2 \rangle^2 \to 0 $. 
In the rest of the paper we will always mean ``field cumulants'' and
will  use the subscript ``$c$'' to tag them.
Expressing the Gram-Charlier coefficients via the field cumulants, we find that for
the orders  $n=3,4,5$  they coincide, e.g. 
\begin{equation}
 \left\langle \zeta^2 q^2 J_1 \right\rangle_\mathrm{{\scriptscriptstyle GC}} = -\left\langle H_2(\zeta) L_1( q^2) H_1( J_1) \right\rangle=
  - \left\langle (\zeta^2-1)(-q^2+1)J_1\right\rangle= \left\langle \zeta^2 q^2 J_1 \right\rangle- \left\langle \zeta^2 J_1 \right\rangle - \left\langle q^2 J_1 \right\rangle = \left\langle \zeta^2 q^2 J_1 \right\rangle_\mathrm{c}.
\end{equation} This property follows from the link between the Gram-Charlier and the Edgeworth expansions (see Appendix~\ref{sec:Edgeworth}).
For the next orders ($n \ge 6$), the Gram-Charlier coefficients are  more complex combinations of cumulants, as explained in Appendix~\ref{sec:sigma2}.

\subsection{The Gram Charlier non-Gaussian  joint probability distribution in three dimensions }
In 3D, the polynomial
invariants of the matrix $x_{ij}$, as written in the coordinate frame, are
\begin{eqnarray}
I_1 &\equiv& \mathrm{Tr}(x_{ij}) = x_{11}+x_{22}+x_{33} 
\,, \quad
I_2 \equiv x_{11}x_{22}+x_{22}x_{33}+x_{11}x_{33}
- x_{12}^2 - x_{23}^2 - x_{13}^2 
\,,\\
I_3 &\equiv& \det\left| x_{ij} \right| = 
x_{11} x_{22} x_{33} 
+ 2 x_{12} x_{23} x_{13} - x_{11} x_{23}^2 - x_{22} x_{13}^2 - x_{33} x_{12}^2 \,.
\end{eqnarray}
The same invariants expressed through eigenvalues are
\begin{equation}
I_1 \equiv \mathrm{Tr}(x_{ij}) =
\lambda_1+\lambda_2+\lambda_3 
\,,\quad
I_2 \equiv 
 \lambda_1 \lambda_2 + \lambda_2 \lambda_3 + \lambda_1 \lambda_3 
\,,  \quad {\rm and} \quad
I_3 \equiv \det\left| x_{ij} \right| = \lambda_1 \lambda_2 \lambda_3 
\,.
\end{equation}
This variables can be made partially independent using the linear combinations
\begin{equation}
J_1  =  I_1 ,\quad
J_2  =  I_1^2 - 3 I_2\quad {\rm and} \quad
J_3  =  I_1^3 - \frac{9}{2} I_1 I_2 + \frac{27}{2} I_3.
\end{equation}
In $N$ dimensions, the Gaussian JPDF, 
$G_{\rm ND}(\zeta, q^2, J_1, J_{s\ge2})$, retains complete factorization
with respect to $\zeta, q^2, J_1$
\begin{equation}
G_{\mathrm{ND}}=
\frac{(\frac{N}{2})^{\frac{N}{2}} q^{N-2}}{2\pi \Gamma\left[\frac{N}{2}\right]} 
\exp\!\left[-\frac{\zeta^2}{2}\! -\! \frac{N q^2}{2}\! -\! \frac{J_1^2}{2}
\right]
{\mathcal G}(J_{s\ge2})\,,
\label{eq:3DG}
\end{equation}
so in the moment expansion this sector always gives rise to Hermite
$H_i(\zeta)$, $H_k(J_1)$
and generalized Laguerre 
$L_j^{(N-2)/2}(N q^2/2)$ polynomials.
However, the distributions of the rest $J_{s\ge2}$ contained in
${\mathcal G}(J_{s \ge 2})$ are coupled. 
Specifically, in 3D, ${\mathcal G} (J_2,J_3) = \frac{25 \sqrt{5}}{6\sqrt{2\pi}} 
\exp[-\frac{5}{2} J_2]$,
and $J_3$ is distributed uniformly between
$-J_2^{3/2}$ and $J_2^{3/2}$:
\begin{equation}
G_{\rm 3D}(\zeta, q^2, J_1, J_2, J_3) = \frac{25 \sqrt{15}}{8 \pi^2} \sqrt{q^2} 
\exp\left[-\frac{1}{2} \zeta^2 - \frac{3}{2} q^2 - \frac{1}{2} J_1^2
- \frac{5}{2} J_2 \right],
\end{equation}
with the ranges that variables span summarized in the normalization integral
\begin{equation}
{\int_{-\infty}^\infty \!\!\!\! {\rm d} \zeta  \int_0^\infty {\rm d} q^2
\int_{-\infty}^\infty  \!\!\!\! {\rm d} J_1 
\int_0^\infty  \!\!\!\! {\rm d} J_2 \int_{-J_2^{3/2}}^{J_2^{3/2}} \!\! d J_3
G_{\rm 3D}(\zeta, q^2, J_1, J_2, J_3) }
= 1\, .
\end{equation}
Let us denote the orthogonal polynomials in the variables $J_2$ and $J_3$, 
$F_{lm}(J_2,J_3)$, where $l$ is the power of $J_2$ and $m$ is the power of 
$J_3$. They obey the orthonormality condition
\begin{equation}
\int_0^\infty  \!\!\!\! {\rm d} J_2 \int_{-J_2^{3/2}}^{J_2^{3/2}} \!\! {\rm d} J_3
{\mathcal G} \, F_{l m}(J_2,J_3) F_{l^\prime m^\prime}(J_2,J_3) 
= \delta_l^{l^\prime} \delta_{m}^{m^\prime}  . \label{eq:ortho}
\end{equation}
We shall not give the full theory of these polynomials here, but note that
one can construct them by a Gram-Schmidt orthogonalization procedure (say using the Mathematica {\tt Orthogonalize} routine using the dot product defined by equation~(\ref{eq:ortho})\footnote{for instance, the orthogonal polynomial of order one in $J_2$ and $J_3$ is  proportional to $
   (5 J_2-11) J_3$.})
to any given order.  Two special cases 
$
F_{l0} = \sqrt{\frac{ 3 \times 2^l \times l!}{(3+2l)!!}} 
L_l^{(3/2)}\left(\frac{5}{2} J_2\right) 
$
and
$
F_{01} =  \frac{5}{\sqrt{21}} J_3
$
are sufficient to obtain the general expression for the Euler characteristic
of the excursion sets of the field to arbitrary order, and to calculate
the critical point and skeleton statistics to quartic order.
Hence, we write the moment expansion for invariant non-Gaussian JPDF,
$P_{\rm 3D}(\zeta, q, J_1, J_2, J_3)$ as a series in the power order of the
field, $n$ in the form
\begin{eqnarray}
P_{\rm 3D} &=&  G_{\rm 3D}
\left[1+
\sum_{n=3}^\infty \sum_{i,j,k,l}^{i+2 j+k+2 l=n} 
\frac{(-1)^{j+l} 3^j 5^l \times 3}{i!\;(1+2j)!!\; k!\; (3+2l)!!} 
\left\langle \zeta^i {q^2}^j {J_1}^k {J_2}^l \right\rangle_\mathrm{{\scriptscriptstyle GC}}
H_i\left(\zeta\right) L_j^{(1/2)}\left(\frac{3}{2} q^2 \right)
H_k\left(J_1\right) L_l^{(3/2)}\left(\frac{5}{2} J_2\right) \right.
+\nonumber \\
 && 
\sum_{n=3}^\infty \sum_{i,j,k}^{i + 2j + k + 3=n} 
\frac{(-1)^j 3^j \times 25}{i!\;(1+2j)!!\; k! \times 21} 
\left\langle \zeta^i {q^2}^j {J_1}^k J_3 \right\rangle_\mathrm{{\scriptscriptstyle GC}}
H_i\left(\zeta\right) L_j^{(1/2)}\left(\frac{3}{2} q^2 \right)
H_k\left(J_1\right) J_3
\nonumber \\
&&+ \left. 
\sum_{n=5}^\infty \sum_{i,j,k,l,m=1}^{i + 2j + k + 2l + 3m = n} 
\frac{(-1)^j 3^j c_{lm} }{i!\;(1+2j)!!\; k! } 
\left\langle \zeta^i {q^2}^j {J_1}^k {J_2}^l {J_3}^m \right\rangle_\mathrm{{\scriptscriptstyle GC}}
H_i\left(\zeta\right) L_j^{(1/2)}\left(\frac{3}{2} q^2 \right)
H_k\left(J_1\right) F_{lm}\left(J_2,J_3\right)
\right]\,.
\label{eq:3DP_general}
\end{eqnarray}
The first non-Gaussian correction term contains $J_2$ but not $J_3$, the second
is linear in $J_3$ and contains no $J_2$, while the last one  
contains all the remaining combinations of
$J_2$ and $J_3$.
In this last term we leave
the normalization
coefficients $c_{lm}$ undetermined, but $c_{01}=0$, so the first
non-zero contribution, $l=m=1$, is of the fifth power of the field 
and proportional to $ J_2 J_3$.
As in 2D, the orthogonality can be used to express the coefficients in terms of moments. For $m=0$:
\begin{equation}
 \left\langle \zeta^i {q^2}^j {J_1}^k {J_2}^l \right\rangle_\mathrm{{\scriptscriptstyle GC}} = 
\frac{(-1)^j j!}{\left({3}/{2}\right)^j}  \frac{(-1)^l l!}{\left({5}/{2}\right)^l}
\left\langle H_i\left(\zeta\right) L_j^{(1/2)}\left(\frac{3}{2} q^2 \right)
H_k\left(J_1\right) L_l^{(3/2)}\left(\frac{5}{2} J_2\right) \right\rangle ;
\label{eq:GCtomoments3D1}
\end{equation}
and for $l=0$ and $m=1$:
\begin{equation}
\left\langle \zeta^i {q^2}^j {J_1}^k J_3 \right\rangle_\mathrm{{\scriptscriptstyle GC}} = \frac{(-1)^j j!}{\left({3}/{2}\right)^j}
\left\langle H_i\left(\zeta\right) L_j^{(1/2)}\left(\frac{3}{2} q^2 \right)
H_k\left(J_1\right) J_3 \right\rangle .
\label{eq:GCtomoments3D2}
\end{equation}
Equations (\ref{eq:2DP_general}) and (\ref{eq:3DP_general}) are the main results
of this section. 
They fully characterize  the joint probability 
function of a field and its derivatives in two and three dimensions, up to second order 
derivatives, to an arbitrary order in the
moment expansion (see also \cite{PGP2009,PPG2011}).

\section{Non-Gaussian critical sets in 2 and 3 D}
\label{sec:application}
Given formulas (\ref{eq:2DP_general}) and (\ref{eq:3DP_general}) 
one can  easily compute any rotation-invariant statistics that depend
exclusively on these descriptors of the field.
In particular, one can compute the filling factor of the field excursion set (Section~\ref{sub:PDF}),
the length of field isocontours in 2D and the area of isosurfaces in 3D  (Sections~\ref{sec:minkov2D}
and~\ref{sec:mink3D}),  the  Euler characteristic of the excursion sets
of the field (Sections~\ref{2Deuler} and \ref{sec:euler3D}), the density of 
extrema (Sections~\ref{sec:count2D} and \ref{sec:count3D}), and the properties of the critical lines that describe the skeleton
of the field (Sections~\ref{sec:skel2D} and \ref{sec:skel3D}). 

\subsection{The filling factor of high excursion set} \label{sub:PDF}
The filling factor, which is the volume fraction occupied by the field with values exceeding the threshold $\nu$,
presents the simplest case  when one considers only the field and not its derivatives.
The JPDF, equations~(\ref{eq:2DP_general}) and (\ref{eq:3DP_general}), is then reduced both in 2D and 3D
to the same expression:
\begin{equation}
 P(x) = G(x) \left[ 1+ \sum_{n=3}^\infty \frac{\langle x^n \rangle_\mathrm{{\scriptscriptstyle GC}}}{n!} H_n(x) \right]. \label{eq:PDF}
\end{equation}
This is the classical one-variable expression for the Gram-Charlier expansion. This expansion does not have good convergence properties \citep{Blinnikov}, but it is very simple to implement. When the origin of non-Gaussianities is known and can be used to predict the amplitude of each coefficient, this expansion can be re-summed to a more physically motivated one. For example, as shown in Appendix~\ref{sec:Edgeworth}, when the non-Gaussianities give rise to a hierarchy of cumulants such that $\langle x^n \rangle_\mathrm{c} \propto \sigma^{p-2}$ (notably in the case of gravitational evolution), the re-summation to a series in terms of $\sigma$ leads to the Edgeworth expansion \citep{Juszkiewicz}:
\begin{equation}
 P(x) = G(x) \left[1 + \left( \frac{\langle x^3 \rangle_\mathrm{c} }{6} H_3(x) \right) + \left(  \frac{\langle x^4 \rangle_\mathrm{c} }{24} H_4(x) + \frac{{\langle x^3 \rangle_\mathrm{c}}^2 }{72} H_6(x) \right) + \dots \right].
\end{equation}
From the non-Gaussian PDF given by equation (\ref{eq:PDF}),
the  filling factor, $f(\nu)$ immediately follows
\begin{equation}
 f(\nu) \equiv \int_\nu^{\infty} P(x) {\rm d}x= 
 \frac{1}{2} \mathrm{Erfc} \left(\frac{\nu}{\sqrt{2}}\right) +G(\nu) \sum_{n=3}^\infty \frac{ \langle x^n \rangle_\mathrm{{\scriptscriptstyle GC}} }{n!} H_{n-1}(\nu).\label{eq:defF}
\end{equation}

\subsection{Non-Gaussian critical sets in two dimensions}
\label{sec:2Dtheory}
\subsubsection{The length of 2D isocontours} 
\label{sec:minkov2D}
The average length of the isocontour $x=\nu$ is one of the two Minkowski functionals 
\citep{Schmalzing1998, Mecke1994, Schmalzing1997} that characterize the 
level sets in 2D. In our formalism this is the simplest statistics for which the invariant
variables appear non-trivially. To compute the average length per unit volume ${\cal L}$
of a $\nu$-isocontour, 
\begin{equation}
 {\cal L}(\nu)=\frac{1}{R_0}\int_{-\infty}^{+\infty} {\rm d}x \int_0^{\infty} {\rm d}q^2 P_{\rm 2D}(x,q^2) \delta^\mathrm{D} (x-\nu) q
=\frac{1}{R_0}\int_0^{\infty} {\rm d}q^2 P_{\rm 2D}(\nu,q^2) q \,,\label{eq:defMin2D}
\end{equation}
one needs JPDF of the field and the magnitude of its gradient.
From equation~(\ref{eq:2DP_general})
\begin{equation}
P_{\rm 2D}(x,q^2)=\frac{1}{\sqrt{2\pi}} e^{-\frac{x^2}{2}-q^2} \left( 1 + \sum_{n=3}^{\infty} \sum_{i,j}^{i+2j=n} \frac{(-1)^j}{i! j!} \left\langle x^i q^{2j} \right\rangle_\mathrm{{\scriptscriptstyle GC}} H_i(x) L_j(q^2) 
\label{eq:defPsimpl}\right)\,,
\end{equation}
yielding
\begin{equation}
 {\cal L}(\nu)=\frac{1}{R_0 \sqrt{2\pi}} e^{-\frac{\nu^2}{2}} \left( \int_0^{\infty} e^{-q^2} q dq^2
 + \sum_{n=3}^{\infty} \sum_{i,j}^{i+2j=n} 
\frac{(-1)^j}{i! j!} \left\langle x^i q^{2j} \right\rangle_\mathrm{{\scriptscriptstyle GC}} H_i(\nu)
 \int_0^{\infty} e^{-q^2} q L_j(q^2) dq^2 \right)\,.
\end{equation}
The identity
 $
 \int_0^{\infty} e^{-q^2} q L_j(q^2) dq^2 = -{\Gamma(j-\frac{1}{2})}/{(4 \Gamma(j+1))}
 $
allows us to write the length of the $\nu$ isocontour to an arbitrary order in non-Gaussianity as
\begin{equation}
 {\cal L}(\nu)=\frac{1}{2\sqrt{2} R_0} e^{-\frac{\nu^2}{2}} 
\left( 1 + \frac{1}{2\sqrt{\pi}} \sum_{n=3}^{\infty} \sum_{i,j}^{i+2j=n} \frac{(-1)^{j+1}}{i! j!}  \frac{\Gamma(j-\frac{1}{2})}{\Gamma(j+1)} \left\langle x^i q^{2j} \right\rangle_\mathrm{{\scriptscriptstyle GC}} H_i(\nu) \right)\,.\label{eq:defCont2Dfinal}
\end{equation}
This result is a generalization  of the  published results to first \citep{Matsubara} and second \citep{Matsubara2010} orders.
Appendix~\ref{sec:MinkovFunc} reorders the expression~(\ref{eq:defCont2Dfinal}) as an Edgeworth expansion.

\subsubsection{The Euler characteristic of non-Gaussian 2D fields}
\label{2Deuler}
%
%
The Euler characteristic of a manifold is defined as the alternating sum of the Betti numbers:
$ \chi = b_0 - b_1 + b_2 - \dots $.
The first Betti numbers have a simple interpretation: $b_0$ is the number of connected components, $b_1$ is the number of two-dimensional holes and $b_2$ is the number of three-dimensional voids. It is also one of the Minkowski functionals \citep{Mecke} and in particular it is additive. For example the Euler characteristic is 1 for a ball and 2 for a sphere (which is a region surrounding a void). The Euler characteristic is an important quantity in topology and plays a central role in many different theories. For example, it can be linked to the geometry of the surface via the Gauss-Bonnet theorem or the extrema of a function in the Morse theory \citep{jost}.

In the specific case of an orientable and closed two-dimensional surface, it can be shown to be topologically equivalent to a sphere with $g$ handles.  One can thus use $g$, called the genus, as an equivalent to the Euler characteristic. There is a direct relation between the Euler characteristic of a closed surface and its genus:
$ \chi=2-2g. $

In cosmology, the genus has been used to characterize the topology of the isocontours. However the definition generally used by cosmologists is that the genus is the number of holes minus the number of isolated regions, which leads to an offset of 1: for an isolated region, the genus is then defined to be the number of holes minus 1 (see e.g. the warning in \cite{Gott06}). The advantage of this shifted definition is due to the fact that the relation between the Euler characteristic $\chi$ and the ``cosmological`` genus $g_c$ in this definition is simply:
$ \chi = -2 g_c. $
One can therefore keep the intuitive definition of the genus (i.e. the number of holes) while inheriting all the good properties from the Euler characteristic (i.e. the Gauss-Bonnet theorem and the results from Morse theory).
Table \ref{tab:genus} gives the values of the genus (in the two definitions) and the Euler characteristic for basic sets.
\begin{table}
 \begin{center}
\begin{tabular}{|l|c|c|c|c|c|c|}
&\hspace{2mm}\includegraphics[width=5mm]{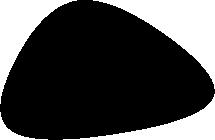} \hspace{2mm}
&\hspace{2mm}\includegraphics[width=1cm]{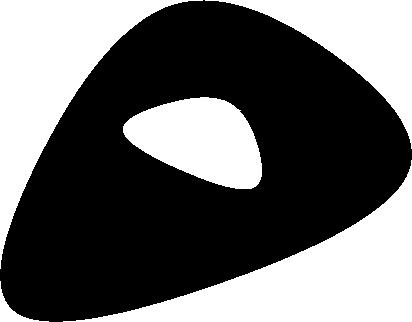}\hspace{2mm}
&\hspace{2mm}\includegraphics[width=1cm]{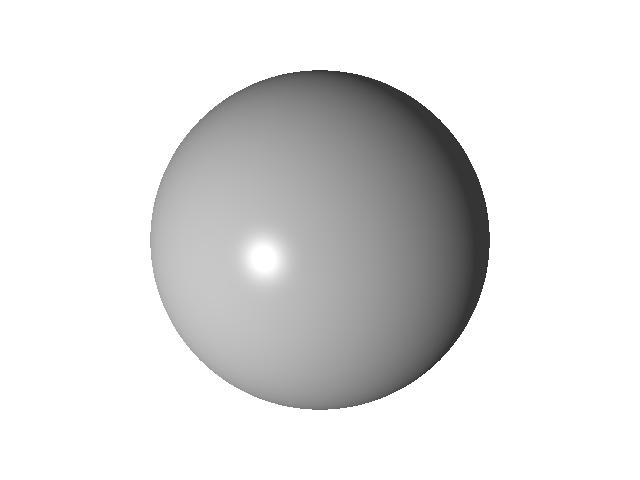}\hspace{2mm}
&\hspace{2mm}\includegraphics[width=1cm]{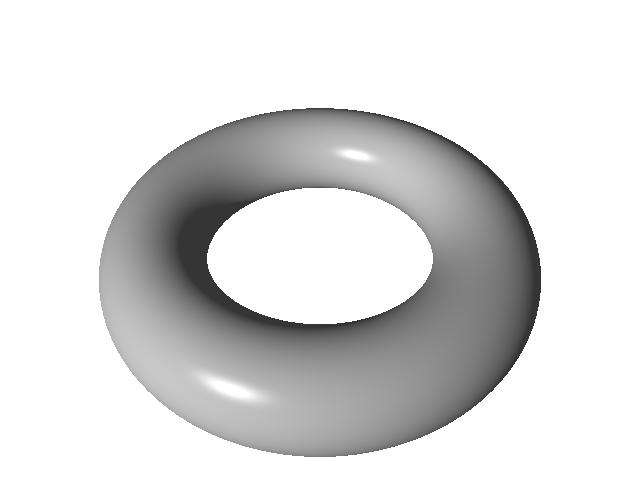}\hspace{2mm}
&\hspace{2mm}\includegraphics[width=1cm]{sphere.jpg}\hspace{2mm}
&\hspace{2mm}\includegraphics[width=1cm]{tore.jpg}\hspace{2mm}\\
&(2D)&(2D)&(surface)&(surface)&(volume)&(volume)\\
$\chi$ & 1 & 0 & 2 & 0 & 1 & 0\\
$g$ & - & - & 0 & 1 & - & - \\ 
$g_c$ & - & - & -1 & 0 & - & - \\ 
\label{tab:genus}
\end{tabular}
\caption{Values of the Euler characteristic and the different definitions of the genus, for basic topologies.}
 \end{center}
\end{table}
In this paper, we will focus on the Euler characteristic of the excursion sets, i.e. the volumes above a given threshold. This approach is equivalent but nonetheless different from looking at the genus of the isocontour surface. However, the difference only affects the prefactor since, for a 3D field, the Euler characteristic of the isocontours is simply twice the Euler characteristic of the corresponding excursion sets \citep{Mecke}:
$ \chi(\partial M) = \chi(M) \left[1+ (-1)^{d-1} \right] $ where $d=\mathrm{dim}(M)$.
To compute the Euler characteristic of a random field, we use a central result from Morse theory \citep{jost}. Morse theory makes the link between the topology of a manifold and the critical points of the functions defined upon it. It shows that the Euler characteristic is simply equal to the alternating sum of the number of critical points of index $i$ (i.e. where the Hessian has $i$ negative eigenvalues):
$ \chi = \sum_i (-1)^i n_i .  $
For a two-dimensional random field,  the Euler characteristic $\chi(\nu)$ 
of the excursion set $x > \nu$ 
is given by:
\begin{equation} \chi(\nu) = n_{\rm max}(\nu) -n_{\rm sad}(\nu) +n_{\rm min}(\nu) \,, 
\label{eq:EC2D}
\end{equation}
where $n_{\rm max}(\nu) $, $n_{\rm sad}(\nu) $ and $n_{\rm min}(\nu) $ are the number of maxima, saddle points and minima above the excursion threshold, $\nu$. The density of Euler characteristic
can thus be computed by integrating the determinant of the Hessian, $I_2$:
\begin{equation}
\chi(\nu) = \frac{1}{{R_*^2}} \int_{-\infty}^\infty {\rm d} J_1 
\int_{ \frac{\nu+\gamma J_1}{\sqrt{1-\gamma^2}}}^{\infty} {\rm d} \zeta
\int_0^\infty {\rm d} J_2 P_\mathrm{ext} (\zeta, J_1, J_2) I_2  ,
\label{eq:euler2D_I2}
\end{equation}
 where $P_\mathrm{ext}$ is the distribution function under the condition of zero gradient:
\begin{equation}  
P_\mathrm{ext}(\zeta, J_1, J_2) =  \int_0^{+\infty} {\rm d}q^2 P_{\rm 2D}(\zeta, q^2, J_1, J_2) \delta_D (x_1) \delta_D (x_2).
\end{equation}
As demonstrated in Appendix~\ref{sec:deriveuler2D}, this integration yields:
\begin{multline}
\chi(\nu) = 
\frac{1}{2 {}} \mathrm{Erfc} \left(\frac{\nu}{\sqrt{2}} \right)
\chi(-\infty) +
\frac{1}{4 \pi \sqrt{2 \pi} {R_*^2}}
\exp\left(-\frac{\nu^2}{2}\right) \times
\left[\vphantom{ \sum_{i,j,k}^{i+2 j+k=n} }
\gamma^2 H_1(\nu) +
\right. \\
+ \left.
\sum_{n=3}^\infty 
\sum_{i,j,k}^{i+2 j+k=n} 
\frac{(-1)^{j+k}}{i!\; j!} 
\left\langle {\zeta^i q^2}^j {J_1}^k \right\rangle_\mathrm{{\scriptscriptstyle GC}}
\left( 1-\gamma^2 \right)^{{i}/{2}}
\sum_{s=0}^{\mathrm{min}(2,k)} 
\frac{ 2! \gamma^{k+2-2s}}{s!(2-s)!(k-s)!} 
H_{i+k+1-2s} \left(\nu\right) 
\right. 
 \\
+\left.  \sum_{n=3}^\infty \sum_{i,j,k}^{i+2 j+k+2=n} 
\frac{(-1)^{j+k+1}}{i!\; j!\; k!} 
\left\langle \zeta^i {q^2}^j {J_1}^k {J_2} \right\rangle_\mathrm{{\scriptscriptstyle GC}}
\left(1-\gamma^2\right)^{{i}/{2}} \gamma^k
H_{i+k-1}\left(\nu\right)
\right].
\label{eq:genus_2D}
\end{multline}
Note that in the rare event  limit, $\nu\gg 1$ or $\nu\ll 1$, equation~(\ref{eq:genus_2D}) represents resp. the differential number of maxima or  minima, 
as the contribution to equation~(\ref{eq:EC2D}) from other extremal points becomes negligible \citep{Aldous}.
In this limit,  equation~(\ref{eq:genus_2D})  yields both quantities to all orders.

A comment  is in order on the value of the Euler characteristic density evaluated at infinitely low threshold, 
$\chi(-\infty)$.  In a finite, topologically non-trivial space\footnote{The only topologically
non-trivial space in 2D that is also maximally symmetric, i.e., homogeneous and isotropic, is the two-sphere
$S^2$, the case relevant to CMB studies. Due to the curvature of $S^2$, the
invariant second derivatives are expressed via covariant derivative tensor with mixed components,
${x_{;i}}^{;j}$, $J_1 = {x_{;i}}^{;i}$ and $ J_2 = J_1^2 - 4 |{x_{;i}}^{;j}|$.  In all spaces
where the global isotropy is broken, as, for example, 
in all topologically compactified flat spaces, like torus, the invariant formalism gives an approximation
which is more accurate smaller the $R_*$ scale is relative to the
size of the manifold.}
$\chi(-\infty)$ evaluates to the Euler characteristic of the manifold itself, divided by its volume.
While the result of Morse theory in equation~(\ref{eq:EC2D}) is exact, the expression via the local
estimate of the extrema densities, equation~(\ref{eq:euler2D_I2}), does not contain this effect as
$\langle I_2 \rangle = 0$. The expansion in equation~(\ref{eq:genus_2D}) shows the appearance
of the boundary term $\frac{1}{2} \mathrm{Erfc}(\nu) \chi(-\infty)$ when the moments of the form 
$\langle J_1^2 q^{2l} \rangle - \langle J_2 q^{2l}\rangle$ 
do no cancel exactly. We conjecture that such boundary term 
 accounts for the topology of the manifold on which the field is considered.
It might perhaps be one of the most important  applications  for   the genus statistics of  masked data sets 
with the mask introducing non-trivial topology. 

When the non-Gaussianities give a hierarchy of cumulants such that a cumulant of $n$ fields is of order
 $\sigma^{n-2}$ (notably for gravitational evolution), the Gram-Charlier series can be reordered as the series in powers of  $\sigma$ (called an Edgeworth expansion, see Appendix~\ref{sec:Edgeworth}):
\begin{equation}
 \chi(\nu) =\frac{1}{2 } \mathrm{Erfc} \left(\frac{\nu}{\sqrt{2}} \right)
\chi(-\infty) +
\frac{1}{4 \pi \sqrt{2 \pi}{R_*^2}}
\exp\left(-\frac{\nu^2}{2}\right) \left( \gamma^2 H_1(\nu) + \sum_{n=1}^\infty \chi^{(n)}(\nu) \right)\,,
\end{equation} 
with $\chi^{(n)} \sim \sigma^n$. The first orders of this  expansion in powers of $\sigma$ are (see Appendix~\ref{sec:sigma2}):
\begin{eqnarray}
 \chi^{(1)}(\nu)&=&\left( 2 \gamma  \left\langle q^2 I_1\right\rangle +4 \left\langle x  I_2\right\rangle  \right) H_0(\nu)-\left( \gamma ^2 \left\langle x q^2 \right\rangle + \gamma  \left\langle x ^2 I_1\right\rangle \right) H_2(\nu)+\frac{\gamma^2}{6} \left\langle x ^3\right\rangle  H_4(\nu), \label{eq:euler2Dsigma}
 \\
  \chi^{(2)}(\nu)&=&\left( \frac{\gamma^2}{2} \langle q^4 \rangle_\mathrm{c} + 2\gamma \langle x q^2 J_1 \rangle_\mathrm{c} + 2\langle x^2 I_2 \rangle_\mathrm{c}  \right) H_1(\nu) - \left( \frac{\gamma^2}{2} \langle x^2 q^2 \rangle_\mathrm{c}  + \frac{\gamma}{3} \langle x^3 J_1 \rangle_\mathrm{c} \right) H_3(\nu) + \frac{\gamma^2}{24} \langle x^4 \rangle_\mathrm{c} H_5(\nu) \nonumber
\\&&- \left( 4 \gamma \langle x q^2 \rangle \langle q^2 J_1 \rangle + \langle q^2 J_1 \rangle \langle x^2 J_1 \rangle + 4\langle x q^2 \rangle \langle x I_2 \rangle \right) H_1(\nu) 
- \left( \frac{\gamma}{6} \langle x q^2 \rangle \langle x^3 \rangle + \frac{\gamma}{6} \langle x^3 \rangle \langle x^2 J_1 \rangle \right) H_5(\nu) + \frac{\gamma^2}{72} \langle x^3 \rangle^2 H_7(\nu)\nonumber \\
&&+ \left(\gamma^2 \langle x q^2 \rangle^2 + \frac{\gamma}{3} \langle x^3 \rangle \langle q^2 J_1 \rangle + \gamma \langle x q^2 \rangle \langle x^2 J_1 \rangle + \frac{1}{4} \langle x^2 J_1 \rangle^2 + \frac{2}{3} \langle x^3 \rangle \langle x I_2 \rangle \right) H_3(\nu) ,
\end{eqnarray}
while the $\sigma^3$ correction, $\chi^{(3)}(\nu)$, is given in
Appendix~\ref{sec:sigma2} by 
equation~(\ref{eq:defchi3}).  
\begin{figure}
 \begin{center}
  \includegraphics[width=0.45\textwidth]{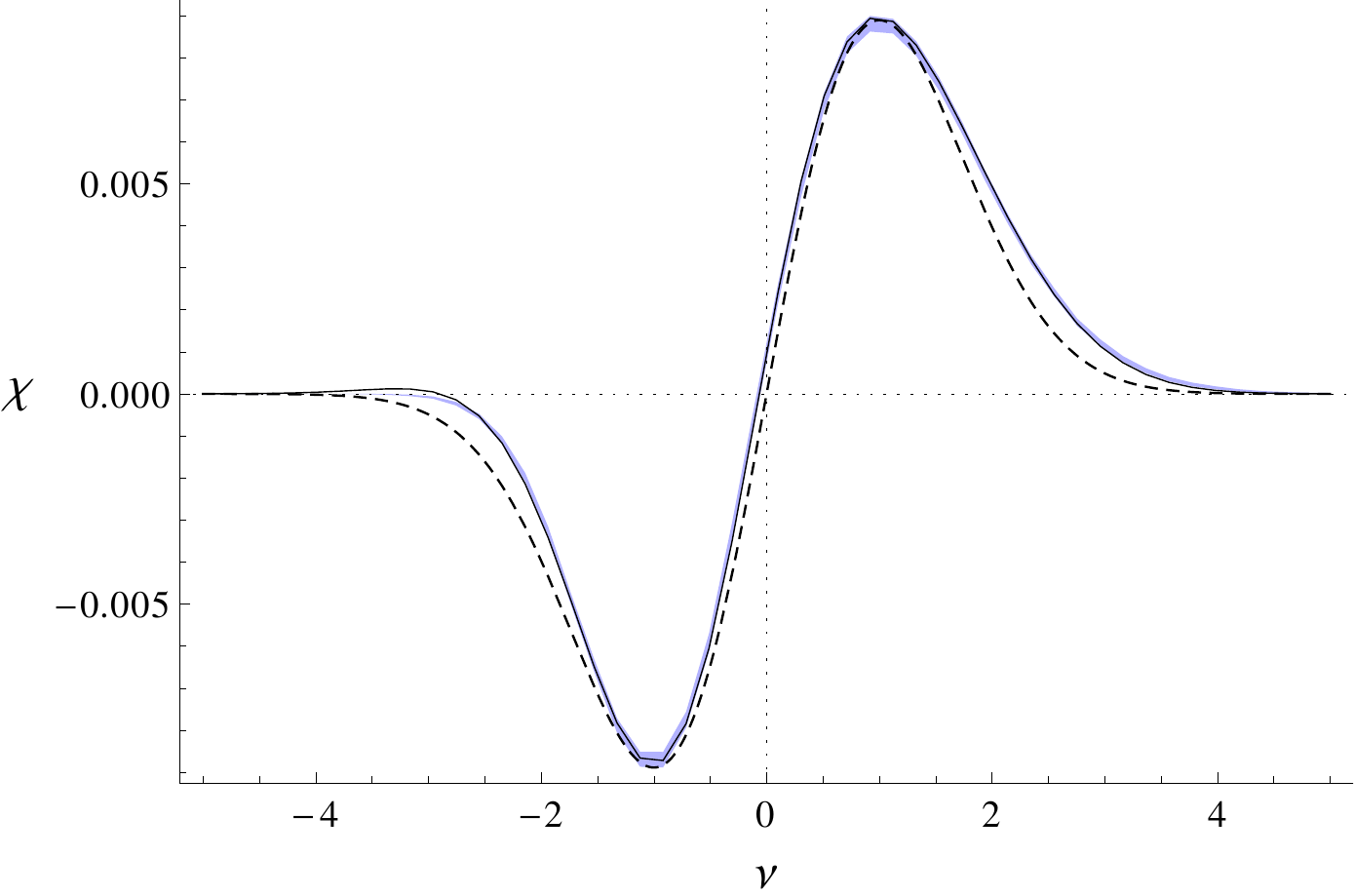}\hskip 0.25cm
\includegraphics[width=0.45\textwidth]{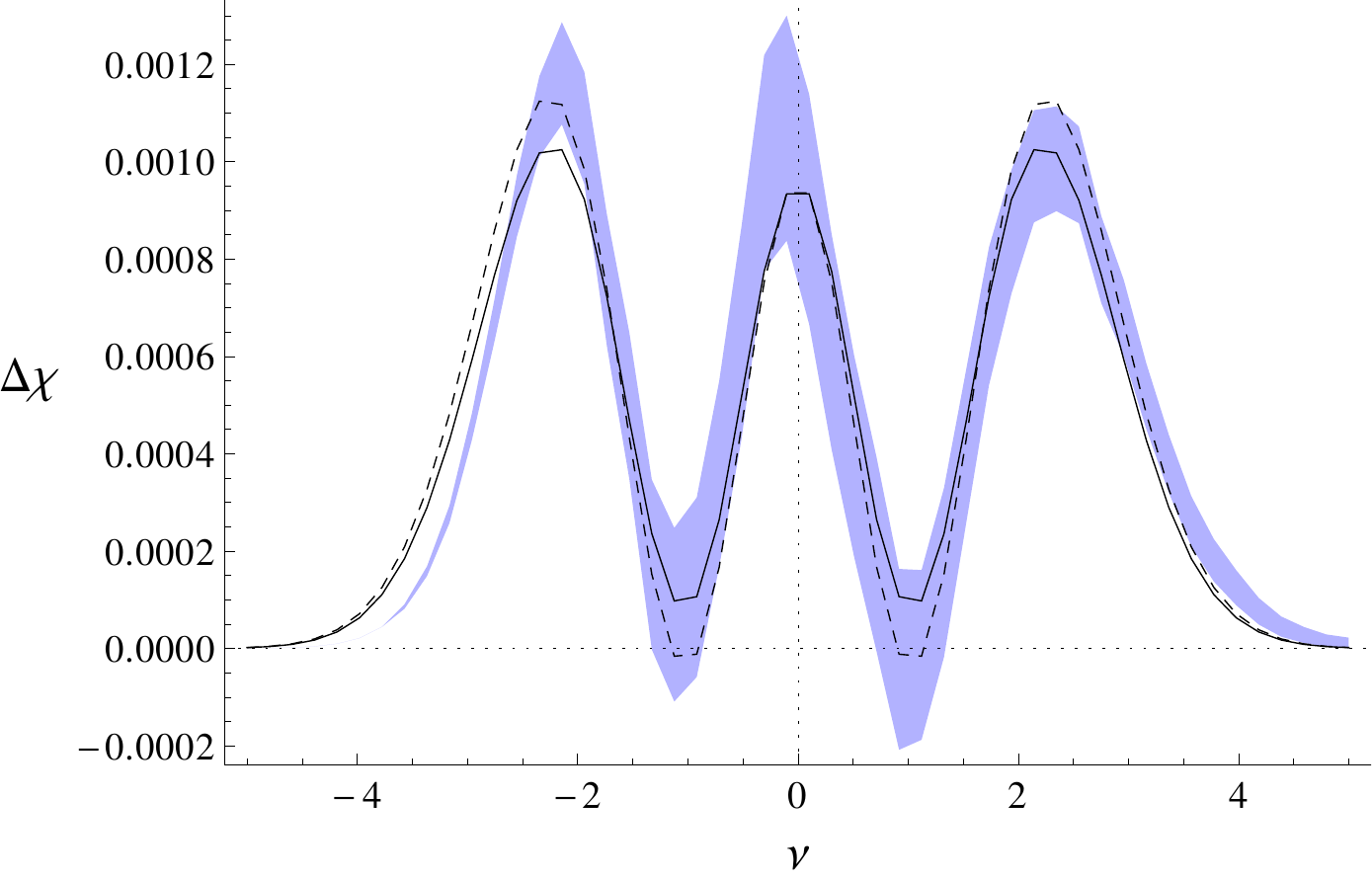}
\caption{\emph{Left:} Euler characteristic in 2D slices of a mildly non-Gaussian
field (gravitational clustering, $\sigma=0.1$). \emph{Dashed line:} Gaussian
approximation, \emph{solid line:} first order prediction using measured
cumulants (following the prescription described in
Section~\ref{sec:measurements}).  The narrow shaded band  corresponds to the
3-sigma dispersion for the measurement of the Euler characteristic in the
simulated fields.
 \emph{Right:} The residual between the
non-Gaussian Euler characteristic and the Gaussian approximation. 
The solid line and the shaded band are the same as in the left panel, minus
the Gaussian contribution.
While the solid line is obtained by using the cumulants measured in the
simulations, the dashed line uses the cumulants as predicted by the
perturbation theory. 
\label{fig:euler2D}
}
 \end{center}
\end{figure}
Figure~\ref{fig:euler2D} compares the first-order
non-Gaussian prediction to simulations of non-Gaussian density field developed
through non-linear gravitational instability. As described in detail in
section~\ref{sec:measurements}, we performed 25 scale-invariant
($n_\mathrm{s}=-1$) $256^3$ N-body simulations which are then cut into 25 slices
each to form 2D fields. The cubic cumulants required to form the non-Gaussian
prediction  (equation \ref{eq:euler2Dsigma}) are then measured from these slices
and the 2D Euler
characteristic is determined using the  suite of algorithms described in
Appendix~\ref{sec:euleralgorithm}. The agreement between the measured Euler
characteristic and the theory is remarkable. Perhaps the most striking fact is
that the non-Gaussian correction indeed arises as a combination
of low order Hermite polynomials.
Alternatively, in this  model the cumulants can be predicted by  perturbation
theory (PT, see section~\ref{sec:PT}). The right panel in
Figure~\ref{fig:euler2D} demonstrates that both approaches give consistent
results\footnote{For illustrative
purposes we restrict the comparison of our all-order expansion
(\ref{eq:genus_2D}) to simulations to first order non-Gaussian
corrections as they can be predicted from first order PT. The second order
terms have been studied in the perturbation theory in \cite{Matsubara2010}.}.
%

\subsubsection{Non-Gaussian total and differential extrema counts in 2D }\label{sec:count2D}
Extrema counts, especially that of the maxima of the field, also have long application to cosmology
\cite[e.g.][]{BBKS},
however  theoretical development  have been mostly restricted
to  Gaussian fields. Given the knowledge of the JPDF (equation~(\ref{eq:2DP_general}), non-Gaussian
extrema can be computed in the similar way as the Euler characteristic
\citep{PPG2011}. For example, the density number of all critical points above
the threshold $\nu$ is 
\begin{equation}
n(\nu) = n_{\mathrm max}(\nu)+n_{\mathrm sad}(\nu) + n_{\mathrm min}(\nu) =
\frac{1}{{R_*^2}} \int_{-\infty}^\infty {\rm d} J_1 
\int_{\frac{\nu+\gamma J_1}{\sqrt{1-\gamma^2}}}^{\infty} {\rm d} \zeta
\int_0^\infty {\rm d} J_2 P_\mathrm{ext} (\zeta, J_1, J_2) |I_2|~,  
\end{equation}
the difference from equation~(\ref{eq:euler2D_I2}) only involve taking the absolute
value of the Hessian determinant $I_2$.
Different types of critical points can be considered separately by restraining
the integration domain
in the $J_1$-$J_2$ plane to ensure the appropriate signs for the eigenvalues.
Specifically, for maxima the integration
range is $J_1 \in [-\infty,0]$, $J_2 \in [0,J_1^2]$, for minima it is $J_1 \in [0, \infty]$, $J_2 \in [0,J_1^2]$,
and for saddle points $J_1 \in [-\infty, \infty]$, $J_2 \in [J_1^2, \infty]$.

The total density number of 
extrema is obtained by integrating over threshold $\nu$.
The effect of the first (cubic) non-Gaussian correction on extrema density
in 2D is given by
\begin{eqnarray}
n_{\rm max} & = & \frac{1}{8 \sqrt{3} \pi {R_*^2}} + 
\frac{18 \left\langle q^2 J_1 \right\rangle - 5\left\langle J_1^3 \right\rangle
+ 6 \left\langle J_1 J_2 \right\rangle}{54 \pi \sqrt{2\pi} {R_*^2}}
\,, \\
n_{\rm sad} &=& \frac{1}{4 \sqrt{3} \pi{R_*^2}} 
+ \quad\quad\quad\quad\quad\quad 0 \quad\quad\quad\quad\quad\quad~
\,,\\
n_{\rm min} & = & \frac{1}{8 \sqrt{3} \pi{R_*^2}} - 
\frac{18 \left\langle q^2 J_1 \right\rangle - 5\left\langle J_1^3 \right\rangle
+ 6 \left\langle J_1 J_2 \right\rangle}{54 \pi \sqrt{2\pi}{R_*^2}}
\,.
\end{eqnarray}
The total number of saddles, as well as the 
total of all the extremal points are preserved in the first
order (the latter following for the former, as $n_{\rm max}-n_{\rm sad}+n_{\rm min}$ is the Euler characteristic of the manifold and is thus constant).

The differential distribution of the number of extrema with the threshold
can be also found analytically in 2D:
\begin{eqnarray}
\frac{\partial n_{\rm max}}{\partial \nu} &=&
\frac{1}{\sqrt{2 \pi}{R_*^2}} \exp\left(-\frac{\nu^2}{2}\right) 
\left[1 + \mathrm{erf}\left(\frac{\gamma \nu}{\sqrt{2(1-\gamma^2)}}\right)
\right] K_1(\nu,\gamma)
+ \frac{1}{\sqrt{2 \pi (1-\gamma^2)}{R_*^2}}
\exp\left(-\frac{\nu^2}{2(1-\gamma^2)}\right) K_3(\nu,\gamma)
\nonumber \\
&& + \frac{1}{\sqrt{2 \pi (1-2/3\gamma^2)}{R_*^2}}
\exp\left(-\frac{3 \nu^2}{6-4 \gamma^2}\right) \left[1 + 
\mathrm{erf}\left(\frac{\gamma \nu}{\sqrt{2(1-\gamma^2)(3-2\gamma^2)}}\right)
\right] K_2(\nu,\gamma), \\
\frac{\partial n_{\rm min}}{\partial \nu}&=&
\frac{1}{\sqrt{2\pi}{R_*^2}} \exp\left(-\frac{\nu^2}{2}\right) 
\left[1 - \mathrm{erf}\left(\frac{\gamma \nu}{\sqrt{2(1-\gamma^2)}}\right)
\right] K_1(\nu,\gamma)
- \frac{1}{\sqrt{2 \pi (1-\gamma^2)}{R_*^2}}
\exp\left(-\frac{\nu^2}{2(1-\gamma^2)}\right) K_3(\nu,\gamma)
\nonumber \\
&& + \frac{1}{\sqrt{2 \pi (1-2/3\gamma^2)}{R_*^2}}
\exp\left(-\frac{3 \nu^2}{6-4 \gamma^2}\right) \left[1 - 
\mathrm{erf}\left(\frac{\gamma \nu}{\sqrt{2(1-\gamma^2)(3-2\gamma^2)}}\right)
\right] K_2(\nu,\gamma)\,, \\
\frac{\partial n_{\rm sad}}{\partial \nu} &=&
\frac{2}{\sqrt{2 \pi (1-2/3\gamma^2)}{R_*^2}}
\exp\left(-\frac{3 \nu^2}{6-4 \gamma^2}\right) K_2(\nu,\gamma), \label{eq:number_counts_2D}
\end{eqnarray}
where $K_1, K_2, K_3$ are polynomials in $\nu$ with coefficients given 
by the cumulants and $\gamma$.  Since
\begin{equation}
\frac{\partial n_{\rm max}}{\partial \nu} + \frac{\partial n_{\rm min}}{\partial \nu} -\frac{\partial n_{\rm sad}}{\partial \nu}
= - \frac{\partial}{\partial\nu} \chi (\nu) = 
\frac{2}{\sqrt{2\pi}{R_*^2}} \exp\left(-\frac{\nu^2}{2}\right) K_1(\nu,\gamma), \label{eq:diffchi}
\end{equation}
we immediately find that $K_1$ is related to the series that appear
in the expression for the Euler characteristic, equations (\ref{eq:genus_2D}) and (\ref{eq:euler2Dsigma}).
Specifically, $K_1$ reads
\begin{multline}
K_1(\nu,\gamma) =
\frac{\gamma^2}{8\pi}
\left[ H_2(\nu) + \left( \frac{2}{\gamma} \left\langle q^2 J_1
\right\rangle + 
\frac{1}{\gamma^2} \left\langle x {J_1}^2 \right\rangle - 
\frac{1}{\gamma^2} \left\langle x J_2 \right\rangle \right) H_1(\nu)
- \left( \left\langle x q^2 \right\rangle + 
\frac{1}{\gamma} \left\langle x^2 J_1 \right\rangle \right) H_3(\nu)
+ \frac{1}{6} \left\langle x^3 \right\rangle H_5(\nu) \right]\,.
\label{eq:2D_K3}
\end{multline}
In the limit $\nu \gg 1$, the 3 terms are dominated by their respective exponentials.
 Since $2(1-\gamma^2)$ and $(6-4\gamma^2)/3$ are smaller than 1, the $K_2$ and $K_3$ terms decrease more quickly
 than the $K_1$ term. Thus in the limit of large $\nu$ the expression of the density of maxima is dominated by
 the $K_1$ term and coincides with the expression of the Euler characteristic, as expected.

\begin{figure}
 \begin{center}
   \includegraphics[width=0.49\textwidth]{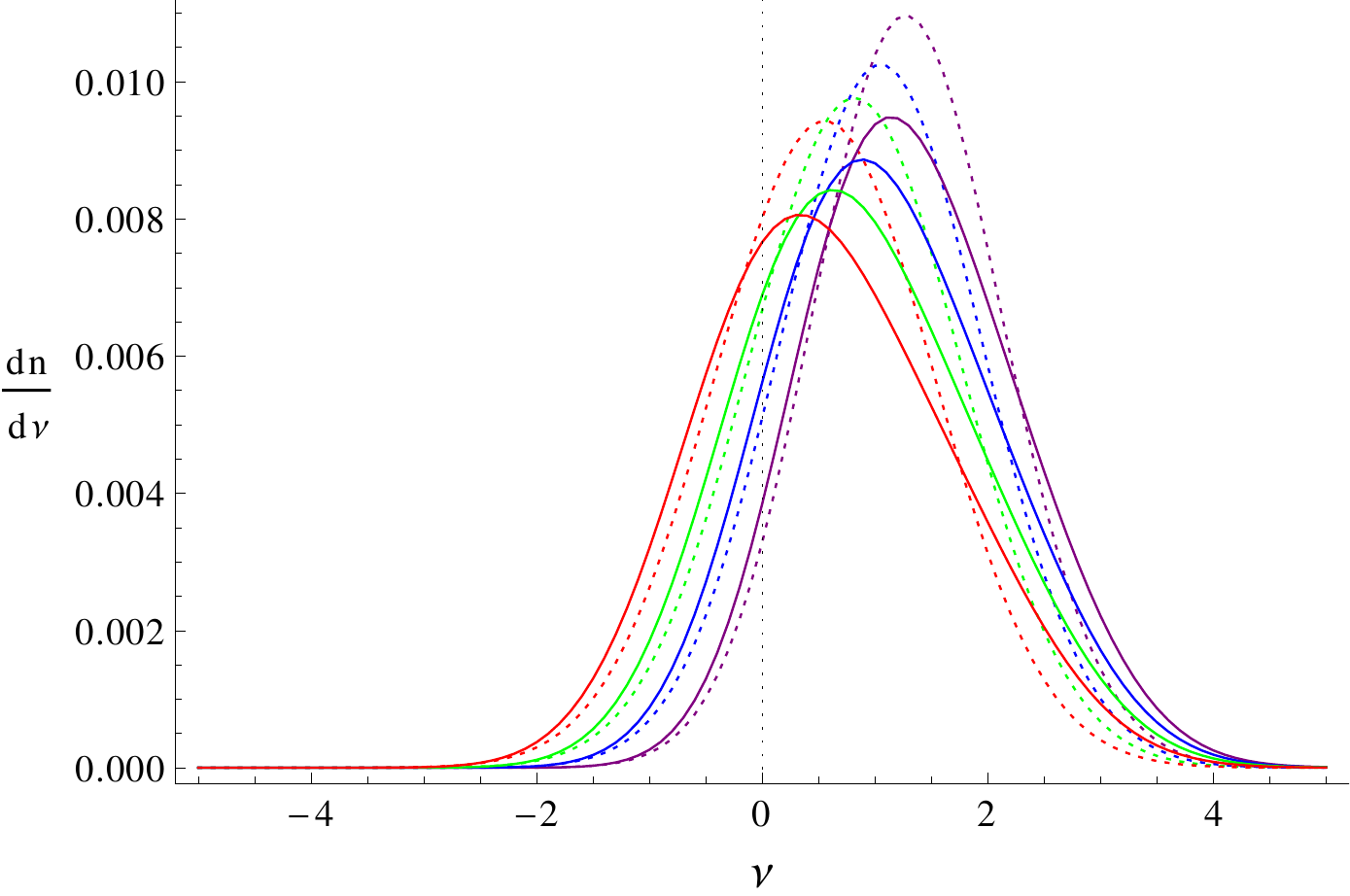}
  \includegraphics[width=0.49\textwidth]{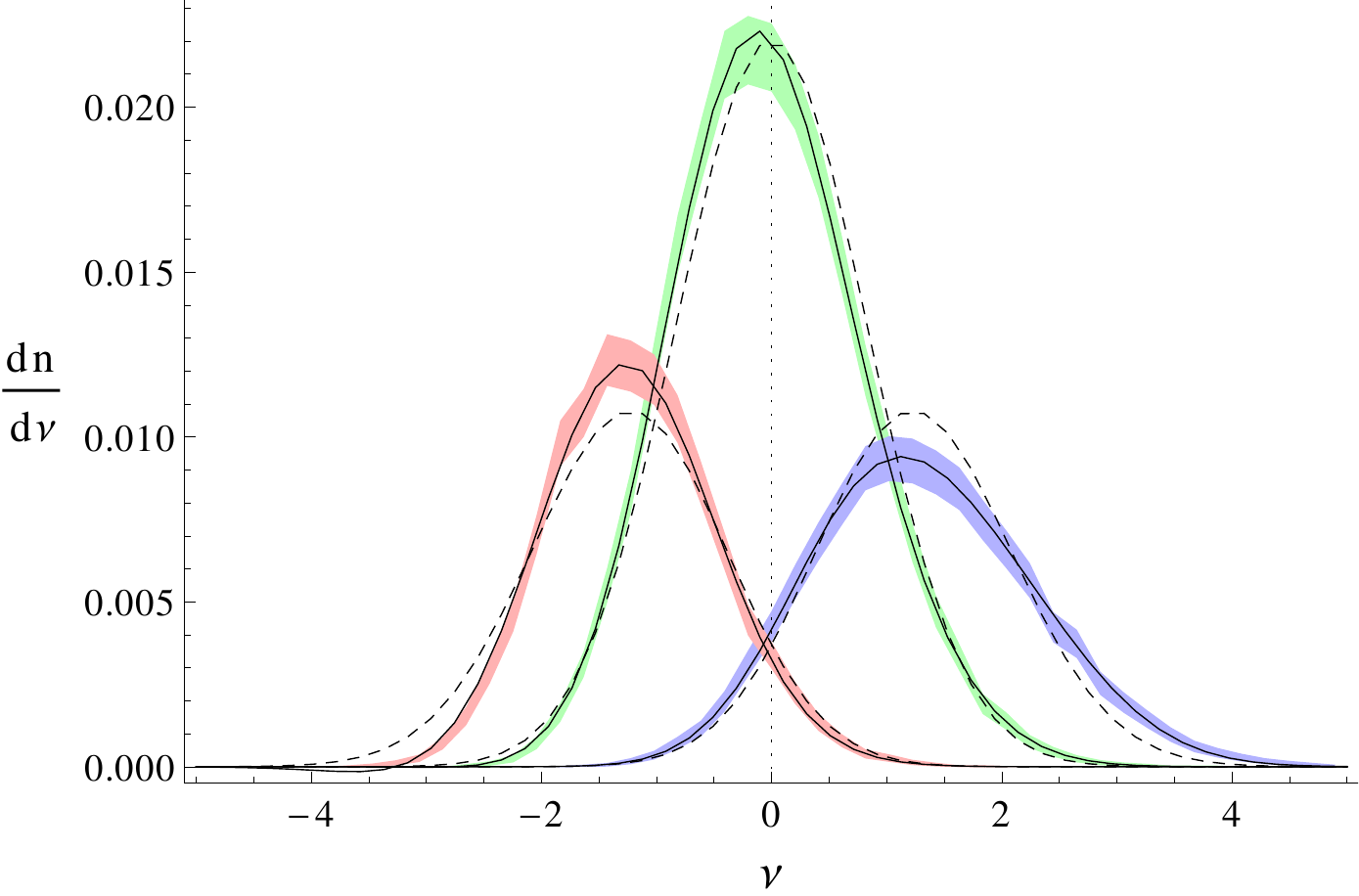}
\end{center}
\caption{{\sl Left}: differential distribution of maxima in 2D slices of 3D
density field evolved to $\sigma=0.1$ for several values of the initial spectral
index (\emph{red:} $n_\mathrm{s}=-1.8$, \emph{green:} $n_\mathrm{s}=-1.5$,
\emph{blue:} $n_\mathrm{s}=-1$,
\emph{purple:} $n_\mathrm{s}=0$).
{\sl Right:} Measured maxima (\emph{blue}), saddle point (\emph{green})
 and minima (\emph{red}) distributions for $n_\mathrm{s}=-1$  versus
Gaussian (dashed lines) and the first order non-Gaussian (solid lines)
predictions.
The shaded region corresponds to the 3-sigma dispersion within the simulation
set.
}
\label{fig:extremanongaussgamma}
\label{fig:extrema2D}
\end{figure}

To write the remaining $K_2$ and $K_3$ more compactly, we introduce two
types of the scaled Hermite polynomials
${\cal H}_n^+(\nu,k)\equiv k^{n} H_n\left(\nu/k\right)$ 
and ${\cal H}_n^-(\nu,k)\equiv k^{-n} H_n\left(\nu/k\right)$,
where the choice of $k$ will be dictated by the exponential prefactor in
the correspondent term. 
The polynomial $K_2$ is given by the straightforward
expansion in ${\cal H}_n^-(\nu, \sqrt{1-2/3\gamma^2})$
\begin{multline}
K_2(\nu,\gamma) =\frac{1}{8\pi \sqrt{3}}
\left[\vphantom{\frac{1}{6}} 1  \right.
- \left( \left\langle x q^2 \right\rangle 
+ \frac{2}{3} \gamma \left\langle q^2 J_1 \right\rangle 
+ \frac{1}{3} \left\langle x {J_1}^2 \right\rangle
- \frac{4}{3} \left\langle x J_2 \right\rangle
+ \frac{2}{9} \gamma \left\langle {J_1}^3 \right\rangle  
- \frac{2}{3} \gamma \left\langle J_1 J_2 \right\rangle \right)
{\cal H}_1^- \left(\nu,\sqrt{1-2/3\gamma^2} \right)
\\
\left. + \frac{1}{6} \left(
\left\langle \nu^3 \right\rangle
+ 2 \gamma \left\langle x^2 J_1 \right\rangle
+ \frac{4}{3} \gamma^2 \left\langle x {J_1}^2 \right\rangle
+ \frac{2}{3} \gamma^2 \left\langle x J_2 \right\rangle
+ \frac{8}{27} \gamma^3 \left\langle {J_1}^3 \right\rangle
+ \frac{4}{9} \gamma^3 \left\langle J_1 J_2 \right\rangle
\right)
{\cal H}_3^- \left(\nu,\sqrt{1-2/3\gamma^2} \right) \right].
\label{eq:2D_K2}
\end{multline}
The knowledge of $K_1$ and $K_2$ is sufficient to describe 
the differential number density of the sum of extrema of all the types
\begin{equation}
\frac{\partial n_{\rm max}}{\partial \nu} +
\frac{\partial n_{\rm min}}{\partial \nu} +
\frac{\partial n_{\rm sad}}{\partial \nu} =
\frac{2}{\sqrt{2 \pi}{R_*^2}} \exp\left(-\frac{\nu^2}{2}\right) K_1(\nu,\gamma)
+
\frac{4}{\sqrt{2\pi (1-2/3 \gamma^2)}{R_*^2}}
\exp\left(-\frac{3 \nu^2}{6-4 \gamma^2}\right) K_2(\nu,\gamma) \, .
\end{equation}
The remaining term, $K_3(\nu)$ is the most complicated. It is the simplest to
express it as the expansion in ${\cal H}_n^+(\nu,\sqrt{1-\gamma^2})$ 
but the expression is still quite long
\begin{multline}
K_3(\nu,\gamma) =\frac{\gamma (1-\gamma^2)}{2 (2 \pi)^{3/2}}
\left[\vphantom{\frac{1}{6}} {\cal H}_1 \left(\nu,\sqrt{1-\gamma^2}\right)
+ \right.
\\
\left(\frac{\left(1-\gamma^2\right)\left(1+2 \gamma ^2\right)}{3-2 \gamma ^2}
\left\langle x q^2 \right \rangle 
+\frac{\gamma^2 \left(1+\gamma^2-26 \gamma^4 + 28 \gamma^6 - 8 \gamma^8 \right)}
{2 \left(3-2 \gamma ^2\right)^3}
\left\langle x ^3\right \rangle
+ \frac{4 (1-\gamma^2)} {\gamma \left(3 -2 \gamma ^2 \right)}
\left\langle q^2 J_ 1\right \rangle
-\frac{2 \gamma^3 \left(13-14 \gamma^2+4 \gamma^4\right)}
{\left(3-2 \gamma^2\right)^3}
\left\langle x^2 J_1 \right\rangle
\right. \\
\left. 
-\frac{\left(24-26 \gamma^2+8 \gamma^4\right)}{\left(3-2 \gamma^2\right)^3}
\left\langle x  J_1^2\right \rangle
-\frac{2 \left(5-6 \gamma^2+2 \gamma^4\right)}
{\gamma \left(3-2 \gamma^2\right)^3}
\left\langle J_1^3\right \rangle
+\frac{\left(15-23 \gamma^2 + 8 \gamma^4\right)}{\left(3-2 \gamma^2\right)^3}
\left\langle x J_2\right \rangle
+\frac{6 \left(1-\gamma^2\right) \left(2-\gamma^2\right)}
{\gamma  \left(3-2 \gamma^2\right)^3}
\left\langle J_1 J_2\right \rangle \right)
H_0^+\left(\nu,\sqrt{1-\gamma^2}\right)
\\
 +\left(-\left\langle x q^2 \right \rangle 
+\frac{\left(-27-36 \gamma^2+224 \gamma^4-192 \gamma^6+48 \gamma^8\right)}
{6 \left(3-2 \gamma ^2\right)^3}
\left\langle x^3\right \rangle 
+\frac{2 \left(-9+27 \gamma^2-18 \gamma^4+4\gamma ^6\right)}
{\gamma \left(3-2 \gamma^2\right)^3}
\left\langle x^2 J_1\right \rangle 
\right.  \\
\left. \left.  +\frac{6}{\left(3-2 \gamma ^2\right)^3}
\left\langle x J_1^2\right \rangle
+\frac{4 \gamma}{3 \left(3-2 \gamma^2\right)^3}
\left\langle J_1^3\right \rangle 
+\frac{3}{\left(3-2 \gamma ^2\right)^3}
\left\langle x J_2\right \rangle
+\frac{2 \gamma} {\left(3-2 \gamma^2\right)^3}
\left\langle J_1 J_2\right \rangle \right)
H_2^+\left(\nu,\sqrt{1-\gamma^2}\right) \right].
\end{multline}
Figure~\ref{fig:extremanongaussgamma} (left panel) shows samples of this
distribution for several values of the spectral parameter $\gamma$ according to
 perturbation theory (see Section~\ref{sec:PT3}).
Figure~\ref{fig:extrema2D} (right panel) shows the very good agreement between
these predictions 
and the measurements of the corresponding number densities in simulations (see
Section~\ref{sec:measurements}).
In 2D, small non-Gaussianities has little effect on the saddle points, slightly
shifting the differential
distribution towards the lower field thresholds, while keeping the total number
of the saddles constant to first order.
In contrast, there are more significant effects on the maxima and
minima. The maxima distribution becomes
broader, developing high $\nu$ excess relative to the Gaussian formula. 
The minima distribution behaves oppositely. It sharpens
around its peak at $\nu \sim -1.5$, while at the same time losing the lowest minima.
This reflects the intuitive picture of 
non-linear gravitational clustering developing high density peaks, while stretching and smoothing low $\nu$ voids.

\subsubsection{The differential length of the skeleton for non-Gaussian fields in 2D}
\label{sec:skel2D}
The skeleton is a tool used to extract the filamentary structure of a field \citep{Novikov2006}. The theory
 of the skeleton for Gaussian fields has been presented in \cite{pogoskel}, which demonstrated that
 its differential length can be approximately (in the so called stiff
approximation) computed as
\begin{equation}\frac{\partial L^\mathrm{skel}}{\partial \nu}
 = \frac{1}{R_*} \int {\rm d}^2 \tilde x_i {\rm d}^3 x_{kl} P(\nu,\tilde x_i,x_{kl}) \delta^\mathrm{D} (\tilde x_2)
 \left| \lambda_2 \right| , \label{eq:skellambda2}
\end{equation}
where the $\tilde x_i$ are the components of the gradient in the Hessian eigenframe.
To separate the part of the skeleton that maps closely the 
high-field value filamentary structures from the other critical lines, in \cite{pogoskel} the notion
of ''primary`` skeleton was introduced, defined by  the constraint
\begin{equation} \lambda_1+\lambda_2 \leq 0.\end{equation}

The skeleton presents an insightful case study of how our invariant formalism can be extended
to the cases in which  the independent rotation symmetry of  the gradient and the Hessian is broken.
The skeleton definition couples the directions of the gradient and the Hessian:
it singles out the direction associated with the highest eigenvalue of the Hessian ($\lambda_1$) 
which is the local direction
of the skeleton. While the components of the
gradient  orthogonal to the skeleton's direction are zero, their parallel component
 should be integrated over.
This does not break the isotropy for the Hessian-related part of the expansion since the integrand
only contains $\lambda_2$ which is a rotation invariant of the Hessian. 
However it means that we cannot independently average over the directions of the gradient and the expansion has
to treat its components separately.

\begin{figure}
 \begin{center}
  \includegraphics[width=0.49\textwidth]{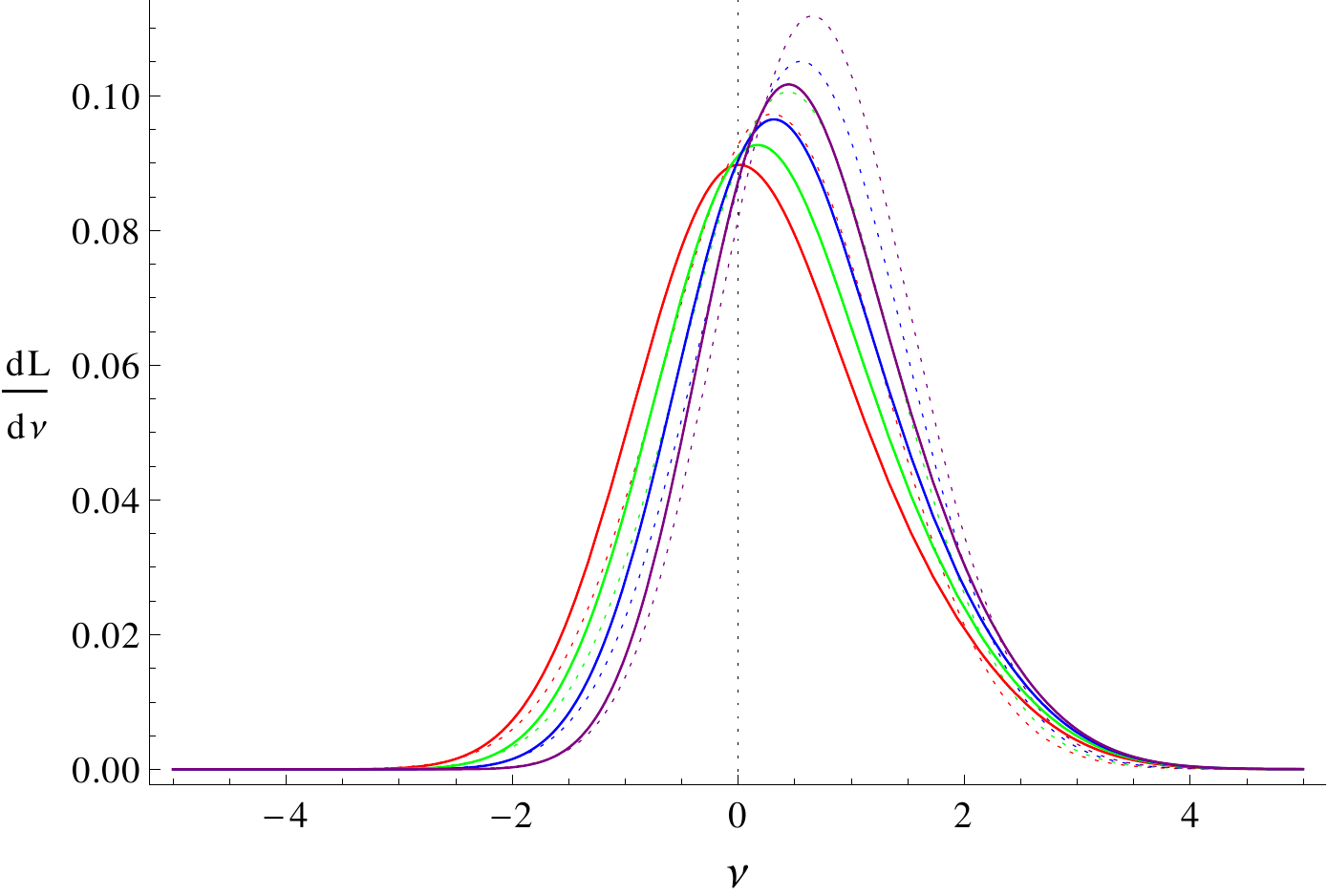}
  \includegraphics[width=0.49\textwidth]{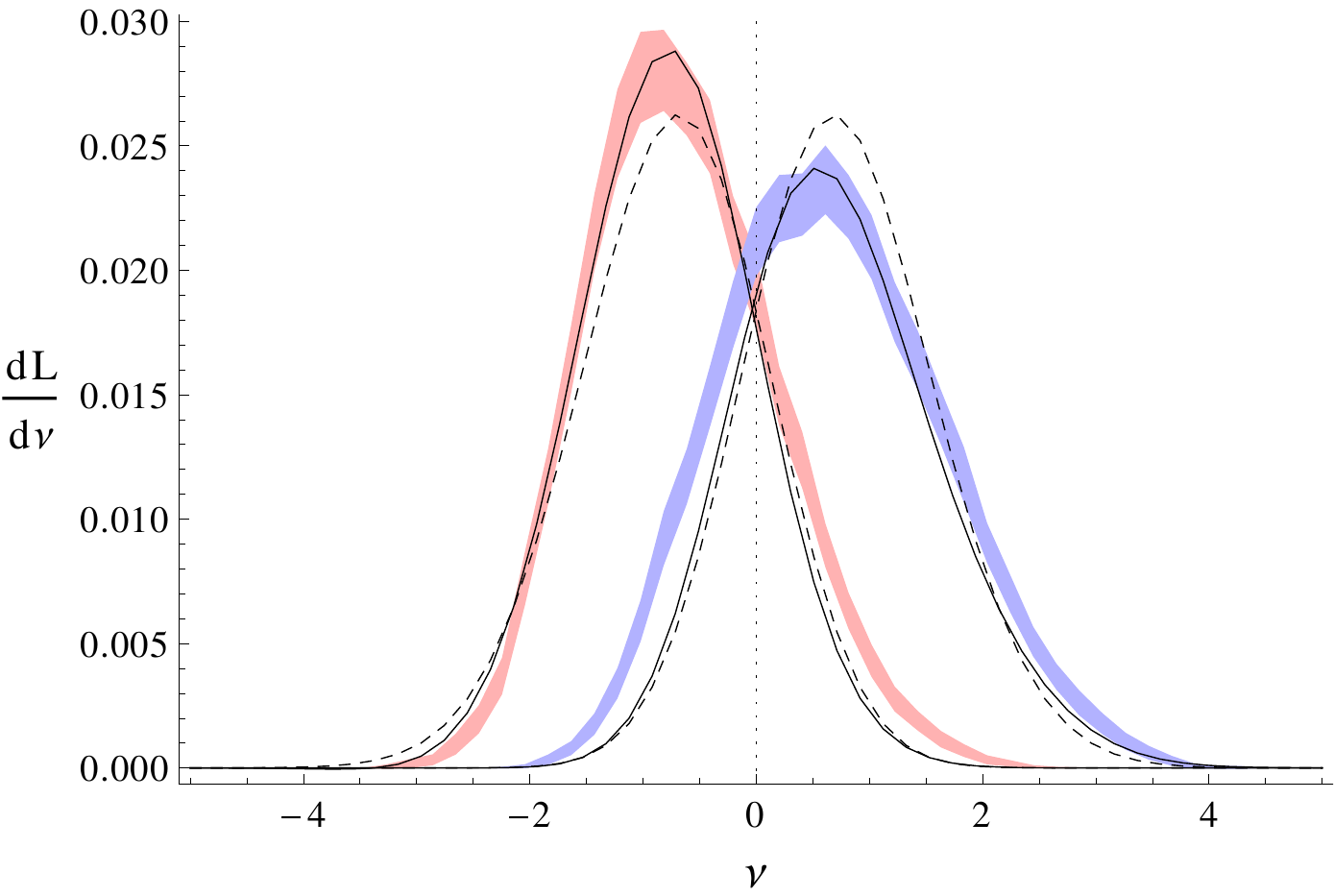}
 \end{center}
\caption{{\sl Left:} 
differential distribution of skeleton length in 2D slices of 3D
density field evolved to $\sigma=0.1$ for several values of the initial spectral
index (\emph{red:} $n_\mathrm{s}=-1.8$, \emph{green:} $n_\mathrm{s}=-1.5$,
\emph{blue:} $n_\mathrm{s}=-1$,
\emph{purple:} $n_\mathrm{s}=0$). The dotted line is the Gaussian approximation,
the solid
line is the first order non-Gaussian prediction. 
{\sl Right:} Measured length distribution of the skeleton (\emph{blue})
 and the anti-skeleton (\emph{red})  for $n_\mathrm{s}=-1$  versus
Gaussian (dashed lines) and the first order non-Gaussian (solid lines) predictions.
The shaded region corresponds to the 3-sigma dispersion within the simulation
set.
}
\label{fig:sklnongaussgamma}
\label{fig:sklnongauss}
\end{figure}

We will first consider that the statistical properties of the components of the gradient are the same in the Hessian eigenframe as in any frame. Appendix~\ref{sec:skelhess} discusses this assumption and shows how to avoid it and that it only has a secondary impact on the results. In practice the gradient part of the Gaussian PDF is
\begin{equation} G(x_1,x_2)=\frac{1}{\pi} \exp({-x_1^2-x_2^2}). \end{equation}
The corresponding orthogonal polynomials needed to build an expansion around the Gaussian kernel
$1/\sqrt{\pi} e^{-x^2}$ are the ``physicists''' Hermite polynomials 
$H^\mathrm{phys}_n(x)=2^\frac{n}{2} H_n(\sqrt{2}x).$
Their orthogonality relation is
\begin{equation} \int_{-\infty}^{+\infty} \frac{1}{\sqrt{\pi}} e^{-x^2} 2^\frac{n}{2} H_n(\sqrt{2}x) 2^\frac{m}{2} H_m(\sqrt{2}x) {\rm d}x= n! \delta_\mathrm{n,m}. \end{equation}
The Gram-Charlier expansion is then
\begin{multline}
P(\zeta, x_1, x_2, J_1, J_2) =  G \Bigg( 1 + 
\sum_{n=3}^\infty \sum_{i,a,b,k,l}^{i+a+b+k+2 l=n} 
\frac{(-1)^{l} 2^\frac{a}{2}  2^\frac{b}{2}}{i!\;a!\;b!\; k!\; l!} 
\left\langle \zeta^i {x_1}^a {x_2}^b {J_1}^k {J_2}^l \right\rangle_\mathrm{{\scriptscriptstyle GC}}
H_i\left(\zeta\right) H_a(\sqrt{2} x_1) H_b(\sqrt{2} x_2)
H_k\left(J_1\right) L_l\left(J_2\right)
\Bigg), \nonumber
\end{multline} with the normalization of the Gram-Charlier coefficients again 
such as to have the highest-degree term to coincide with the corresponding moment:
\begin{equation}
 \langle {x_1}^n \rangle_\mathrm{{\scriptscriptstyle GC}} = \frac{1}{2^\frac{n}{2}} \langle H_n(\sqrt{2} x_1) \rangle. 
\end{equation}
Using the variable $x$ instead of $\zeta$ allows to introduce a threshold more easily. By definition of $\zeta=(x+\gamma J_1)/\sqrt{1-\gamma^2}$, the Jacobian of the transformation is
\(
{\rm d}\zeta {\rm d}J_1 = {1}/{\sqrt{1-\gamma^2}} \; \mathrm{d}x \mathrm{d} J_1
\) and the Gaussian PDF becomes
\begin{equation}
 G(x,x_1,x_2,J_1,J_2)=\frac{1}{2\pi \sqrt{1-\gamma^2} } \exp\left({-\frac{(x+\gamma J_1)^2}{2 (1-\gamma^2)}-{x_1}^2-{x_2}^2 -\frac{{J_1}^2}{2} - J_2 }\right).
\end{equation} The expression of the non-Gaussian PDF in these variables is then
\begin{multline}
P(x, x_1, x_2, J_1, J_2) =  G(x, x_1, x_2, J_1, J_2) \Bigg( 1 + 
\sum_{n=3}^\infty \sum_{i,a,b,k,l}^{i+a+b+k+2 l=n} \\
\frac{(-1)^{l} 2^\frac{a}{2}  2^\frac{b}{2}}{i!\;a!\;b!\; k!\; l!} \frac{1}{\left(1-\gamma^2\right)^{{i}/{2}}}
\left\langle \left(x+\gamma J_1\right)^i {x_1}^a {x_2}^b {J_1}^k {J_2}^l \right\rangle_\mathrm{{\scriptscriptstyle GC}}
H_i\left(\frac{x+\gamma J_1}{\sqrt{1-\gamma^2}}\right) H_a(\sqrt{2} x_1) H_b(\sqrt{2} x_2)
H_k\left(J_1\right) L_l\left(J_2\right)
\Bigg).
\end{multline} 
To express the integrand $|\lambda_2|$ via the Hessian invariants and determine the appropriate limits of integration
we note that
 the skeleton constraint, $J_1=\lambda_1 + \lambda_2 \leq 0$,  implies that $\lambda_2 \leq 0$, 
since  eigenvalues are sorted, $\lambda_1 \geq \lambda_2$ .
The relation between the integrand and the invariants is thus
\begin{equation} 
|\lambda_2|=\frac{1}{2} \left(\sqrt{J_2}-J_1 \right) \, ,
\end{equation}
and the differential length of the skeleton is given by
\begin{equation}
 \frac{\partial L^\mathrm{skel}}{\partial \nu} = \frac{1}{2 R_*} \int_0^{\infty} {\rm d}J_2 \int_{-\infty}^0 {\rm d}J_1 
\int_{-\infty}^{+\infty} {\rm d}x_1  \int_{-\infty}^{+\infty} {\rm d}x_2 \delta^\mathrm{D} (x_2) 
\left(\sqrt{J_2}-J_1 \right) P(x=\nu,x_1,x_2,J_1,J_2).
\end{equation}
The integration over $x_2$ is trivial
%
while the integration over $x_1$ corresponds to the projection on $H_0(\sqrt{2} x_1)$ that selects only the $a=0$ term:
\begin{multline}
 \frac{\partial L^\mathrm{skel}}{\partial \nu} = \frac{1}{2 R_*} \int_0^{+\infty} {\rm d}J_2 \int_{-\infty}^0 {\rm d}J_1
 \left(\sqrt{J_2}-J_1 \right) G(x=\nu,0,J_1,J_2) \Bigg( 1 + \\
\sum_{n=3}^\infty \sum_{i,b,k,l}^{i+b+k+2 l=n} \frac{(-1)^{l} 2^\frac{b}{2} }{i!\;b!\; k!\; l!} 
\left\langle \zeta^i {x_2}^b {J_1}^k {J_2}^l \right\rangle_\mathrm{{\scriptscriptstyle GC}}
H_i\left(\frac{\nu+\gamma J_1}{\sqrt{1-\gamma^2}}\right) H_b(0)
H_k\left(J_1\right) L_l\left(J_2\right) \Bigg)
.
\end{multline}
The result of the remaining integrations takes the form
\begin{equation}
  \frac{\partial L^\mathrm{skel}}{\partial \nu} =  \frac{1}{\sqrt{2\pi} R_*} e^{-\frac{\nu^2}{2}} \left[ A \exp \left(-\frac{\gamma^2 \nu^2}{2(1-\gamma^2)} \right) + B \left( 1+ \mathrm{erf} \left( \frac{\gamma \nu}{\sqrt{2(1-\gamma^2)}} \right) \right) \right]. \label{eq:skel_2D}
\end{equation}
In the Gaussian case, the coefficients $A$ et $B$ were given in \cite{pogoskel}:
%
 \(A^{(0)}= {\sqrt{1-\gamma^2}}/{(2\sqrt{2}\pi)}$ and $B^{(0)}= {1}/{(8\sqrt{\pi})} \left( \sqrt{\pi} + 2 \gamma \nu\right). \)
%
At the $\sigma$ order, the non-Gaussian correction are:
\begin{multline}
 A^{(1)}= \frac{\sqrt{1-\gamma^2}}{2\sqrt{2}\pi} \Bigg[
\frac{ \sqrt{\pi} (-2\gamma^3+\gamma^5)+2(3\gamma^2-5\gamma^4+2\gamma^6) H_1(\nu) - \sqrt{\pi} (3\gamma-3\gamma^3+\gamma^5) H_2(\nu) + 2 (1-\gamma^2)^3 H_3(\nu) }{ 12 (1-\gamma^2)^3 } \langle x^3 \rangle \\
+ \frac{ -\sqrt{\pi} \gamma^2 + 2(2\gamma-3\gamma^3+\gamma^5) H_1(\nu) - \sqrt{\pi} H_2(\nu) }{ 4 (1-\gamma^2)^3 } \langle x^2 J_1 \rangle
+ \frac{ -\sqrt{\pi} \gamma^3 - 2 H_1(\nu) - \sqrt{\pi} \gamma H_2(\nu) }{ 4 (1-\gamma^2)^3 } \langle x {J_1}^2 \rangle \\
+ \frac{ \sqrt{\pi} (1-2\gamma^2) -2 \gamma H_1(\nu) -\sqrt{\pi} \gamma^2 H_2(\nu) }{ 12 (1-\gamma^2)^3} \langle {J_1}^3 \rangle
- \frac{ \sqrt{\pi} \gamma }{4 (1-\gamma^2) } \langle x J_2 \rangle
- \frac{ \sqrt{\pi} }{4 (1-\gamma^2) } \langle J_1 J_2 \rangle \\
+ \left(\frac{ \sqrt{\pi} \gamma}{2(1-\gamma^2)} - H_1(\nu) \right) \langle x {x_2}^2 \rangle
+ \frac{ \sqrt{\pi}}{2(1-\gamma^2)} \langle {x_2}^2 J_1 \rangle
\Bigg]\,,
\end{multline} 
\begin{equation}
 B^{(1)} = \frac{1}{8\sqrt{\pi}} \Bigg[
\left( \frac{\sqrt{\pi}}{6} H_3(\nu) + \frac{\gamma}{3} H_4(\nu) \right) \langle x^3 \rangle
- H_2(\nu) \langle x^2 J_1 \rangle
+ \frac{\sqrt{\pi}}{2} H_1(\nu) \langle x J_2 \rangle
- \left( \sqrt{\pi} H_1(\nu) + 2\gamma H_2(\nu) \right) \langle x {x_2}^2 \rangle
+ 2 \langle {x_2}^2 J_1 \rangle
\Bigg].
\end{equation}
Figure~\ref{fig:sklnongaussgamma} (left panel) displays the variation of this
differential length w.r.t. the underlying power-index, $n_\mathrm{s}$.
The dynamical evolution tends to lower the total length of the skeleton.
 This corresponds to filaments merging. At the same time,  at the very
high density tail $\nu > 2$ notice an increase in
the length of the skeleton per
unit volume, which reflects the compression of the
high-density regions.
These predictions can easily be adapted to the anti-skeleton that maps the
troughs of the field by simply replacing both  fields and cumulants with their opposite,
\emph{i.e.} replacing $\nu$ with $-\nu$ and the odd moments by their opposite.
 Figure~\ref{fig:sklnongauss} (right panel) shows a reasonable agreement between
these predictions and the
 measured differential length.  The discrepancies between the predictions and
the measurements could
 come from the approximation used to predict the differential length  of the skeleton
  \citep[the ``stiff approximation'', as explained in][]{pogoskel} or from differences between the analytical
 definition of the skeleton and its numerical implementation \citep{Sousbie2008}.
Note that it is  the total skeleton length that is most sensitive to a possible mismatch of the exact definition
of the ``primary'' skeleton in the theory versus the techniques used in numerical
simulations, while 
the differential behaviour at high threshold values is much more robust, as was
pointed out in \cite{pogoskel}.
Accordingly, when comparing
equation~(\ref{eq:skel_2D}) to the data in the right panel of
Figure~\ref{fig:sklnongauss}
we correct the overall amplitude by a factor
 $L_{\rm tot, measured}/L_{\rm tot, predicted} =0.23$ for $n_\mathrm{s}=-1$.
As non-Gaussianities distort the skeleton statistics without much changing its
total length,
this empirical correction does not prevent us from estimating $\sigma$
from the shape of the differential length,
as is seen in Figure~\ref{fig:sklnongauss}.
Incidently, in 3D, we found the amplitude to be correctly predicted 
(modulo some change in the condition for the definition of ridges), so no such correction is required (see below Sec.~\ref{sec:skel3D}).

\subsection{Non-Gaussian critical sets in three dimensions}

In 3D dimensions the basis of our consideration is equation~(\ref{eq:3DP_general}). The expressions for the critical sets are somewhat more complex, while cosmology provides
 a context in which we can directly predict 
the value of the relevant cumulants within the framework of gravitational perturbation theory
 (see Sec.~\ref{sec:PT}). 
\subsubsection{The area of 3D isosurfaces}
\label{sec:mink3D}
Let us again start with a 3D Minkowski functional,
the surface, ${\cal A}(\nu)$, of the $x=\nu$ isocontour, which  is given by
\begin{equation}
 {\cal A}(\nu)= \frac{1}{R_0} \int_{-\infty}^{+\infty} {\rm d}x \int_0^{\infty}
{\rm d}q^2 P(x,q^2) \delta^\mathrm{D} (x-\nu) q
=\frac{1}{R_0} \int_0^{\infty} {\rm d}q^2 P(\nu,q^2) q \label{eq:defA}
\end{equation}
where the joint PDF of the field and its gradient, $P(\nu,q^2)$ follows from equation~(\ref{eq:3DP_general})
 after integration over the $J_i$'s:
\begin{equation}
 P(x,q^2)=\frac{3\sqrt{3}}{2\pi} \sqrt{q^2}
\exp\left({-\frac{x^2}{2}-\frac{3}{2} q^2}\right) \left( 1 + \sum_{n=3}^{\infty}
\sum_{i,j}^{i+2j=n} \frac{(-3)^j}{i! (1+2j)!!} \left\langle x^i
q^{2j}\right\rangle_\mathrm{{\scriptscriptstyle GC}} H_i (x) L_j^{(1/2)}
(\frac{3}{2}q^2) \right).
\end{equation}
Performing integration using
$\int_0^{\infty} \exp({-{3}/{2} q^2}) q^2  L_j^{(1/2)}(\frac{3}{2}q^2) dq^2
 = - {2 \Gamma(j-\frac{1}{2}) } /({9\sqrt{\pi} \Gamma(j+1) })$
yields
\begin{equation}
 {\cal A}(\nu)=\frac{2}{\sqrt{3}\pi R_0} e^{-\frac{\nu^2}{2}} 
\left( 1 + \frac{1}{2\sqrt{\pi}} \sum_{n=3}^{\infty} \sum_{i,j}^{i+2j=n} \frac{(-3)^{j+1}}{i! (1+2j)!!}  \frac{ \Gamma(j-\frac{1}{2}) }{\Gamma(j+1) } \left\langle x^i q^{2j}\right\rangle_\mathrm{{\scriptscriptstyle GC}} H_i (\nu)   \right)\,.\label{eq:defCont3Dfinal}
\end{equation}
For instance, to first order in the non-Gaussian expansion, we have:
\begin{equation}
 {\cal A}(\nu)=\frac{2}{\sqrt{3}\pi R_0} e^{-\frac{\nu^2}{2}} \left( 1 + \frac{1}{6} \left\langle x^3 \right\rangle H_3(\nu) + \frac{1}{2} \left\langle x q^2 \right\rangle H_1(\nu) \right).
\end{equation}
Appendix~\ref{sec:MinkovFunc} demonstrates how to re-order  expansion~(\ref{eq:defCont3Dfinal}) as an Edgeworth expansion.
\subsubsection{The Euler characteristic of non-Gaussian 3D fields}
\label{sec:euler3D}
\begin{figure}
  \includegraphics[width=0.495\textwidth]{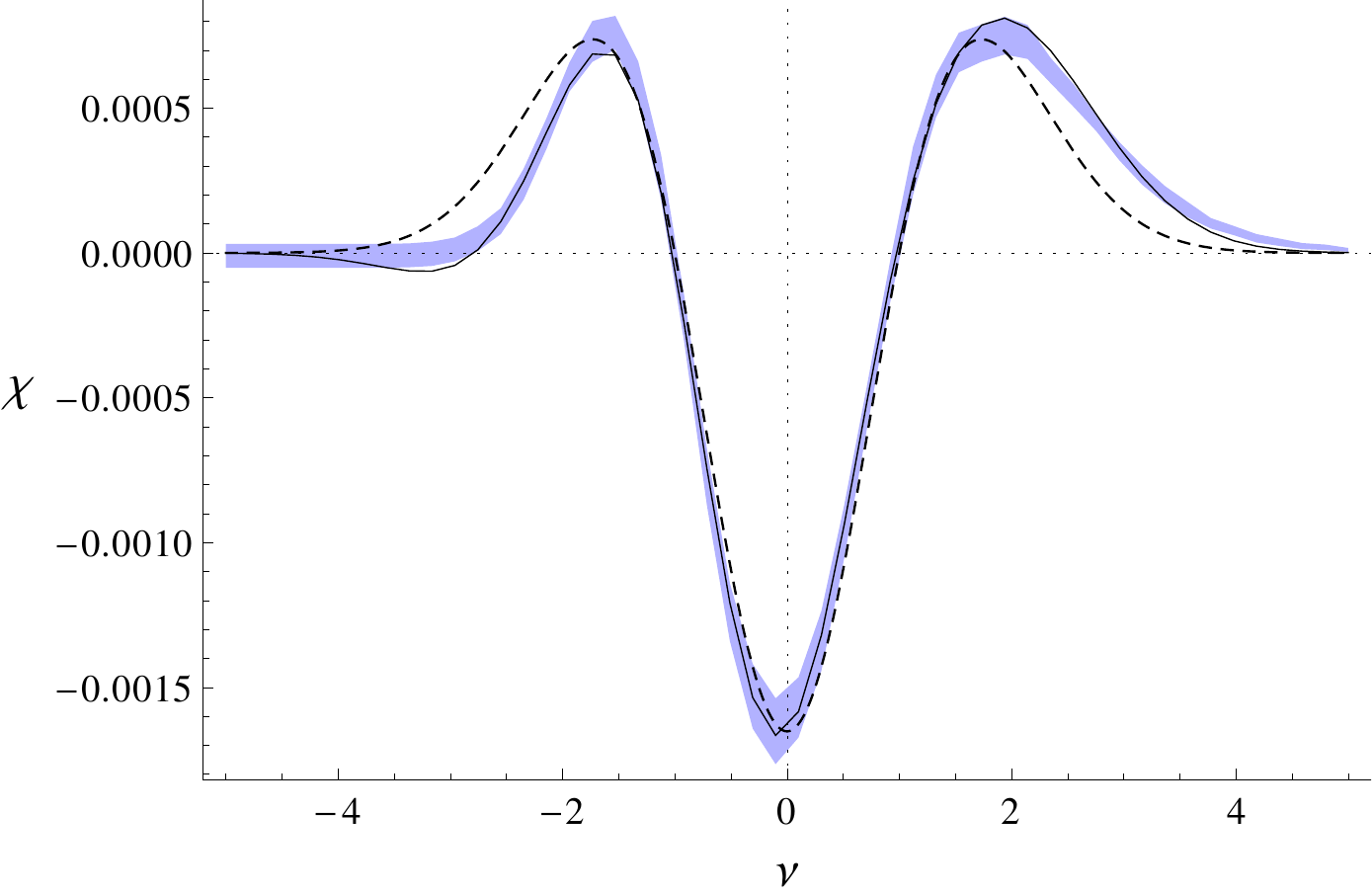}
  \includegraphics[width=0.495\textwidth]{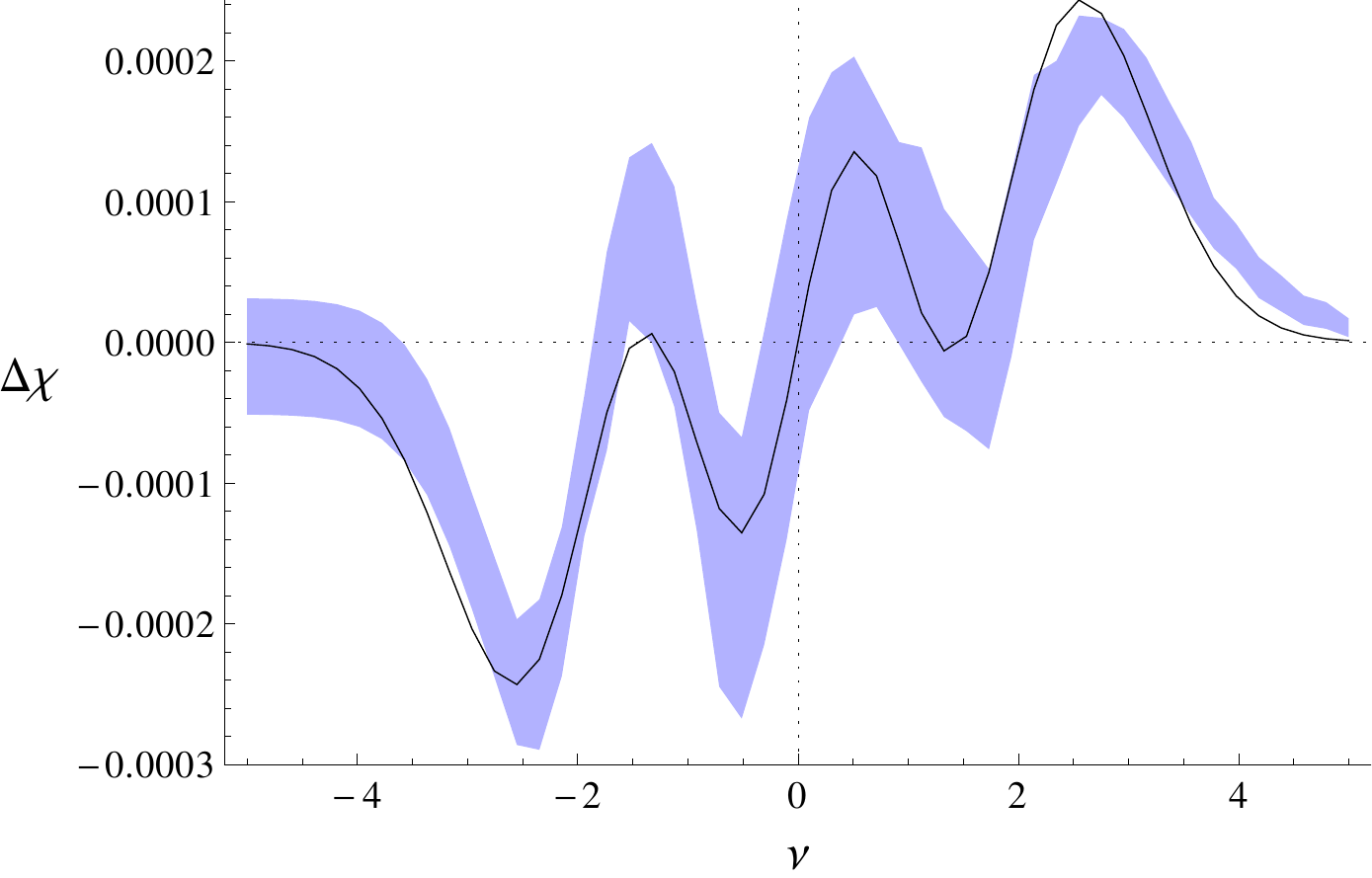}
\caption{{\sl Left:} Euler characteristic in 3D for a mildly non-Gaussian field ($\sigma=0.1$).
 \emph{Dashed line:} Gaussian prediction, \emph{solid line:} first order prediction.
 The shaded band  corresponds to the
3-sigma dispersion for the measurement of the Euler characteristic in the
simulated fields.
\emph{Right:} The residual between the
non-Gaussian Euler characteristic and the Gaussian approximation. 
The solid line and the shaded band are the same as in the left panel, minus
the Gaussian contribution.
}
\label{fig:euler3D}
\end{figure}

To compute the Euler characteristic in 3D, following the 2D case, one averages $-I_3=-\left( {J_1}^3 - 3 J_1 J_2 + 2 J_3 \right)/{27}$
over the full range of the second derivatives:
\begin{equation}
\chi(\nu) = -  \frac{1}{{R_*^3}}
\int_{-\infty}^\infty  \!\!\!\! {\rm d} J_1 
{\int_{\frac{\nu+\gamma J_1}{\sqrt{1-\gamma^2}}}^\infty \!\!\! {\rm d} \zeta 
\int_0^\infty  \!\!\!\! {\rm d} J_2 \int_{-J_2^{3/2}}^{J_2^{3/2}} \!\! {\rm d} J_3
P_{\rm ext}(\zeta, J_1, J_2, J_3) } I_3.
\label{eq:genus3D_I3}
\end{equation}
As shown in Appendix~\ref{sec:deriveuler3D}, the resulting expression is
\begin{multline}
\chi(\nu) = \frac{1}{2} \mathrm{Erfc} \left(\frac{\nu}{\sqrt{2}} \right)
\chi(-\infty) + \frac{1}{27 {R_*^3}} \left(\frac{3}{2\pi}\right)^{3/2}
\frac{1}{\sqrt{2\pi}} \exp\left(-\frac{\nu^2}{2} \right) 
\left[ \vphantom{ \sum_{n=3}^\infty} \gamma^3 H_2\left(\nu\right) + \right. 
\\
+ 
\sum_{n=3}^\infty \sum_{i,j,k}^{i+2 j+k=n} 
\frac{(-1)^k }{i!\;j!}  \left({\textstyle-\frac{3}{2}} \right)^j  
\left\langle \zeta^i {q^2}^j {J_1}^k \right\rangle_\mathrm{{\scriptscriptstyle GC}}
\left( 1 -\gamma^2 \right)^{{i}/{2}}
\sum_{s=0}^{\mathrm{min}(3,k)} 
\frac{ 3! \gamma^{k+3-2s} }{s!(3-s)!(k-s)!} 
H_{i+k+2-2s}\left(\nu\right) 
\\
+  
\sum_{n=3}^\infty \sum_{i,j,k}^{i+2 j+k+2=n} 
\frac{(-1)^{k+1} 3 }{i!\; j!}  \left({\textstyle-\frac{3}{2}} \right)^j  
\left\langle \zeta^i {q^2}^j {J_1}^k {J_2} \right\rangle_\mathrm{{\scriptscriptstyle GC}}
\left( 1 -\gamma^2 \right)^{{i}/{2}}
\sum_{s=0}^{\mathrm{min}(1,k)} 
\frac{ \gamma^{k+1-2s} }{(k-s)!} 
H_{i+k-2s}\left(\nu\right) 
\\
+ \left.
\sum_{n=3}^\infty \sum_{i,j,k}^{i+2j + k + 3=n} 
\frac{(-1)^{k+1} 2  }{i!\; j!\; k!}  \left({\textstyle-\frac{3}{2}} \right)^j  
\left\langle \zeta^i {q^2}^j {J_1}^k J_3 \right\rangle_\mathrm{{\scriptscriptstyle GC}}
\left( 1 -\gamma^2 \right)^{{i}/{2}}
\gamma^k H_{i+k-1}\left(\nu\right)
\right].
\label{eq:genus_3D}
\end{multline}
As in 2D, in the rare event  limit, $\nu\gg 1$ or $\nu\ll 1$, equation~(\ref{eq:genus_3D}) also represents resp. the general differential number of maxima or  minima. 
When equation~(\ref{eq:genus_3D}) is reordered in $\sigma$, the first corrections are (see Appendix~\ref{sec:sigma2}) :
\begin{equation}
 \chi^{(1)}=\left( \frac{9}{2} \gamma^2 \langle q^2 J_1 \rangle + 9 \gamma \langle x I_2 \rangle \right) H_1(\nu) - \left( \frac{3}{2} \gamma^3 \langle x q^2 \rangle + \frac{3}{2} \gamma^2 \langle x^2 J_1 \rangle \right) H_3(\nu) + \frac{1}{6} \gamma^3 \langle x^3 \rangle H_5(\nu)\,, \label{eq:genus3Dfirst}
\end{equation} and
\begin{multline}
 \chi^{(2)}= - \left( \frac{27}{2} \gamma \langle q^2 I_2 \rangle_\mathrm{c} + 27 \langle x I_3 \rangle_\mathrm{c} \right) H_0(\nu) + \left( \frac{9}{8} \gamma^3 \langle q^4 \rangle_c + \frac{9}{2} \gamma^2 \langle x q^2 J_1 \rangle_\mathrm{c} + \frac{9}{2} \gamma \langle x^2 I_2 \rangle_\mathrm{c} \right) H_2(\nu) \\- \left( \frac{3}{4} \gamma^3 \langle x^2 q^2 \rangle_\mathrm{c} + \frac{1}{2} \gamma^2 \langle x^2 J_1 \rangle_\mathrm{c} \right) H_4(\nu) + \frac{1}{24} \gamma^3 \langle x^4 \rangle_\mathrm{c} H_6(\nu)
 + \left( \frac{135}{16} \gamma \langle q^2 J_1 \rangle^2 + \frac{27}{2} \langle q^2 J_1 \rangle \langle x I_2 \rangle + \frac{81}{2} \langle x q^2 \rangle \langle I_3 \rangle \right) H_0(\nu) \\- \left( \frac{45}{4} \gamma^2 \langle x q^2 \rangle \langle q^2 J_1 \rangle + \frac{9}{2} \gamma \langle q^2 J_1 \rangle \langle x^2 J_1 \rangle + \frac{27}{2} \gamma \langle x q^2 \rangle \langle x I_2 \rangle + \frac{9}{2} \langle x^2 J_1 \rangle \langle x I_2 \rangle + \frac{9}{2} \langle x^3 I_3 \rangle \right) H_2(\nu) \\+ \left(\frac{15}{8} \gamma^3 \langle x q^2 \rangle^2 + \frac{3}{4} \gamma^2 \langle x^3 \rangle \langle q^2 J_1 \rangle + \frac{9}{4} \gamma^2 \langle x q^2 \rangle \langle x^2 J_1 \rangle + \frac{3}{4} \gamma \langle x^2 J_1 \rangle^2 + \frac{3}{2} \gamma \langle x^3 \rangle \langle x I_2 \rangle \right) H_4(\nu) \\- \left( \frac{1}{4} \gamma^3 \langle x q^2 \rangle \langle x^3 \rangle + \frac{1}{4} \gamma^2 \langle x^3 \rangle \langle x^2 J_1 \rangle \right) H_6(\nu) + \frac{1}{72} \gamma^3 \langle x^3 \rangle^2 H_8(\nu). 
  \label{eq:genus3Dsecond}
\end{multline}
Figure~\ref{fig:euler3D} shows that the first order expansion provides
 a good match to the measured 3D Euler characteristic, matching closely the  functional form of the
non-Gaussian correction, antisymmetric in 3D.
Measurements are performed by computing the alternating sum of the counts of critical points.

In cosmological applications we do not expect our Universe to have complex topology at the large-scale structure scales,
if at all, and, thus $\chi(-\infty) = 0$. This matches exactly $\langle I_3 \rangle = 0$ 
from equation~(\ref{eq:genus3D_I3}), i.e, the following relation between the moments of the 3D 
Hessian invariants holds
\begin{equation}
\langle J_1^3 \rangle - 3 \langle J_1 J_2 \rangle + 2 \langle J_3 \rangle = 0 \quad .
\end{equation}

\subsubsection{Non-Gaussian total and differential extrema counts in 3D }
\label{sec:count3D}

\begin{figure}
\center \includegraphics[width=0.495\textwidth]{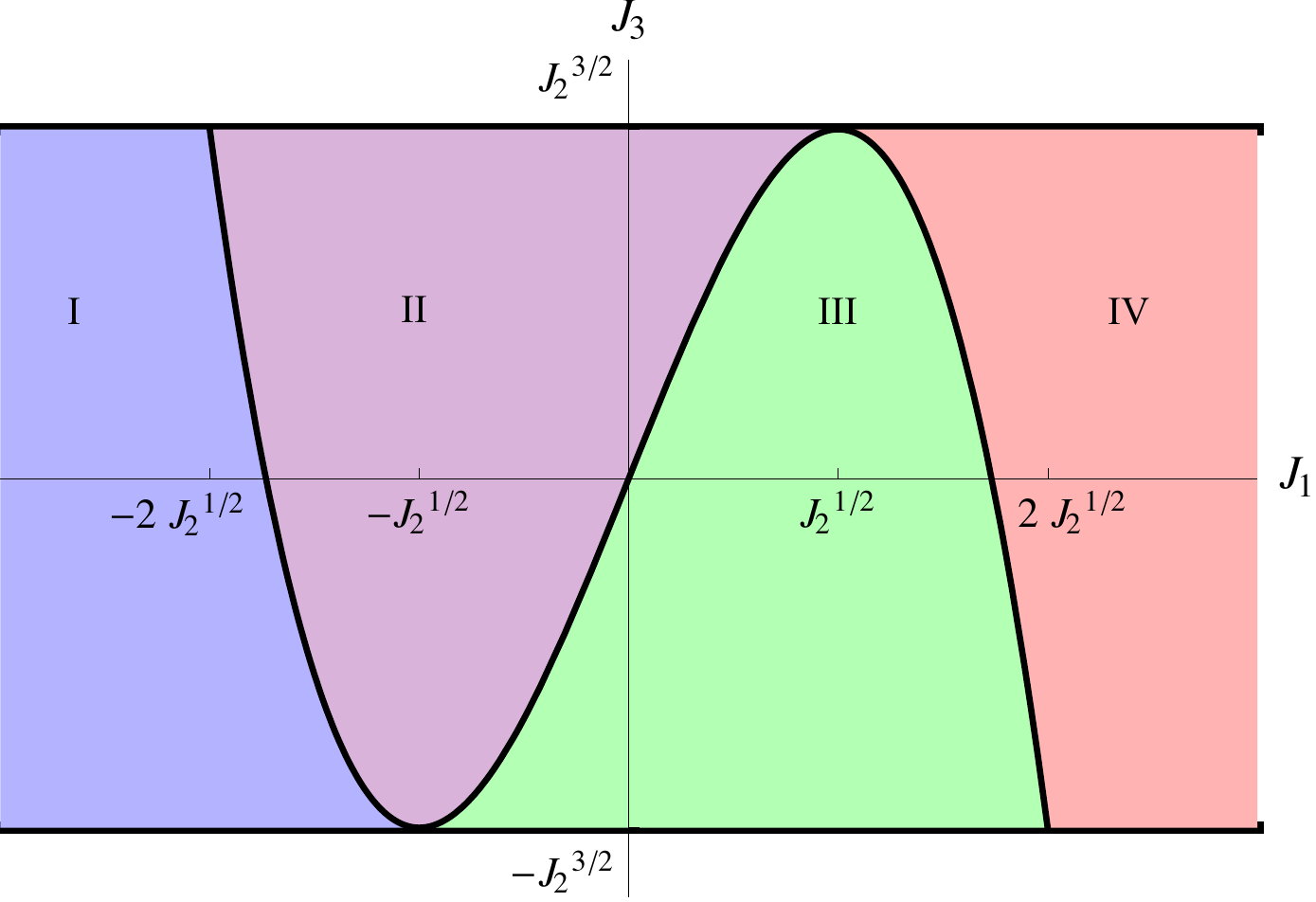}
\caption{Integration regions in $(J_1,J_3)$ space for extrema of different types. I - maxima, II - saddle points
of filamentary type ($\lambda_1 > 0, \lambda_3 \le \lambda_2 < 0 $), III - saddle points of pancake type 
($\lambda_1 \ge \lambda_2 > 0, \lambda_3 < 0 $) and IV - minima. The boundary
curve is 
$J_3 = -\frac{1}{2} J_1^3 + \frac{3}{2} J_1 J_2$.
} 
\label{fig:3Dboundary}
\end{figure}

The total number of critical points in 3D is given by
\begin{equation}
 n_\mathrm{ext} = \frac{1}{{R_*^3}} \int_{-\infty}^{+\infty} {\rm d}J_1 \int_\frac{\nu+\gamma J_1}{\sqrt{1-\gamma^2}}^{+\infty} {\rm d}\zeta \int_0^{+\infty} {\rm d}J_2 \int_{-{J_2}^{3/2}}^{{J_2}^{3/2}} {\rm d}J_3  P_\mathrm{ext}(\zeta,J_1,J_2,J_3) |I_3|.
\end{equation}
For a specific set of extrema, sign constraints must be imposed on the sorted eigenvalues
 $\lambda_1 \ge \lambda_2 \ge \lambda_3$ which establish
the non-overlapping ranges of integration in $(J_1,J_2,J_3)$ space, defining the maxima ($- - -$), the minima ($+ + +$) and
the saddle points of  ``filamentary'' ($+ - -$) and ``pancake'' ($+ + -$) types. While $J_2$ continues to span
unrestricted range of positive values,
for each fixed $J_2$ value, the boundaries between the different extrema types are set in the $(J_1, J_3)$ subspace
by the cubic curve $I_3 = 0$, i.e $2 J_3 = - J_1^3 + 3 J_1 J_2$, and by the limits
$-J_2^{3/2} \le J_3 \le  J_2^{3/2}$. In Figure~\ref{fig:3Dboundary}, the ranges
of integration are shown for each extrema type.

The total number of extrema can be found analytically at all orders of the non-Gaussian expansion.
Here we give only the first corrections that read 
\begin{eqnarray}
n_{\mp --} &=& \frac{29 \sqrt{15} \mp 18 \sqrt{10}}{1800 \pi^2 {R_*^3}} +
\frac{5 \sqrt{5}}{24 \pi^2 \sqrt{6 \pi}  {R_*^3}}
\left( \left\langle q^2 J_1 \right\rangle - 
\frac{8}{21} \left\langle {J_1}^3 \right\rangle
+ \frac{10}{21} \left\langle J_1 J_2 \right\rangle
\right), 
\nonumber \\
n_{++\pm} &=&  \frac{29 \sqrt{15} \mp 18 \sqrt{10}}{1800 \pi^2  {R_*^3}} -
\frac{5 \sqrt{5}}{24 \pi^2 \sqrt{6 \pi}  {R_*^3}}
\left( \left\langle q^2 J_1 \right\rangle - 
\frac{8}{21} \left\langle {J_1}^3 \right\rangle
+ \frac{10}{21} \left\langle J_1 J_2 \right\rangle
\right). \label{eq:extrema_3D}
\end{eqnarray}
signifying that the first non-Gaussian corrections are equal in magnitude for all extrema types
acting in the same direction for the maxima and the ''filamentary`` saddle points,
and opposite to that for the ``pancake'' saddles and the minima.

\begin{figure}
 \begin{center}
 \includegraphics[width=0.495\textwidth]{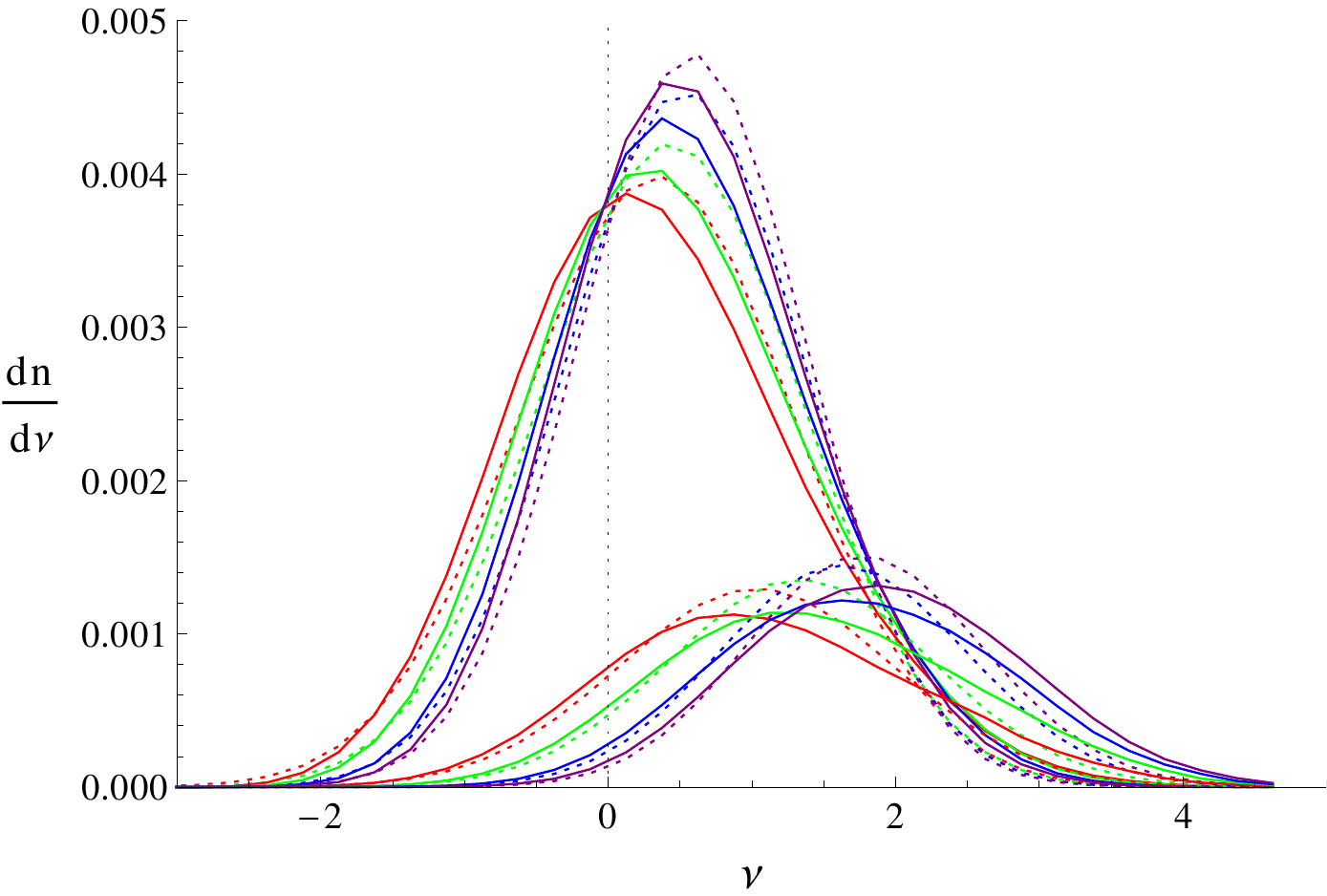} \label{fig:extrema3Dgamma}
\includegraphics[width=0.495\textwidth]{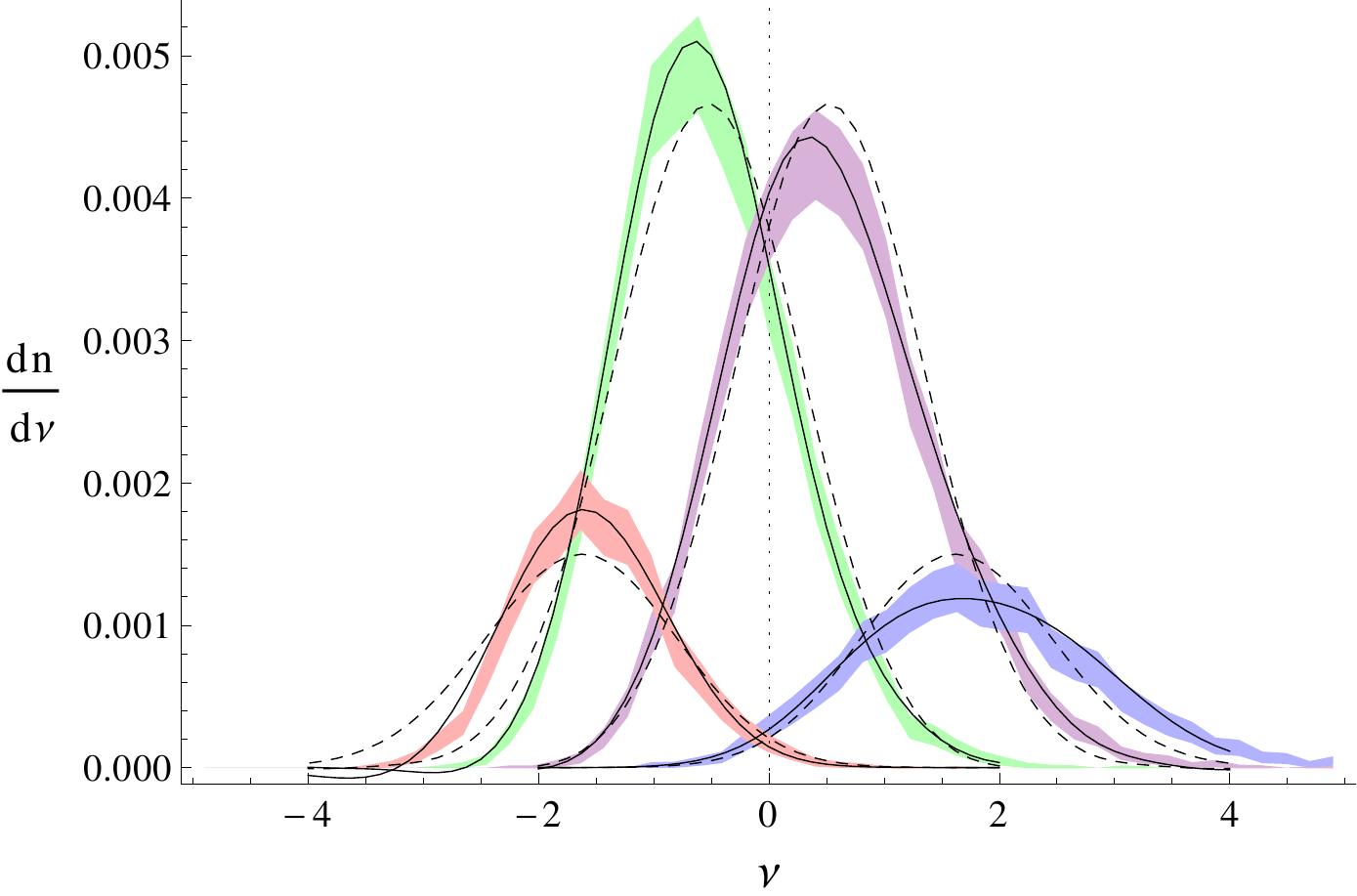}
\caption{
{\sl Left:} Sample differential distributions of maxima (right group of curves) and filament-type saddle points (left group)
in 3D for several values  of the spectral index (\emph{red:}
$n_\mathrm{s}=-2.5$, \emph{green:} $n_\mathrm{s}=-2$, \emph{blue:}
$n_\mathrm{s}=-1$,
\emph{purple:} $n_\mathrm{s}=0$). Dotted lines are the Gaussian approximation,
solid lines are the
 non-Gaussian prediction for $\sigma=0.1$ computed via Monte-Carlo integration.
{\sl Right:} Measured differential distribution of critical points in 3D for a mildly non-Gaussian field
from $n_\mathrm{s}=-1$ simulations at $\sigma=0.1$ versus the
 first order non-Gaussian prediction (\emph{solid}) and the Gaussian approximation (\emph{dashed}).
}
\label{fig:extrema3D}
 \end{center}
\end{figure}

For the differential distributions of critical points, no analytical expression is known even
for  Gaussian fields and we thus proceed numerically. We use a  semi Monte-Carlo integration:
we draw values of the field and its derivatives according to the corresponding Gaussian distribution
and then integrate the integrand times the Edgeworth correction over the domain of integration consistent with the
required sign for the eigenvalues\footnote{This technique is inspired by importance sampling \citep{Robert98montecarlo}.}.
The left panel of Figure~\ref{fig:extrema3D} shows the results of the numerical integration for non-Gaussian 
maxima and filamentary saddles of the density fields
that are obtained by non-linear gravitational evolution from several scale-invariant initial conditions
with power-law spectral index $n_\mathrm{s}\in[-2.5,0]$ to $\sigma=0.1$.
The numerical results for non Gaussian distribution of maxima are accurately
fitted by the  expressions given in Appendix~\ref{sec:fit}.

In the right panel of Figure~\ref{fig:extrema3D} the first order perturbation correction is compared
to the actual measurements of the differential number densities of extremal
points for  $n_\mathrm{s}=-1$  scale invariant power spectrum.
The theoretical predictions fit the measurements very well for  $\sigma=0.1$, correctly capturing
the main qualitative features -  the development of an excess count for the high amplitude peaks and filamentary saddles,
with a simultaneous decrease in the number of  peaks and filamentary saddles for  average contrasts 
($\nu \sim 2$ for peaks and $\nu \sim 1$ for saddles of this type) and the opposite trend for minima and ``pancake''
saddles, namely, depopulation of the underdense tails of their distribution compensated by an increase in numbers
at contrasts near the mean ($\nu \sim -2$ for minima and $\nu \sim -1 $ for ``pancake'' saddles). 
We note that according to equation~(\ref{eq:extrema_3D}) and the values of the cumulants from Table~\ref{tab:S3}, the 
total number of extrema has changed very insignificantly at $\sigma=0.1$, by just over 2\% for maxima and minima
and $\sim 0.07\%$ for the saddle points. This shows that the differential distribution of extrema counts
is a much more sensitive tool to probe non-Gaussianities.

\subsubsection{The differential length of the skeleton for non-Gaussian fields in 3D}\label{sec:skel3D}

\begin{figure}
\includegraphics[width=0.495\textwidth]{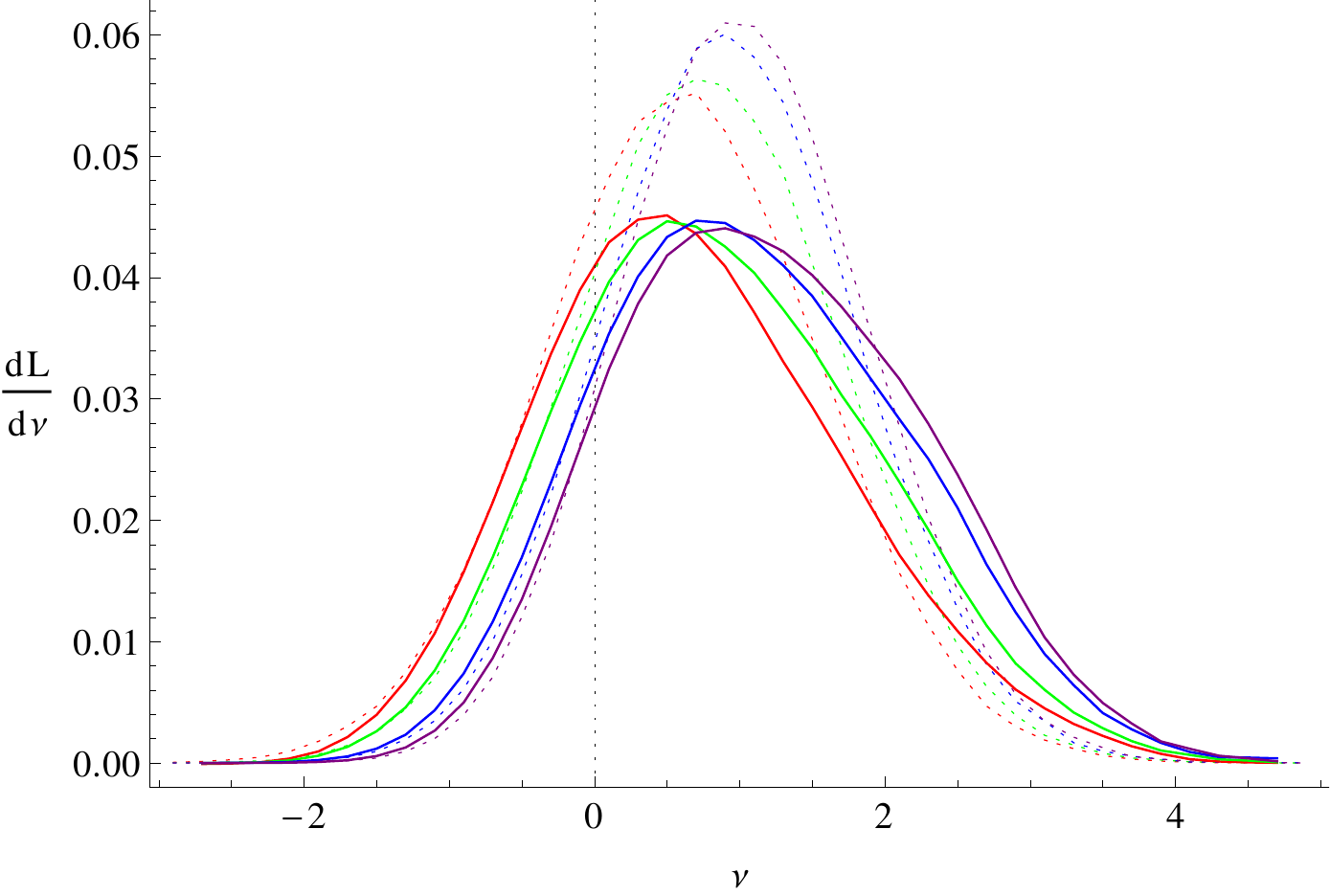}
\includegraphics[width=0.495\textwidth]{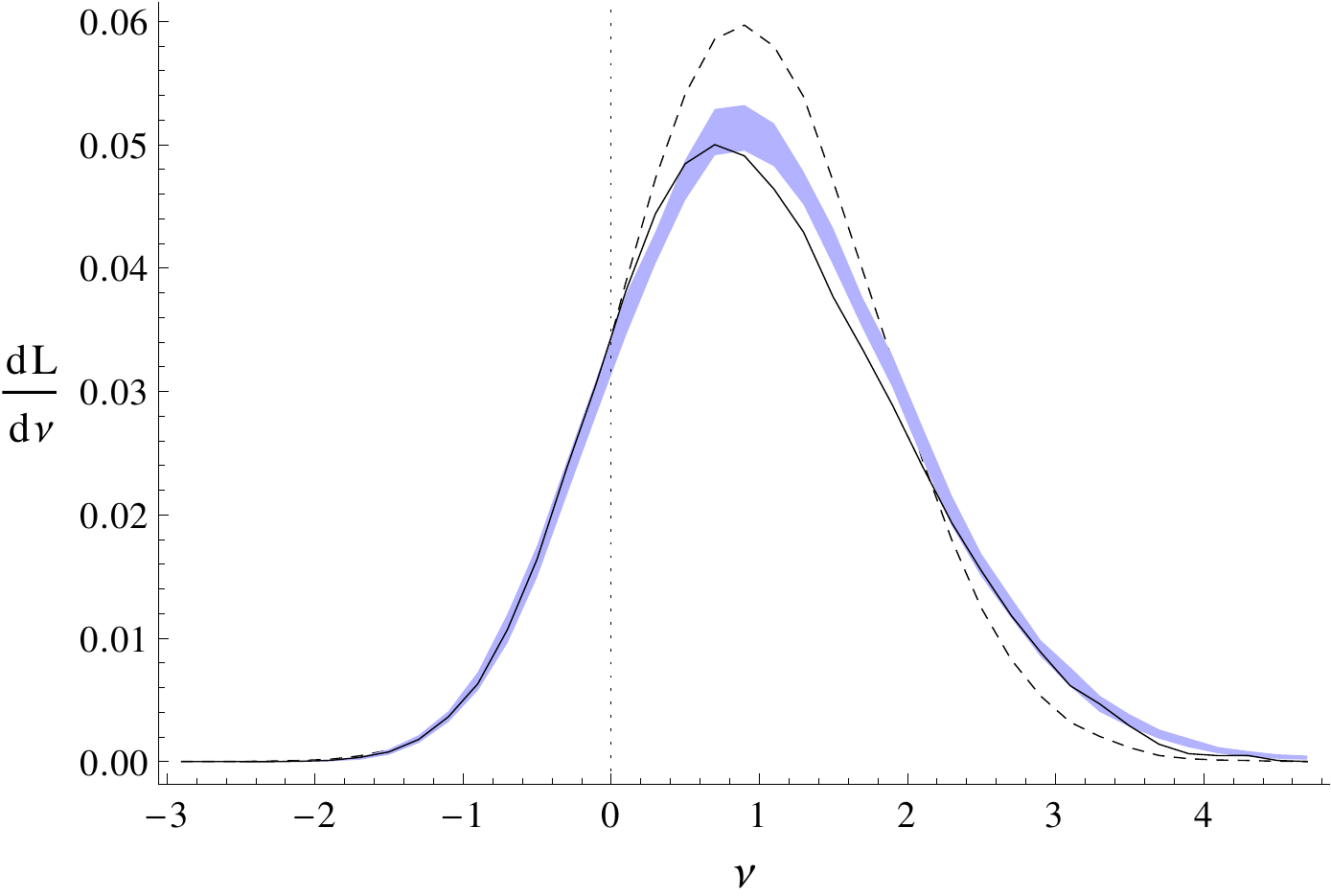}
\caption{{\sl Left:} differential length of the skeleton in 3D for several values of the spectral index
 (\emph{red:} $n_\mathrm{s}=-2.5$, \emph{green:} $n_\mathrm{s}=-2$, \emph{blue:}
$n_\mathrm{s}=-1$, \emph{purple:} $n_\mathrm{s}=0$).
 The dotted lines are the Gaussian case, the solid lines are the non-Gaussian prediction for $\sigma=0.1$.
{\sl Right:} differential length of the skeleton (\emph{blue})  in 3D for a mildly non-Gaussian field ($\sigma=0.1$); \emph{solid line:} Gaussian prediction; \emph{dashed line:} first order prediction. The match  between the two curves is significantly improved, compared to the 2D skeleton.   }
\label{fig:skel3D}
\end{figure}

In 3D, the length of the skeleton is given in the stiff approximation by:
\begin{equation}\frac{\partial L^\mathrm{skel}}{\partial \nu} = \frac{1}{{R_*^2}} \int {\rm d}^3 \tilde x_i {\rm d}^6 x_{kl} P(\nu,\tilde x_i,x_{kl}) \delta^\mathrm{D} (\tilde x_2)\delta^\mathrm{D} (\tilde x_3) \left| \lambda_2 \lambda_3\right| , \label{eq:skellambda2} \end{equation} with the constraint $\lambda_1+\lambda_2+\lambda_3 \leq 0$.
As in 2D, in order to compute the differential length of the skeleton we need to separate the component of the gradient aligned with the highest eigenvalue of the Hessian from the others. Let $Q$ be the normalized norm of the projection of the gradient in the hyperplane orthogonal to this direction:
\begin{equation}
 Q^2 = \frac{q^2 -x_1^2}{\langle q^2 - x_1^2 \rangle}.
\end{equation}
In the Gaussian case, we have
$
 Q^2 = \frac{3}{2} \left( x_2^2 + x_3^2 \right)
$. The Gaussian PDF for $q^2$ is
$
 G(q^2)=3\sqrt{\frac{3}{2\pi}} q e^{-\frac{3}{2} q^2}
$
which leads to the PDF for $x_1$ and $Q^2$:
\begin{equation}
 G(x_1,Q^2) {\rm d}Q^2=\sqrt{\frac{3}{2\pi}} e^{-\frac{3}{2}x_1^2} e^{-Q^2} {\rm d}Q^2.
\end{equation}
After integration over $x_1$, the Gram-Charlier expansion of the PDF is thus:
\begin{multline}
P(x,Q^2,J_1,J_2,J_3) =  G(x,Q^2,J_1,J_2,J_3)
\Bigg[1+ 
\sum_{n=3}^\infty \sum_{i,j,k}^{i + 2j + k + 3=n} 
\frac{(-1)^j \times 25}{i!\;j!!\; k! \times 21} 
\left\langle \zeta^i {Q^2}^j {J_1}^k J_3 \right\rangle_\mathrm{{\scriptscriptstyle GC}}
H_i\left(\zeta\right) L_j\left(Q^2 \right)
H_k\left(J_1\right) J_3
\\
+\sum_{n=3}^\infty \sum_{i,j,k,l}^{i+2 j+k+2 l=n} 
\frac{(-1)^{j+l} 5^l \times 3}{i!\;j!\; k!\; (3+2l)!!} 
\left\langle \zeta^i {Q^2}^j {J_1}^k {J_2}^l \right\rangle_\mathrm{{\scriptscriptstyle GC}}
H_i\left(\zeta\right) L_j\left(Q^2 \right)
H_k\left(J_1\right) L_l^{(3/2)}\left(\frac{5}{2} J_2\right)+\\
 \left. 
\sum_{n=5}^\infty \sum_{i,j,k,l=0,m=1}^{i + 2j + k + 2l + 3m = n} 
\frac{(-1)^j c_{lm} }{i!\;j!\; k! } 
\left\langle \zeta^i {Q^2}^j {J_1}^k {J_2}^l {J_3}^m \right\rangle_\mathrm{{\scriptscriptstyle GC}}
H_i\left(\zeta\right) L_j\left(Q^2 \right)
H_k\left(J_1\right) F_{lm}\left(J_2,J_3\right)
\right]\,.
\end{multline}
The differential length of the skeleton is given by
\begin{equation}\frac{\partial L^\mathrm{skel}}{\partial \nu} =  \frac{1}{ {R_*^2}} \int_{-\infty}^0 {\rm d}J_1 \int_{\frac{\nu+\gamma J_1}{\sqrt{1-\gamma^2}}}^{+\infty} {\rm d}\zeta \int_0^{+\infty} {\rm d}J_2 \int_{-{J_2}^{3/2}}^{{J_2}^{3/2}} {\rm d}J_3 \frac{1}{\pi} P(\zeta, Q^2=0, J_1, J_2, J_3)\left| \lambda_2 \lambda_3 \right|.
\label{eq:skel_3D}
\end{equation}
The integrand $|\lambda_2 \lambda_3|$ is a non-trivial combination of the invariant variables: 
one would have to find the sorted roots of the third-degree characteristic polynomial
 $\lambda^3 - I_1 \lambda^2 + I_2 \lambda - I_3 =0$. 
Furthermore \cite{pogoskel} has shown that the differential length of the skeleton does not have an analytical
 expression in 3D. It is thus easier to use a numerical integration, as was described for the extrema.
 {Note that we use $\lambda_1+\lambda_2+\lambda_3 \leq 0$ instead of
$\lambda_1+\lambda_2 \leq 0$ (that was
adopted in \cite{pogoskel})  as the constraint for defining the skeleton; this choice yields a  perfect match to the total
length of the skeleton as measured in simulations (in
Figure~\ref{fig:skel3D} the curves are not normalized by the total length). }

For  $n_s\in [-2.5,0]$, the Gaussian and first order prediction for the differential skeleton length are given in Appendix~\ref{sec:fit}.
Figure~\ref{fig:skel3D} shows the variation of a non Gaussian 3D skeleton with the spectral index, and illustrates the match between the predicted and the measured skeleton.
%

\section{Applications:  CMB primordial non-Gaussianities,   dark energy in the LSS}
\label{sec:applications}

With the full knowledge of the departure from Gaussianity of critical sets (resp. in 2D equations~(\ref{eq:genus_2D})
for the Euler characteristic, (\ref{eq:number_counts_2D}) for extrema, (\ref{eq:skel_2D}) for the skeleton, in 3D, 
equations~(\ref{eq:genus_3D}), (\ref{eq:extrema_3D}) and (\ref{eq:skel_3D})), let us know illustrate how these estimates
can be used to address  problems arising in cosmology. 
The first example uses 2D extrema counts to estimate primordial non-Gaussianities in
synthetic CMB maps, while the second example estimates the density dispersion, $\sigma(z)$ of dark matter simulations from  differential extrema counts, the Euler characteristic and skeleton length as a function of redshift.
Note that alternative illustrations can be found in \cite{PGP2009,PPG2011}, including models involving  second order corrections.

\subsection{Primordial non-Gaussianities in the cosmic microwave background}
\label{sec:CMB}

\begin{figure}
    \includegraphics[width=1.05\textwidth]{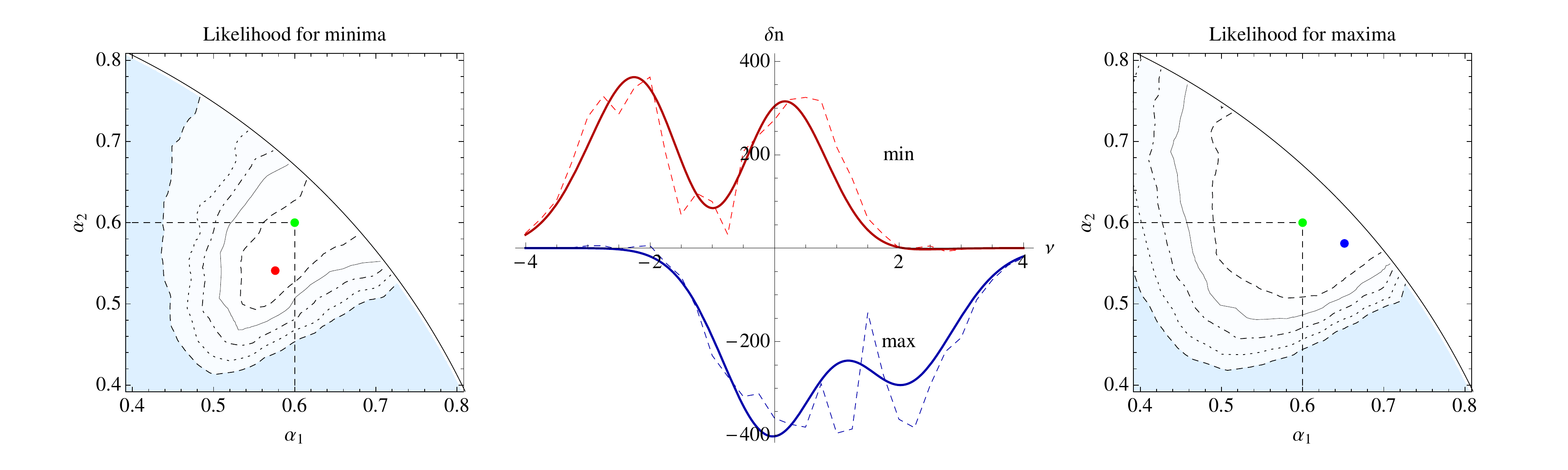}
\caption{
{\sl Middle panel}: the predicted ({\sl solid line}) excess number of
maxima (bottom), and minima (top) in $\Delta\nu=0.25$ bins as a function of the
threshold, $\nu$, on top of the measured 
count (dashed line) from a {\sl single} realization full-sky {\sc \small nside}=1024 {\sc \small
HEALPix} map ({\sl histogram}). The temperature field is smoothed with
the Gaussian filter of 10 arcmin FWHM, resulting in $R_* \approx 11 $ arcmin
$\approx 3$ pixels.  {\sl Left and right panels}: likelihood contours in the parameter space of  the  Harmonic Oscillator model of non-Gaussianity given by equation~(\ref{eq:defharm}); the input model corresponds to $\alpha_1=0.6$, $\alpha_2=0.6$, represented by a green dot in the likelihood map. 
The five contours correspond to 1,2,3,4  and 5 sigma likelihood contours, while the blue and red  dot mark the most likely estimate for maxima and minima respectively.
 Hence different models of non-Gaussianity can be well distinguished by their effects on differential
extrema counts. 
}
\label{fig:CMB}
\end{figure}

Let us generate sets of parameterized non-Gaussian maps
using the package {\tt sky-ng-sim} \citep{Rocha} of  {\sc \small HEALPix}. 
In this so called harmonic model, the PDF, $f_T(T)$ of the pixels' temperature, $T$ is given by
 \begin{equation}
 f_T(T)=
 \exp\left(-\frac{T^2}{2 {\tilde \sigma_0}^2}\right) \left| \sum_{i=0}^n \alpha_i c_i 
H_i\left(\frac{T}{{\tilde \sigma_0}}\right)\right|^2\,, \label{eq:defharm}
 \end{equation}
 where the $c_i$ are normalization constants.  In this paper, we use 
 {\sc \small nside}=2048, $\ell_{\max}=4096$, $n=2$, ${\tilde \sigma_0}=1$, $\alpha_{0}=0$ and
 vary $\alpha_1$ and  $\alpha_2$ on a $50\times50$ polar grid in the range $[0.1,0.8]$ subject to $\alpha_1^2+\alpha_2^2<1$.   
For each set of maps, we compute its derivatives, and arithmetically  average
 the corresponding cumulants, using the code {\tt map2cum} \citep{PPG2011} for which the 
invariant variables $J_1$ and $J_2$  on a sphere are
defined via the mixed tensor of covariant derivatives $J_1 = {x_{;i}}^{;i}$ and
$J_2 = J_1^2 - 4 \left| {x_{;i}}^{;j} \right|$.
The differential counts  are then evaluated for a range of 
threshold, $\nu\in[-5,5]$. For our reference map, $\alpha_1=\alpha_2=0.6$, the number of extrema is
computed by the procedure {\tt hotspot} \citep{hivon05}. Figure~\ref{fig:CMB} illustrates our hability to 
recover input model from a given set of maxima and minima extracted from a given realization of the sky.
Here the residual departure from the Gaussian number count is shown.
 Indeed the maximum likelihood solution, solid lines,  
corresponds to a very good match to the  measured residual differential number counts.  
Note that this class of model  has an 
 interesting feature  that $\left\langle x^3\right\rangle$ is  significantly smaller than say, $ \left\langle J_1^3\right\rangle $.
 Indeed for this model, 
$
\left\langle x^3\right\rangle =-0.076,
 \left\langle x^2 J_1\right\rangle =0.072,
 \left\langle q^2 x\right\rangle =-0.091,
 \left\langle x J_1^2\right\rangle =-0.15,
 \left\langle x J_2\right\rangle =-0.087, 
 \left\langle q^2 J_1\right\rangle =0.054, 
 \left\langle J_1^3\right\rangle =0.31,
 \left\langle J_1 J_2\right\rangle =0.023 $. 
 This simple minded experiment should be improved to account for, e.g. more realistic and physically motivated non-Gaussian maps (such as $f_{\rm NL}$ models), realistic galactic cuts and hit maps, variable instrumental response. The main challenge is to model incompleteness in observed samples in order to weight accordingly the range of relevant thresholds. 

\subsection{Constraining the dark energy equation of state via the geometry of the LSS}
\label{sec:app3D}

At supercluster scales where non-linearity is mild, we can expect the non-Gaussianity of the matter density to be also mild, but still essential for a quantitative understanding of the filamentary Cosmic Web in-between the galaxy clusters.
Indeed, e.g. Figure~\ref{fig:extrema3D} shows that the interplay of gravity and (possibly accelerated) expansion will induce some level of distortion in the PDF of the differential counts.  Importantly, this distortion is not sensitive to the overall number-count as the PDF are normalized.  Fitting measured counts to the distorted theoretical expectation should therefore allow us to
robustly  estimate some features of the underlying dark energy equation of state. 

\subsubsection{Extension of the gravitational  Perturbation Theory to critical
set cumulants}
\label{sec:PT}
When the field is the 3D density field of the large-scale structure of the Universe and for Gaussian initial conditions, the non-Gaussianities appear because of the non-linearities of  gravitational clustering. As mentioned in the introduction,
equations~(\ref{eq:genus_3D}), (\ref{eq:extrema_3D}) and (\ref{eq:skel_3D}) have  in common that they can be re-arranged  as
\begin{equation}
{\cal E}_R(\nu,z)= \sum_n {\cal C}_n(\nu,R) \sigma_0^n(R) D(z)^n\,, \label{eq:calE}
\end{equation}
where $\sigma_0^2$ is today's variance of the cosmic density field  at the corresponding smoothing scale, $R$,  and $D(z)$ is the growth factor as a function of redshift.
For instance, for the Euler characteristic, ${\cal C}_1 \sigma_0 D(z)$ is given by equation~(\ref{eq:genus3Dfirst}) while  ${\cal C}_2  \sigma_0^2 D(z)^2$ is given by equation~(\ref{eq:genus3Dsecond}).
When gravity is generating mildly non-Gaussianities, we will show that it is possible to predict the emerging statistical properties of the field using a perturbative theory. 
As we will demonstrate below, all the cumulants involved in the Edgeworth expansion entering the ${\cal C}_n$s  may then be predicted from first principle.
While the geometrical analysis of catalogs, say using codes like {\tt DisPerSE} \citep{persistent} allow us to measure ${\cal E}_R(\nu,z)$, and provided that i) we consider large enough scales so that equation~(\ref{eq:calE}) converges,  and 
ii) that gravity is the driving cause of non-Gaussianity, equation~(\ref{eq:calE}) can be inverted for $D(z)$.
In turn, measuring $D(z)$ below $z\sim1$ probes the dark energy equation of state. 
One of the difficulties in implementing our statistics to observed surveys is  that  $\sigma$ is not known, hence we cannot
normalize $\nu\equiv\rho/\sigma$. Following \cite{Gott1988}, a work-around is to introduce
  $\nu_f$, the density contrast corresponding to a given filling factor, $f$, instead of $\nu$ which are related via the identity 
\begin{equation}
f \equiv \int^{\infty}_{\nu_{{}}} {\rm d }\nu 
 { P}(\nu) =
 \int^{\infty}_{\nu_{f}} {\rm d }\nu 
 { G}(\nu) \,, \label{eq:filling}
\end{equation}
where $f(\nu)$  is given by equation~(\ref{eq:defF}) and  $G$ stands for the corresponding Gaussian PDF,
so that ${\cal E}_R(\nu[\nu_f],z)$ is a measurable quantity  encoding the departure from Gaussianity which is not already encoded in $P(\nu)$.
In practice we will not make use of equation~(\ref{eq:filling}) below and we will assume perfect knowledge of $\nu$ in these experiments.
%
%
%
\subsubsection{Prediction of the 3rd order  cumulants in Perturbation Theory}
\label{sec:PT3}
In the formalism of perturbation theory  \citep{Bernardeau}, it is easy to see that the $p$-th cumulant of the field is of order $\sigma^{p-2}$, if the field is normalized to a constant variance. Using this prescription, one can re-sum the Gram-Charlier expansion to the so-called Edgeworth expansion. The Edgeworth expansion is an expansion in $\sigma$ and its order is thus physically motivated. In practice it means its convergence properties can be justified on physical grounds, that it can thus be truncated more easily \citep{Blinnikov}. Appendix~\ref{sec:Edgeworth} shows that the $\sigma$ term from the Edgeworth expansion only comes from the third order term of the Gram-Charlier expansion, while the $\sigma^2$ term comes from both the fourth and part of the sixth order terms.
We will therefore restrict our predictions of the involved cumulants to third order. At this order, moments, cumulants and Gram-Charlier coefficient coincide. For the next order in $\sigma$, one would have to extend the predictions to quartic cumulants and to decompose sixth order Gram-Charlier coefficients in cumulants (see Appendix~\ref{sec:sigma2}). For third order terms, the calculation of all the cumulants is similar to the well-known prediction of the skewness \citep{Peebles1980, Fry1984, Bernardeau1993, Bernardeau}. We will only consider an Einstein-de Sitter (EdS) universe since other cosmologies can be described as corrections to this prediction.

For a three dimensional field, the PT  cumulants involving the field and its derivatives can be written as:
\[ {\cal C}_R= \left\langle ( \partial_1^{\alpha_1} \partial_2^{\alpha_2} \partial_3^{\alpha_3} \delta_R ) ( \partial_1^{\beta_1} \partial_2^{\beta_2} \partial_3^{\beta_3} \delta_R ) ( \partial_1^{\gamma_1} \partial_2^{\gamma_2} \partial_3^{\gamma_3} \delta_R ) \right\rangle \,,\] where $\partial_i^n $ is the $n$-th derivative with respect to the $i$-th coordinate,
and $\delta_R$ is the field smoothed on scale $R$.
Using gravitational  perturbation theory, we show in Appendix~\ref{PTderiv}   that $ {\cal C}_R$ obeys
\begin{equation}
 {\cal C}_R=
2 \int {\rm d}^3 \mathbf{k_1} {\rm d}^3 \mathbf{k_2} 
\left[ \mathcal{F}_{\alpha,\beta,\gamma} (\mathbf{k_1},\mathbf{k_2})+ \mathcal{F}_{\beta,\gamma,\alpha} (\mathbf{k_1},\mathbf{k_2})+ \mathcal{F}_{\gamma,\alpha,\beta} (\mathbf{k_1},\mathbf{k_2})  \right] P(k_1) P(k_2) W(k_1 R)W(k_2 R)  W(|\mathbf{k_1}+\mathbf{k_2}| R)\,,
 \end{equation}
 where the so-called generalized geometric factors \citep{Bernardeau},
$\mathcal{F}_{\alpha,\beta,\gamma} (\mathbf{k_1},\mathbf{k_2})$ 
(here $\alpha$ is the short hand for the triplet
$(\alpha_1,\alpha_2,\alpha_3)$,  etc. ) read:
  \begin{eqnarray}
  \mathcal{F}_{\alpha,\beta,\gamma} (\mathbf{k_1},\mathbf{k_2})&=&F_2 (\mathbf{k_1},\mathbf{k_2}){\cal G}(\mathbf{k_1},\mathbf{k_2})\, i^{\alpha_1+\alpha_2+\alpha_3}\, i^{\beta_1+\beta_2+\beta_3}\, (-i)^{\gamma_1+\gamma_2+\gamma_3}  \, , \,\,\, {\rm with} \,\,\,
  F_2(\mathbf{k_1},\mathbf{k_2}) = \frac{5}{7} + \frac{\mathbf{k_1}\cdot \mathbf{k_2}}{{k_1}^2} + \frac{2}{7} \frac{\left( \mathbf{k_1}\cdot \mathbf{k_2}\right)^2}{{k_1}^2 {k_2}^2}\,,
  \nonumber\\ 
   {\cal G}(\mathbf{k_1},\mathbf{k_2})&=&
\left( \mathbf{k}_1^{[1]} \right)^{\alpha_1} \left( \mathbf{k}_1^{[2]} \right)^{\alpha_2} \left( \mathbf{k}_1^{[3]} \right)^{\alpha_3}
\left( \mathbf{k}_2^{[1]} \right)^{\beta_1} \left( \mathbf{k}_2^{[2]} \right)^{\beta_2} \left( \mathbf{k}_2^{[3]} \right)^{\beta_3}
\left( \mathbf{k}^{[1]}_1+\mathbf{k}^{[1]}_2   \right)^{\gamma_1} \left(  (\mathbf{k}^{[2]}_1+\mathbf{k}^{[2]}_2   \right)^{\gamma_2}
 \left(\mathbf{k}^{[3]}_1+\mathbf{k}^{[3]}_2  \right)^{\gamma_3}. \label{eq:defcalG}
 \end{eqnarray} 
The smoothing of the field over the scale $R$ is taken care of  via the window function, $W(k R)$.
As explained in Appendix~\ref{PTderiv}, for a Gaussian filtering and a scale invariant power spectrum, these integrals can be completed and the resulting cumulants are given 
explicitly in terms of the smoothing length and the power index in Appendix~\ref{sec:PTresults}.
For instance, for a $n_\mathrm{s}=-3$ power law index, we generalize the classical result
\begin{eqnarray}
     \left\langle x^{3}\right\rangle = \frac{34}{7} \,, \quad {\rm to} \quad
\left\langle x x_{12}^{2}\right\rangle = \frac{34}{7} \frac{2}{3\cdot5!!}\,, \quad
 \left\langle x x_{11}x_{22}\right\rangle  =\frac{34}{7}  \frac{2}{3\cdot5!!} \,,\quad
  \left\langle x x_{11}^{2}\right\rangle =\frac{34}{7}  \frac{2}{5!!}  \quad {\rm and} \quad
    \left\langle x x_{1}^{2}\right\rangle =\frac{34}{7}  \frac{2}{3^{2}} \,. 
\end{eqnarray}

In Section~\ref{sec:2Dtheory} we also use the perturbation theory to predict 2D
statistics. Since our 2D fields are the slices of the gravitationally evolved 3D
density field, the cumulants in field variables that involve only the
derivatives restricted to the slice are the same as predicted by the 3D
perturbations theory and are given in Table~\ref{tab:S3}.  However, to apply to
2D slices, one needs to combine these cumulants of the field variables into 2D
invariants.

\subsubsection{Comparison to measurements of the critical set cumulants in Monte Carlo simulations}
\label{sec:measurements}

Measurements are carried on sets of simulations produced using {\sc Gadget-2} \citep{Gadget} code and scale-invariant initial conditions generated with {\sc Mpgrafic} \citep{Mpgrafic}. We use $256^3$ particles and 10, 25 and 10 realizations for respectively $n_\mathrm{s}=0$, -1 and -2. To be consistent with the perturbation theory used, we set an Einstein-de Sitter cosmology ($\Omega_m=1$, $\Omega_\Lambda=0$). The 2D results are obtained by cutting these 3D simulations into 25 slices which are averaged along the shortest dimension to give 2D fields.
Following \cite{Percolation}, we benefit from the scale-invariance of our simulations to combine them. For each simulation, we compute the density field with a cloud-in-cell procedure and we smooth on lengths of $R=5$, 10, 15, 20 and 25 pixels.
To eliminate the snapshots contaminated by initial transients or incorrect large-scale mode couplings, we compute the correlation length $l_0$, defined by $\sigma^2 (l_0) =1$. We only keep the snapshots verifying \citep{Colombi}:
\[ \frac{L}{N^{1/3}} \ll l_0 \ll L\,, \] where $L$ is the size of the box and $N$ the number of particles. The first condition ensures that the grid effects have relaxed, while the second condition limits the coupling of the modes modified by the finite size of the box. In practice we choose \[ 2 \frac{L}{N^{1/3}} < l_0 < \frac{1}{20} L. \]

Even within a selected snapshot, all the scales cannot be trusted. The smoothing length has to be large to probe scales for which the discreteness of the simulation is not important, but small enough compared to the size of the box. A detailed study shows that the smallest scale containing information is not given by the initial inter-particle distance, as one could naively think. Indeed the resolution is improved in clustered regions, where the density is higher than the density of the initial conditions. The pertinent characteristic length can be computed using the quantity $N_c(R) = \bar{n} \bar{\xi}(R)$, where $R$ is the smoothing length, $\bar{n}$ the average number of particles per cell and $\xi$ the correlation function. $N_c$ characterizes the average number of neighbors \citep{BalianSchaeffer}. The associated characteristic length is then $l_c$ such as $N_c(l_c)=1$. We keep the smoothing lengths for which  \citep{Colombi}:
\[ l_c \ll R \ll L .\]
In practice, we choose $2l_c < R < {L}/{20} $.
The rotational invariants of the field are computed for each field (using Fourier transforms to estimate the derivatives), 
 and are averaged over realizations once the relevant cumulants are generated. 
The results for the 3rd order moments of the field are given in Table~\ref{tab:S3} and are compared to predictions from first order perturbation theory. Table \ref{tab:tab3rdinv} re-expresses them in terms of invariants and Table~\ref{tab:cumulants} gives the 4th order cumulants.

\begin{table}
\begin{center}
\begin{tabular}{|l|cc|cc|cc|}
\hline
 & \multicolumn{2}{c|}{$n_\mathrm{s}=0$} & \multicolumn{2}{c|}{$n_\mathrm{s}=-1$} & \multicolumn{2}{c|}{$n_\mathrm{s}=-2$}\\ \hline
 & prediction & measurement & prediction & measurement & prediction & measurement\\ \hline
$\langle x^3\rangle / \sigma$ & $3.144$ & $3.08\pm 0.08$ & $3.468$ & $3.5\pm 0.1$ & $4.022$ & $4.4\pm0.4$\\ \hline
$\langle x q^2\rangle / \sigma$ & $2.096$ & $2.05\pm 0.03$ & $2.312$ & $2.39\pm0.09$ & $2.681$ & $3.2\pm0.3$ \\ \hline
$\langle x^2 J_1 \rangle / \sigma$ & $-3.248$ & $-3.15\pm 0.06$ & $-3.270$ & $-3.3\pm0.1$ & $-3.096$ & $-4.0\pm0.4$ \\ \hline
$\langle x {J_1}^2\rangle / \sigma$ & $3.871$ & $3.75\pm 0.06$ & $3.877$ & $4.1\pm0.2$ & $3.847$ & $5.3\pm0.6$\\ \hline
$\langle x J_2 \rangle / \sigma$ & $1.545$ & $1.54\pm 0.02$ & $2.037$ & $2.06\pm0.05$ & $2.625$ & $3.1\pm0.3$ \\ \hline
$\langle q^2 J_1 \rangle / \sigma$ & $-1.335$ & $-1.28\pm 0.02$ & $-1.157$ & $-1.28\pm0.08$ & $-0.941$ & $-1.5\pm0.2$\\ \hline
$\langle {J_1}^3 \rangle / \sigma$ & $-4.644$ & $-4.50\pm 0.08$ & $-4.412$ & $-4.8\pm0.3$ & $-3.960$ & $-7\pm1$ \\ \hline
$\langle J_1 J_2 \rangle / \sigma$ & $-0.679$ & $-0.65\pm 0.01$ & $-0.955$ & $-0.92\pm0.02$ & $-1.142$ & $-1.5\pm0.1$ \\ \hline
$\langle J_3 \rangle / \sigma$ & $1.304$ & $1.28\pm 0.03$ & $0.774$ & $1.0\pm0.1$ & $0.267$ & $1.0\pm0.3$ \\ \hline
\end{tabular}
\caption{Predicted and measured 3D third order invariant cumulants  for three values of the power spectrum index. The method to estimate these cumulants in simulations is 
described in Sec.~\ref{sec:measurements}. The predictions follow from the coordinate cumulants
 computed in Table~\ref{tab:S3} given in Appendix~\ref{sec:PTresults}. According to perturbation theory, the dimensionless third order cumulants are proportional to $ \sigma$.
\label{tab:tab3rdinv}
}
\end{center}
\end{table}
\begin{table}
\begin{center}
\begin{tabular}{|l|c|c|c|}
\hline
 & $n_\mathrm{s}=0$ & $n_\mathrm{s}=-1$ & $n_\mathrm{s}=-2$ \\ \hline
$\langle {J_1}^4 \rangle_\mathrm{c} / \sigma^2$ & $39\pm2$ & $53\pm8$ & $140\pm60$\\ \hline
$\langle {J_1}^2 J_2 \rangle_\mathrm{c} / \sigma^2$ & $3.6\pm 0.2$ & $7.0 \pm 0.6$ & $21 \pm 7$ \\ \hline
$\langle {J_2}^2\rangle_\mathrm{c} / \sigma^2$ & $4.4\pm0.2$ & $7.4\pm0.8$ & $21\pm8$\\ \hline
$\langle J_1 J_3 \rangle_\mathrm{c} / \sigma^2$ & $-1.6\pm0.1$ & $-2.0\pm0.3$ & $-4\pm2$\\ \hline
$\langle x {J_1}^3 \rangle_\mathrm{c} / \sigma^2$ & $-30\pm2$ & $-40\pm5$ & $-88\pm36$\\ \hline
$\langle x J_1 J_2 \rangle_\mathrm{c} / \sigma^2$ & $-3.6\pm0.2$ & $-6.5\pm0.5$ & $-17\pm5$\\ \hline
$\langle x J_3 \rangle_\mathrm{c} / \sigma^2$ & $2.4\pm0.2$ & $4.0\pm0.8$ & $7\pm4$\\ \hline
$\langle x^2 {J_1}^2 \rangle_\mathrm{c} / \sigma^2$ & $24\pm1$ & $32\pm4$ & $62\pm22$\\ \hline
\end{tabular}
\begin{tabular}{|l|c|c|c|}
\hline
 & $n_\mathrm{s}=0$ & $n_\mathrm{s}=-1$ & $n_\mathrm{c}=-2$ \\ \hline
$\langle x^2 J_2 \rangle_\mathrm{c} / \sigma^2$ & $5.1\pm0.3$ & $9.3\pm0.9$ & $22\pm6$\\ \hline
$\langle x^3 J_1 \rangle_\mathrm{c} / \sigma^2$ & $-20\pm1$ & $-26\pm3$ & $-45\pm14$\\ \hline
$\langle x^4 \rangle_\mathrm{c} / \sigma^2$ & $17\pm1$ (17.5) & $23\pm3$ (21.9) & $40\pm12$ (30.4)\\ \hline
$\langle q^2 {J_1}^2 \rangle_\mathrm{c} / \sigma^2$ & $16\pm5$ & $17\pm5$ & $16\pm5$\\ \hline
$\langle q^2 J_2 \rangle_\mathrm{c} / \sigma^2$ & $13\pm5$ & $14\pm5$ & $13\pm5$\\ \hline
$\langle x q^2 J_1 \rangle_\mathrm{c} / \sigma^2$ & $-7.3\pm0.4$ & $-9\pm1$ & $-19\pm7$\\ \hline
$\langle x^2 q^2 \rangle_\mathrm{c} / \sigma^2$ & $8.5\pm0.5$ & $12\pm2$ & $24\pm8$\\ \hline
$\langle q^4 \rangle_\mathrm{c} / \sigma^2$ & $6.8\pm0.4$ & $10\pm1$ & $24\pm9$\\ \hline
\end{tabular}
\end{center}
\caption{Measured 3D 4th order cumulants as in Table~\ref{tab:tab3rdinv}. The corresponding predictions, which would involve extending second order PT to  invariants is left to future work. Those for $\langle x^4 \rangle_\mathrm{c}$ \citep[from][]{Lokas} are indicated in brackets. According to perturbation theory, the dimensionless third order cumulants are proportional to $ \sigma^2$.}
\label{tab:cumulants}
\end{table}

\subsubsection{Growth factor measurement experiment: towards a possible dark energy probe?}
\label{sec:growth}
\begin{figure}
\center{
\subfigure[Minima]{\includegraphics[width=0.3\textwidth]{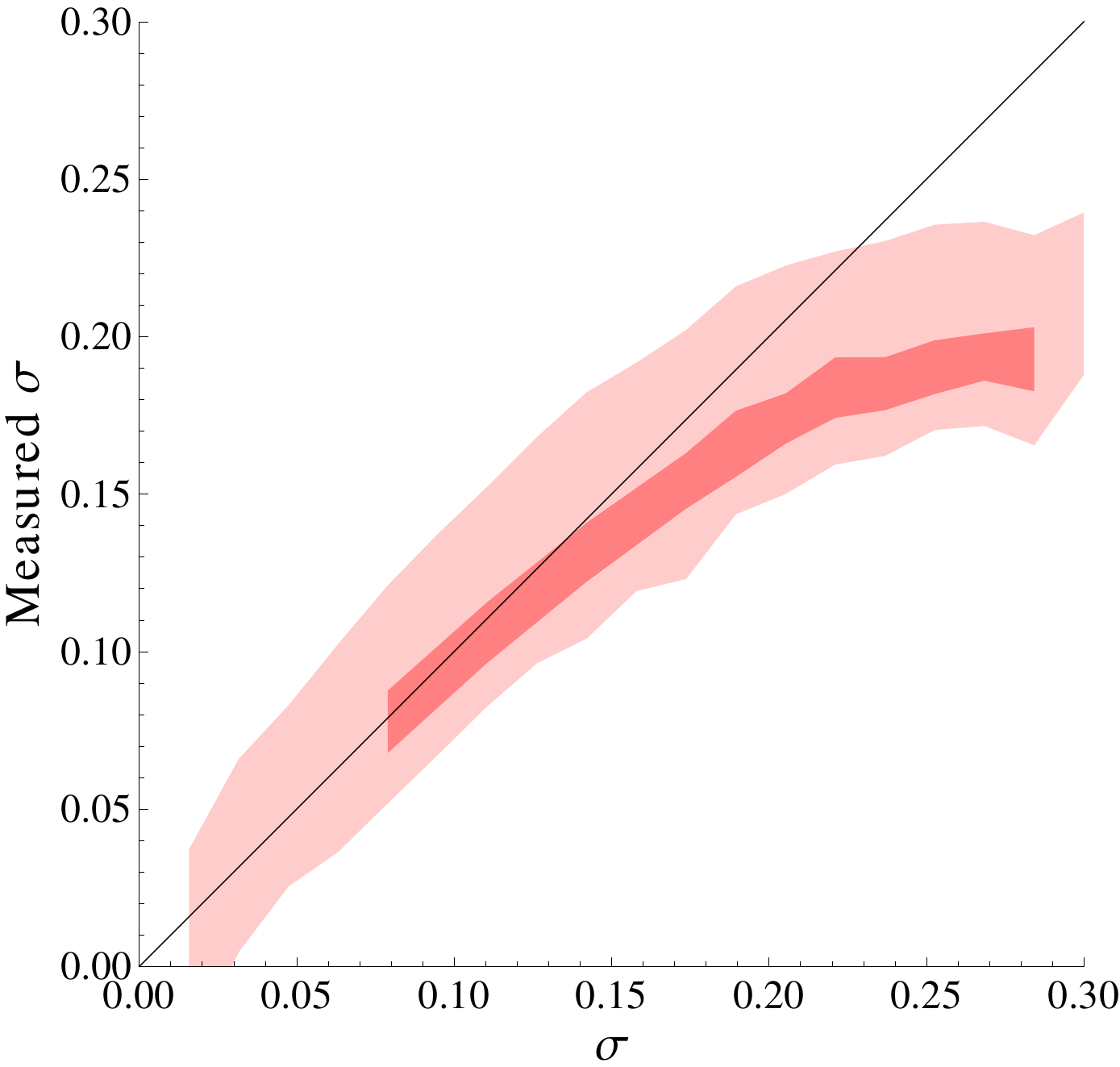}}
\hfil
\subfigure[Pancake-type saddle points]{
\includegraphics[width=0.3\textwidth]{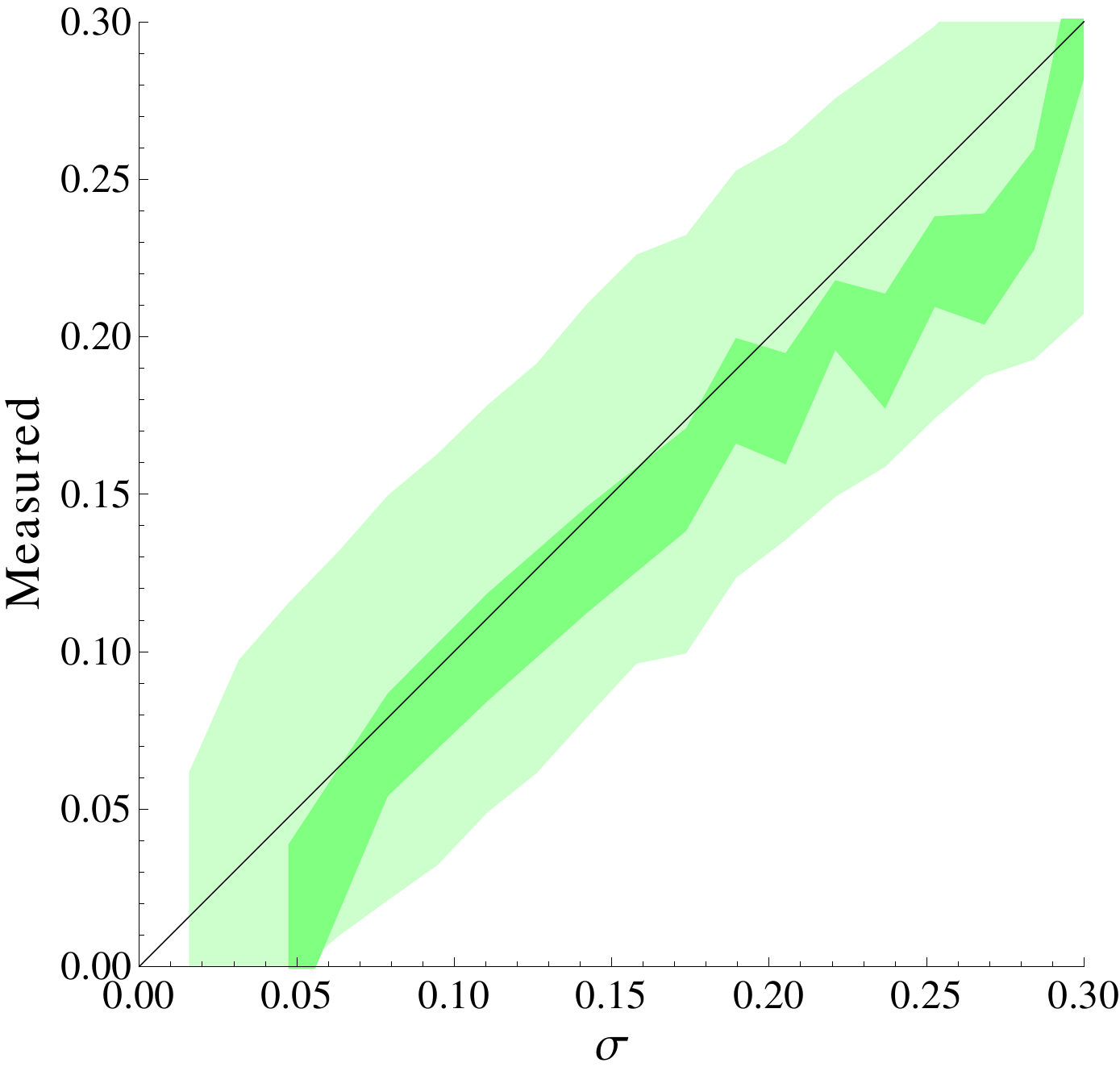}}\hfil
\subfigure[Filament-type saddle points]{
\includegraphics[width=0.3\textwidth]{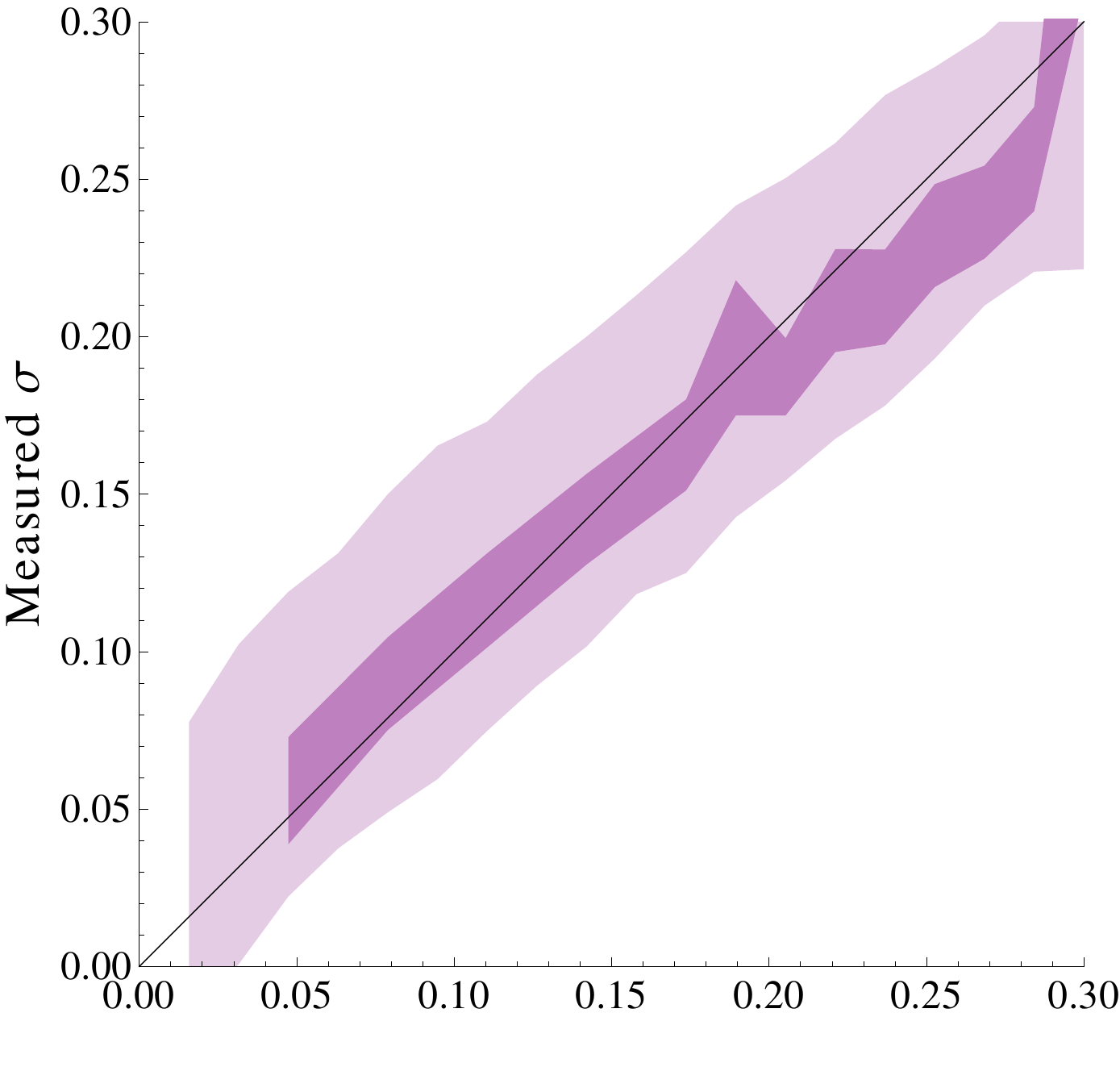}}
}
\subfigure[Maxima]{\includegraphics[width=0.3\textwidth]{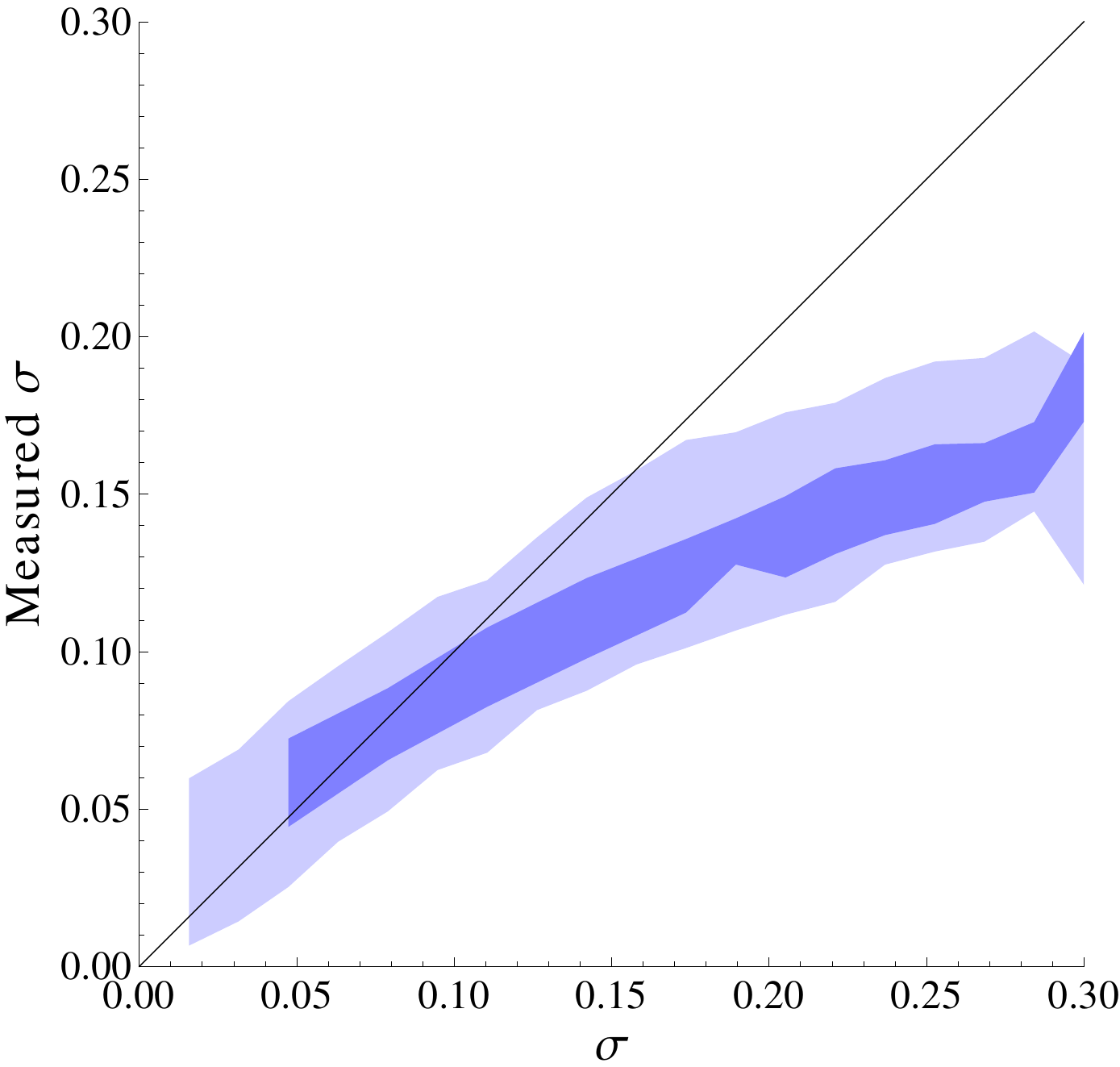}}
\hfil
\subfigure[Euler characteristic]{
\includegraphics[width=0.3\textwidth]{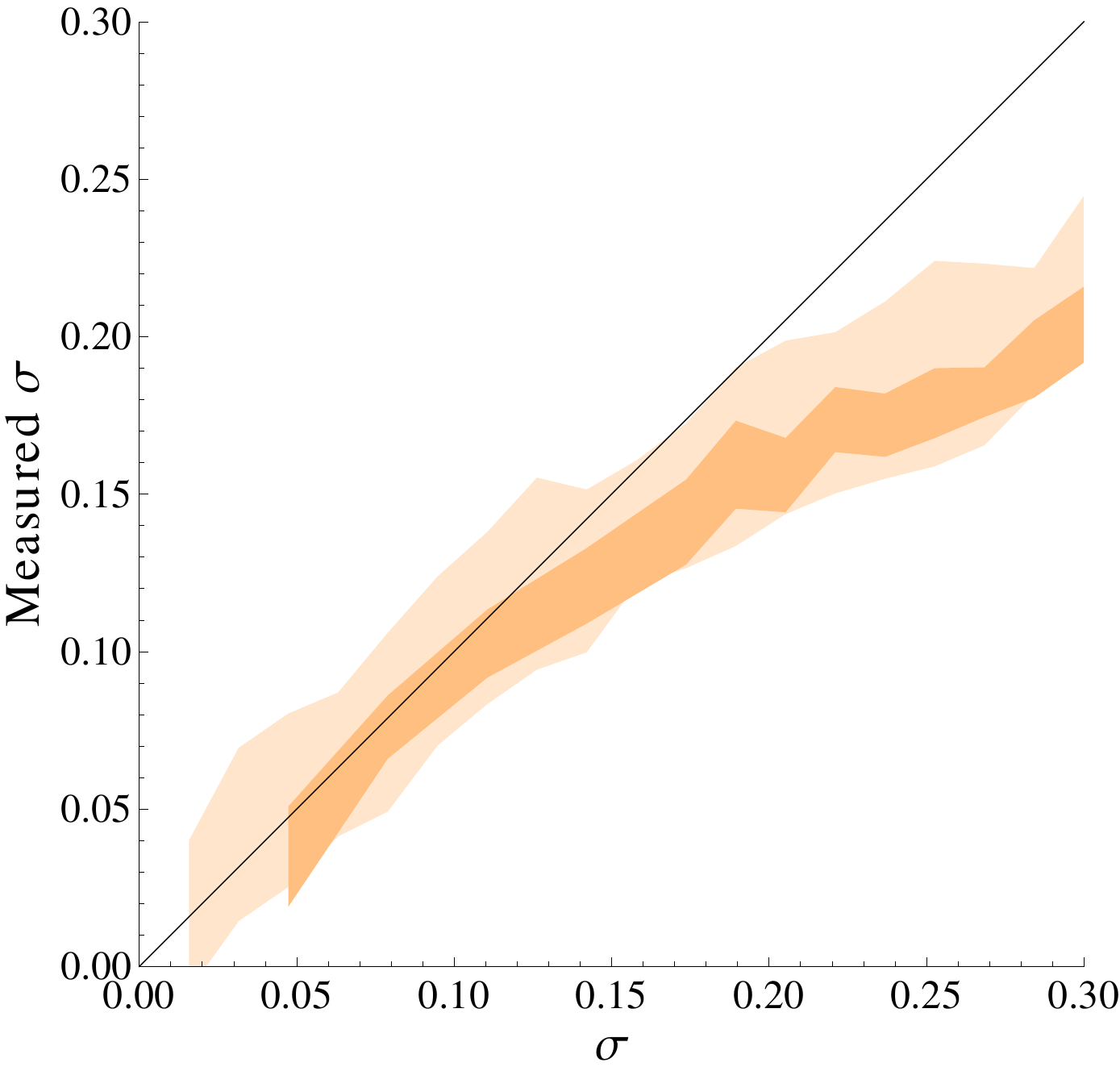}}\hfil
\subfigure[Skeleton]{  
\includegraphics[width=0.3\textwidth]{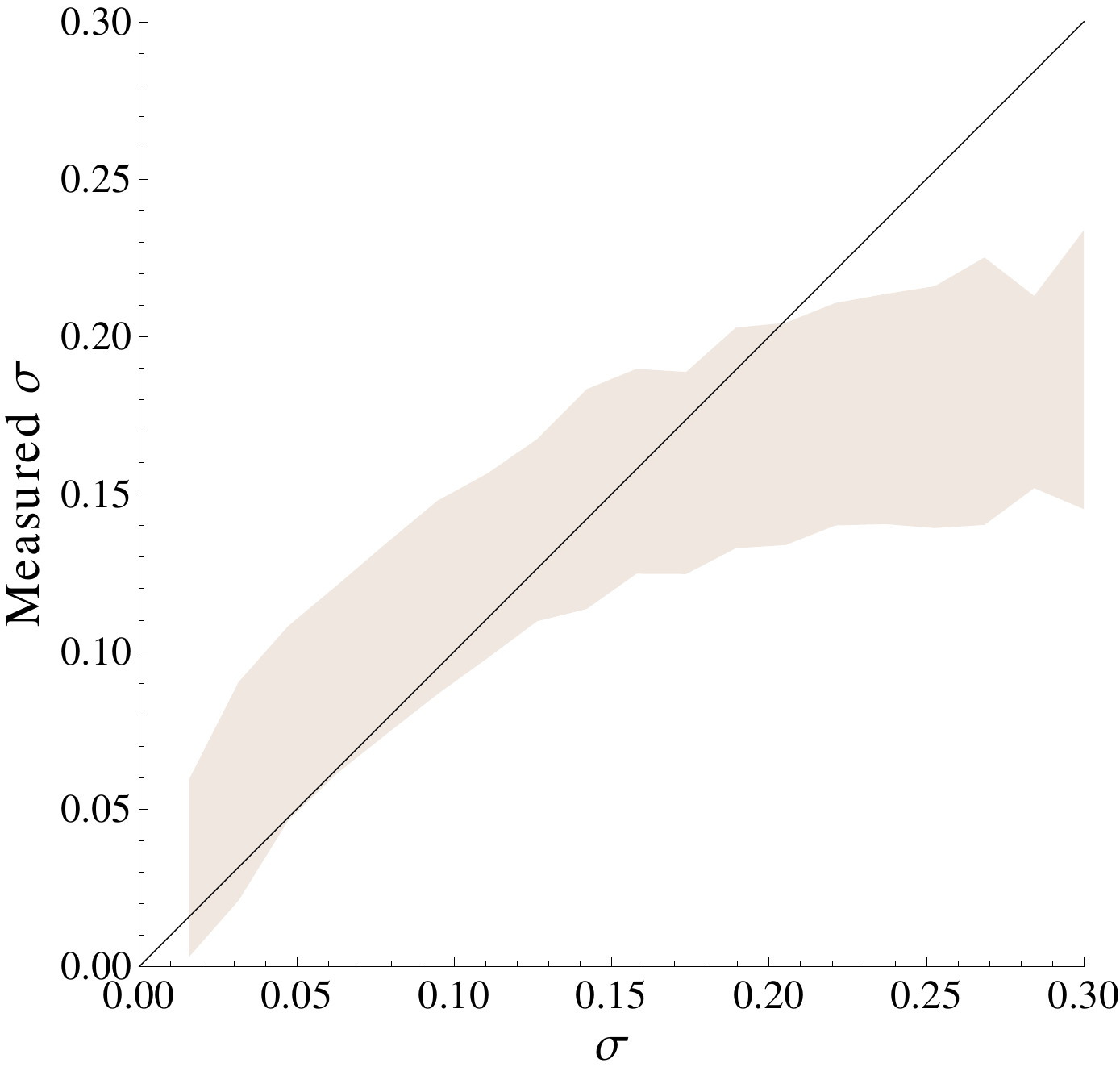}}
 \caption{Fitting $\sigma$ using the Gram-Charlier  and the perturbation theory
results, for the minima distribution (\textit{top left}), the pancake-type
saddle points distribution (\textit{top middle}), the filament-type saddle
points distribution (\textit{top right}), the maxima distribution
(\textit{bottom left}), the Euler characteristic (\emph{bottom middle}) and the
skeleton length (\textit{bottom right}), to a set of  $256^3$, $n_\mathrm{s}=-1$
scale-invariant simulation smoothed over $R=20$ pixels (which corresponds to
approximately 230 maxima) for the larger and lighter contours and $R=10$ pixels
(1800 maxima) for the small and darker ones. The skeleton is smoothed only for
$R=20$ to ensure sufficient smoothness. 
The shaded area corresponds to one standard deviation (computed from 25
simulations).
 Note
that the critical sets which are most sensitive to the higher thresholds tend to
depart from the non-Gaussian linear correction earlier, 
   which is expected in the context of gravitational clustering. 
   \label{fig:fitsigma}
}
\end{figure}

To demonstrate the potential of our non-Gaussian critical set
statistical estimates  to map the evolution of the growth factor $D(z)$ (equations~(\ref{eq:genus_3D}), (\ref{eq:extrema_3D}) and
(\ref{eq:skel_3D})),
let us  discuss a very idealized topological dark energy probe experiment. 
For this purpose, we generate a cosmic sequence of  a set of  25   $n_\mathrm{s}=-1$  scale-invariant simulations with
$256^3$ particle smoothed over $R=10$ and $20$
pixels and 
measure for each redshift the corresponding differential number of extrema,
${\cal E}^{\rm ext}_R(\nu,z)$ (resp. max, min, filament-saddle and pancake
saddle), Euler characteristic ${\cal E}^{\rm euler}_R(\nu,z)$, and skeleton
length,  ${\cal E}^{\rm skel}_R(\nu,z)$.
We  denote the set of such measurements as  $\{{\cal E}^{k}_R(\nu,z)\}_k$, where
$k$ stands for the type of  statistics and $R$ the smoothing scale (c.f. equation~\ref{eq:calE}).  We then
match these  histograms to the 
predictions of the first order Gram Charlier PT model and deduce the best fit
$\hat \sigma$ for  a range of redshifts, using for now a least-squares fit with uniform
weights. This estimated ``geometric'' $\hat \sigma$ is then plotted as a
function of underlying $\sigma$
for the six critical set statistics (4 types of extrema, Euler characteristic
and skeleton length\,\footnote{The skeleton length is only fitted on the range
$\nu<0$ or $\nu>2$ where the agreement between predictions and measurements is
the best.}) and shown in Figure~\ref{fig:fitsigma}. On average,  $\hat \sigma$
is a good match to $\sigma$ (when $\sigma \lwsim 0.2$), with a dispersion
$\sigma^k_{\hat\sigma}$ 
which scales with the smoothing length $R$ and the size of the survey $L$ as 
\begin{equation}
\sigma^k_{\hat\sigma}= \sigma^k_c \left[\left(L_c/R_c \right) \left(R / L \right)\right]^{3/2}\,,
\label{eq:defprefactor}
\end{equation} where the values $\sigma^k_c$ that were measured in our numerical
survey ($L_c=256$ pixels) are given in Table~\ref{tab:sigmas} for two values of
the smoothing scale, $R_c=20$
and $R_c=10$ pixels.
\begin{table}
\begin{center}
\begin{tabular}{|c|c|c|c|c|c|c|}
\hline
$R_c=20$ &
$\sigma^{\rm max}_c \sim0.030\,,$ &
$ \sigma^{\rm min}_c \sim0.034\,,$ & 
$\sigma^{\rm sad-fil}_c \sim0.049\,,$ &
$ \sigma^{\rm sad-pan}_c \sim0.064\,, $& 
$ \sigma^{\rm skel}_c \sim0.033\,,$ & 
$  \sigma^{\rm Euler}_c \sim0.027\,, $ \\
\hline
$R_c=10$ &
$\sigma^{\rm max}_c \sim0.012\,, $&
$ \sigma^{\rm min}_c \sim0.010\,, $& 
$\sigma^{\rm sad-fil}_c \sim0.017\,, $& 
$\sigma^{\rm sad-pan}_c \sim0.017\,, $&   
$-$&
$\sigma^{\rm Euler}_c \sim0.012\,. $\\
\hline
\end{tabular}
\end{center}
\caption{Measured dispersion around the mean value extracted from the scale
invariant experiment described in Figure~\ref{fig:fitsigma} ($L_c = 256$
pixels) for smoothing lengths $R_c=20$ and $10$ pixels.
}
\label{tab:sigmas}
\end{table}%
%
 %
 %
The deviation at larger $\sigma$ seen on  Figure~\ref{fig:fitsigma} could be explained as an effect of the higher order non-Gaussianities
 and should be reduced by going beyond a first order description. This is corroborated by the observations
 that the critical sets which are least sensitive to the higher and, presumably, more non-linear thresholds (e.g. minima and saddle points)
 tend to depart from the non-Gaussian linear correction later.
On the other hand, our statistics are less sensitive at very low $\sigma$'s due to the universal behaviour of the 
geometrical descriptors in the Gaussian limit. Hence the optimal level for the recovery of the variance $\hat \sigma$ is
around $\sigma=0.1$, according to Figure~{\ref{fig:fitsigma}. Thus, 
varying the smoothing scale  with redshift, $R(z)$, in such a way as to ensure a level of non-linearity near
$\sigma(z) \approx 0.1$ may help  the   reconstruction of $D(z)$. 
Note finally and importantly that the overall spread in the measured $\hat
\sigma$ reflects cosmic variance and could be reduced in proportion to the
square root of the volume of the survey, following equation~(\ref{eq:defprefactor}). For instance, a naive extrapolation of
the cluster CFHTLS Wide counts\footnote{http://www.astromatic.net/iip/w1.html} 
\citep{Durret2011} to a full sky experiment with 40 bins of redshift below
$z=1.15$ would yield a noise-to-signal ratio of 1\%  per bin for  $D(z)$ using
the Euler characteristic alone.

  Let us now briefly discuss how these very idealized measurements can be 
extrapolated to anticipate the accuracy of  a  dark energy experiment based on critical sets.
A dark energy probe may directly attempt to estimate the so-called equation of state parameters $(w_a,w_0)$ from these critical sets while relying on
 the cosmic model for the growth rate \citep{Glazebrook}, 
\begin{equation}
D(z|w_0,w_a)=\frac{5\Omega_0 H_0^2}{2} H(a)\int_0^a \frac{{\rm d} a'}{a'^3 H^3(a')}\,, \,\, 
H^2(a)=H_0^2\left[ \frac{\Omega_0}{a^3}+ \Omega_\Lambda \exp\left(3 \int_0^z \frac{1+w(z')}{1+z'} {\rm d} z'  \right)
 \right], \, \label{eq:cosmo}
\end{equation}
 with $\Omega_m$, $\Omega_\Lambda$ and $H_0$ resp. the dark matter and dark energy  densities and the Hubble constant at redshift $z=0$,  $a\equiv1/(1+z)$ the expansion factor, and with the equation of state $w(z)=w_0+ w_a {z}/({1+z})$.
Given equations~(\ref{eq:calE}) 
and~(\ref{eq:cosmo}), optimizing  the probability of observing all counts at all redshifts  with respect to  $(w_a,w_0)$  
yields a maximum likelihood estimate for the dark energy equation of state parameters\footnote{Here we assume for simplicity that the different counts are uncorrelated; an improvement would be to use multinomial statistics.
}:
 \begin{equation}
 (\hat w_0,\hat w_a)= \arg \max_{w_0,w_a} \left\{\prod_{z,\nu_i,k }   {\rm Poisson}\left( {\cal E}^k_{R}[\nu_i,D(z| w_a,w_0)] , {E}^k_{z,i}
 \right) \right\} \,, \label{eq:defDop2t}
 \end{equation}
where  the predicted ``number-counts'', ${\cal E}^k_{R}[\nu_i,D(z| w_a,w_0)] $ 
are given by equation~(\ref{eq:calE}), 
 given equation~(\ref{eq:cosmo})\footnote{As the Euler characteristic is 
cumulative, the model needs to be differentiated since 
equation~(\ref{eq:defDop2t}) assumes that the counts are independent. In other
words,
 ${\cal E}^{\rm Euler}_{R}(\nu,z)\equiv \partial \chi/\partial \nu$ given by  equation~({\ref{eq:diffchi}}).}.
Here $E^k_{z,i}$ is the {\sl measured} $k$ statistics in the
threshold bin $i$ at redshift $z$, and $ {\rm Poisson}(\mu,x)$ is the PDF of a
Poisson process, $x$, of mean $\mu$.
   A joint analysis of all  critical sets, $\{{\cal E}^k(\nu,z)\}_k$, is achieved via the product on  $k$ (assuming somewhat optimistically that all estimators are independent).
   
 We shall not carry such a maximization in full  here. 
 Instead, Figure~\ref{fig:like3D} displays (very naive) likelihood contours for  an experiment corresponding to the CFHT-LS extrapolation \citep{Durret2011} of the cluster counts  to a 1/4 sky experiment ($2 \%$ error on $D(a)$ for 40 bins,  which corresponds to the appropriate rescaling of equation~(\ref{eq:defprefactor}) given the prefactors of Table~\ref{tab:sigmas}).  Here we simply assume that 
 the fit to the critical sets yields access to a noisy $D(a|w_0,w_a)$ via equation~(\ref{eq:cosmo})  (for a Gaussian noise with a SNR of 50= 1/0.02); 
 for a grid of models obeying equation~(\ref{eq:cosmo}) we therefore compute the likelihood of a given draw. The corresponding $3$-sigma contours suggest that critical sets 
 could constrain $w_0$ to $\sim 5 \%$ and $w_a$ to $\sim 10 \%$.

Beyond the obvious over-simplifications of this model,
here are a few limitations/caveats which need to be discussed. (i) the accuracy of the predictions,  ${\cal E}^k(\nu,z)$ is as good 
as PT; it therefore requires a galaxy sample large enough to probe the larger scales, so that $\sigma\lwsim 0.2$.
On scales of the order of $30$ Mpc$/h$,
 given that $\sigma(R,z) =\sigma_0(R) D(z)$, with $\sigma_0(R)$ the r.m.s of the cosmic density field at redshift zero  smoothed over $R$, if we rely on the estimates for $\sigma$ given by equation~(\ref{eq:defprefactor}), we may expect to probe $D(z)$ near $z\sim 0.5$\footnote{Note that the invariants moments $q^2$, $J_1$ or $J_2$ involve higher order derivatives, and are therefore sensitive to smaller, less linear scales.};
 (ii)  for scale invariant power spectra, the cumulants entering  
equations~(\ref{eq:genus_3D}), (\ref{eq:extrema_3D}) and (\ref{eq:skel_3D}) and the shape parameter, $\gamma$ depend on  the power-law index $n_\mathrm{s}$ (see Appendix~\ref{sec:PTresults}). For realistic $\Lambda$CDM cosmologies, following \cite{Bernardeau} one strategy could be to let the  index $n_\mathrm{s}$ be a  (redshift dependent) parameter to be also estimated while fitting the counts\footnote{Alternatively, cumulants like equation~(\ref{eq:finalcum}) can be computed numerically for realistic power spectra.}; 
 (iii)
 one could also estimate the optimal weighting strategy to best constrain $(w_0,w_a)$ using combinations of critical set estimators, varying threshold and redshifts weights. This is achieved by replacing 
  $   {\rm Poisson}\left( {\cal E}^k_{R}(\nu_i,z) , {E}^k_{z,i}
 \right) $ in equation~(\ref{eq:defDop2t}) by  $   {\rm Poisson}\left( w^z_{i,k} {\cal E}^k_{R}(\nu_i,z) ,w^z_{i,k} {E}^k_{z,i}
 \right) $ and minimizing the error on the estimated dark energy parameters w.r.t. the positive weights $w^z_{i,k}$;
 (iv) in this paper, we   ignored all issues regarding completeness,
masking, etc. (see Section~\ref{sec:conclusion}).
In closing, 
 note that the 2D statistics, equations~(\ref{eq:genus_2D}), (\ref{eq:number_counts_2D}) and (\ref{eq:skel_2D}) could also be implemented on non-linear projected data, given some dynamical model for the {\sl projected} cumulants, e.g. in the context of weak lensing  
 \citep[see e.g. ][]{Matsubara,Kratochvil,Kratochvil2,Marian}.
 
 Carrying out the road map sketched in this section while addressing these issues could be one of the target of the upcoming  surveys that have been planned specifically
to probe dark energy, either from ground-based facilities (eg VST-KIDS, %
DES, %
 Pan-STARRS, %
LSST\footnote{\texttt{http://www.astro-wise.org/projects/KIDS/},\, \texttt{https://www.darkenergysurvey.org/},\,\texttt{http://www.lsst.org/}}) 
or space-based observatories (EUCLID, SNAP and JDEM\footnote{\texttt{http://sci.esa.int/euclid/},\,\texttt{http://snap.lbl.gov/},\,\texttt{http://jdem.lbl.gov/}}). 

In this paper, we have chosen to focus mainly on the non-Gaussianities arising from gravitational instability. It should be noted however that our formalism is more general.
 The Gram-Charlier expansion is valid for any mildly non-Gaussian field, while perturbation theory only enters the prediction of the cumulants. 
  An important example of situation where the cumulants can be computed independently is the $f_\mathrm{NL}$ parametrisation of non-Gaussianities in the CMB. In this parametrisation, cumulants can be computed and our formalism can be adapted to give predictions of the evolution of the properties of the field with the value of $f_\mathrm{NL}$. For example, \cite{Fedeli2011} recently used the skeleton to constrain $f_\mathrm{NL}$ on two-dimensional weak-lensing maps. Their approach was numerical and our formalism could provide the theoretical framework for this type of investigation.

\begin{figure}
 \includegraphics[width=0.4\textwidth]{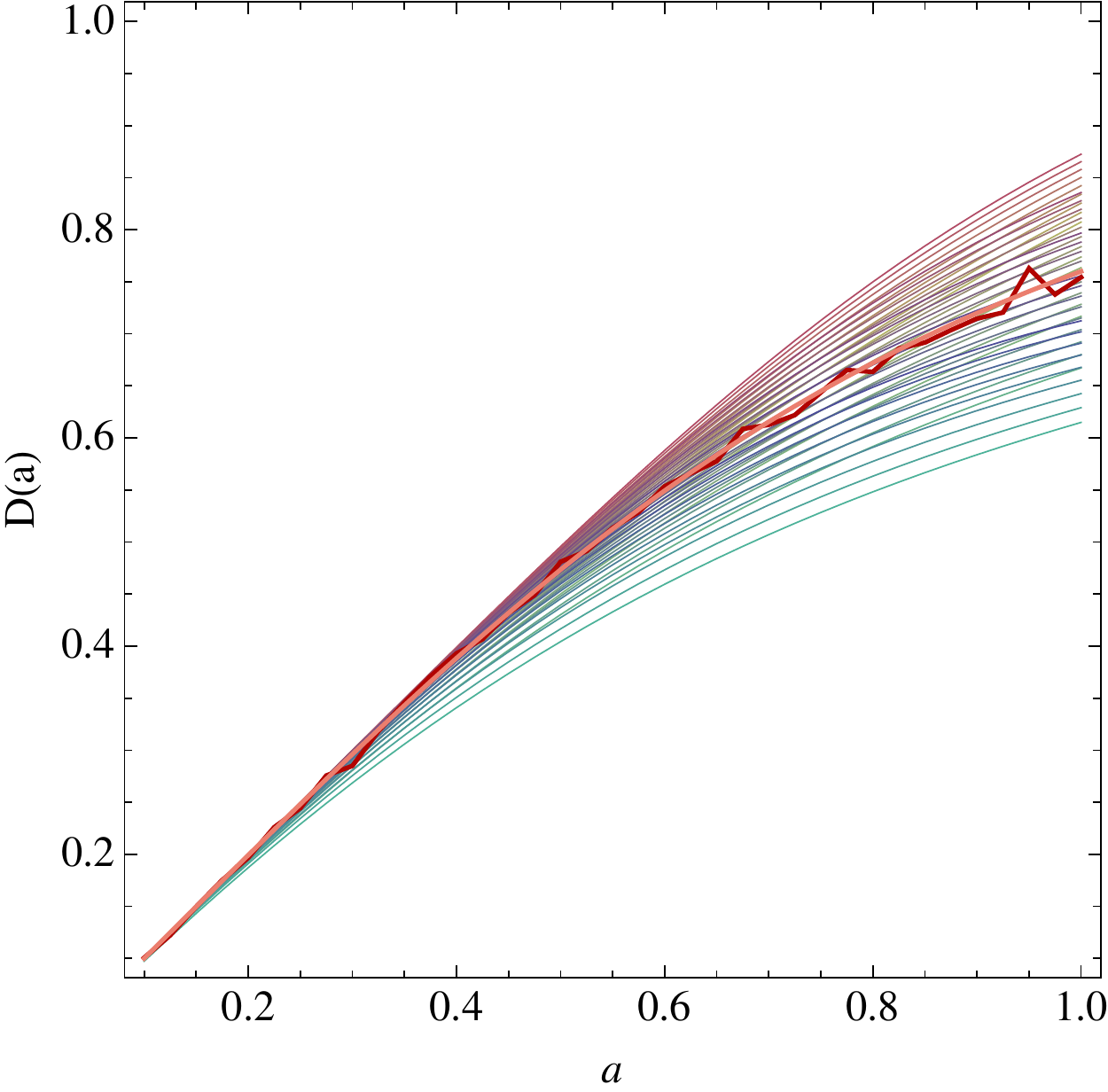} \hskip 2cm
  \includegraphics[width=0.4\textwidth]{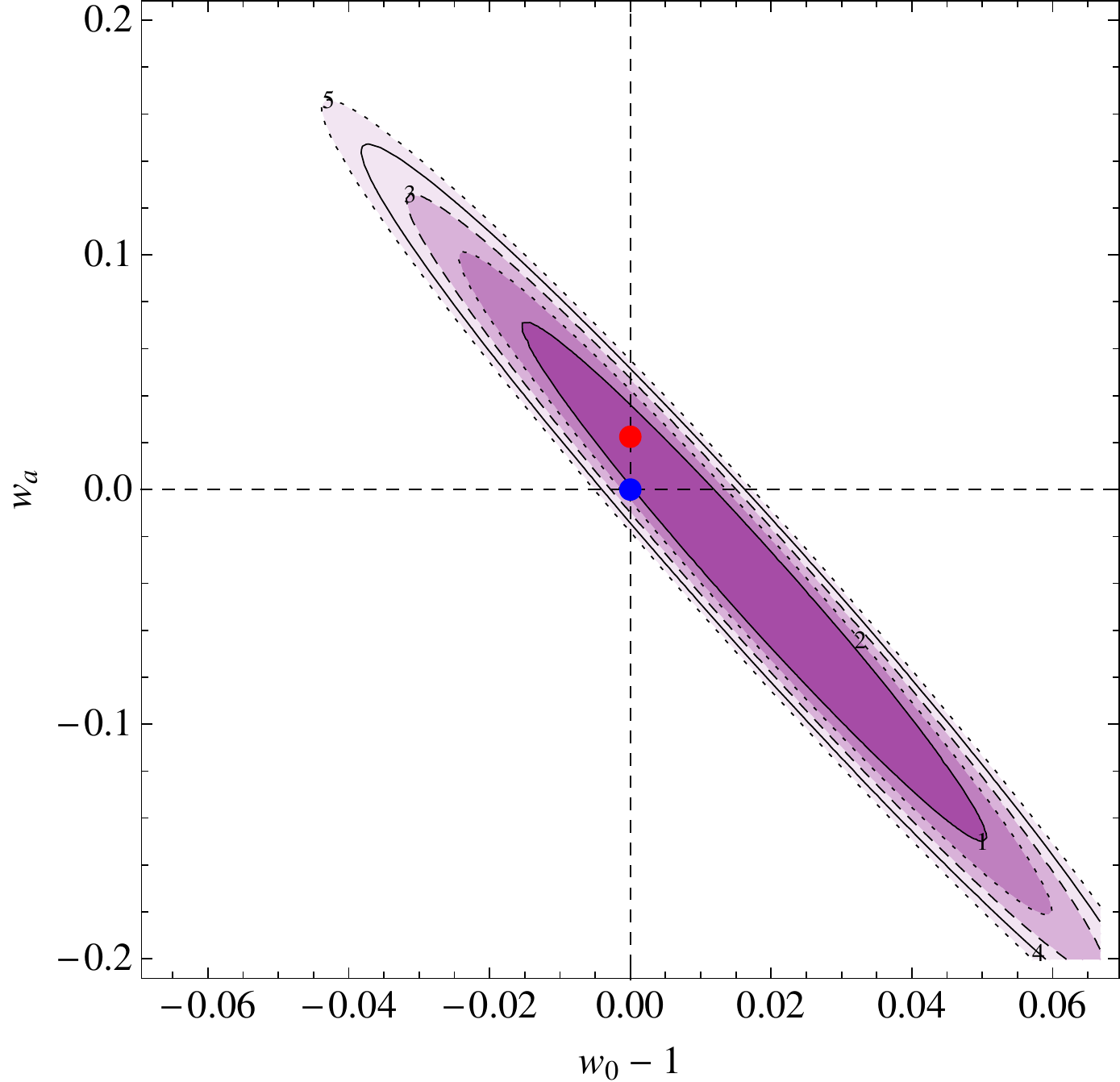}
 \caption{{\sl Left:} $D(a)$ versus $a$ for the pseudo data together with a bundle of models with different values of $w_0$ and $w_a$; {\sl right:} one, two, three, four and five sigmas likelihood contour in the $(w_0,w_a)$ plane for a  one quarter sky experiment (consistent with a significant fraction of  the number of clusters found in the CFHT-LS wide survey) leading to a 2 \% Gaussian relative error on $D(a)$ for 40 bins in $a$ between 0.1 and 1. The red and blue dots correspond to the recovered most likely solution and the input value. For such an experiment,
 5 \% on $w_0$ and 10 \% on $w_a$ seems achievable.
   \label{fig:like3D}
}
\end{figure}

\section{Conclusion and discussion}
\label{sec:conclusion}
We have introduced the Gram-Charlier expansion to describe the point probability distribution  of a field and its derivatives (equations (\ref{eq:2DP_general}) and (\ref{eq:3DP_general})) in the mildly non-Gaussian regime in two and three dimensions.
We have derived from these 2D and 3D PDF the non-Gaussian expressions for the Euler characteristic, equations~(\ref{eq:genus_2D}) and (\ref{eq:genus_3D}),  for differential  extrema counts, equations (\ref{eq:number_counts_2D}) and (\ref{eq:extrema_3D}),  and for the skeleton differential length, equations (\ref{eq:skel_2D})  and (\ref{eq:skel_3D}).
For the Minkowski functionals, including the Euler characteristic (and therefore the rare event tail of the differential number counts) the expansion is valid to arbitrary order in Gram-Charlier and is also given explicitly to third order in r.m.s of the  density field. 
All differential distributions were checked against Monte Carlo realizations of the field and its critical sets: excellent agreement was found for extrema counts (and therefore for the Euler characteristic) for dispersion up to $\sim 0.2$. 
For the skeleton, some level of discrepancies still arises for  some intermediate contrasts (see Figures~\ref{fig:sklnongauss} and \ref{fig:skel3D}). The origin of this disagreement 
 may originate either from the stiff approximation behind the theory, or residual biases in the measurements.   As argued in Section~\ref{sec:growth}, the range of contrasts corresponding to this mismatch may in practice be masked out when fitting observed differential lengths.  
\\
 Furthermore, using gravitational clustering perturbation theory, we have shown how to predict all the 3D cumulants that appear in the Gram-Charlier expansion from  non-linear gravitational dynamics.
  We  can therefore predict the evolution of quantities such as the Euler characteristic, the density of extrema and the differential length of the skeleton as a function of r.m.s of the cosmic density field. The evolution of these critical sets can thus be related to the evolution of the underlying dark matter field. This connection is achieved via the gravitational distortion
  of quantities which can be measured independently of any monotonic bias.
We have sketched a possible dark energy  experiment based on the statistical analysis of these critical sets  in order to constrain
its equation of state, which in the very idealized regime we investigated, seems promising.

Possible extension of this work beyond the scope of this paper involve i)
implementing the dark energy estimation of section~\ref{sec:growth} first on
realistic mock catalogs extracted from large scale simulations such as 
\cite{teyssier2008}, and eventually on the above cited surveys, while accounting for incomplete, magnitude limited samples and modeling the corresponding biases; 
 one remaining task would be to demonstrate that  the statistical analysis of geometrical critical sets can be competitive in practice to constrain this equation of state, compared to e.g. weak lensing \citep{Kaiser,Bridle2007,euclid,aski} or Baryonic acoustic oscillations   probes  \citep{Zunckel2010}; 
ii) deriving the invariant JPDF for higher order derivatives and/or anti-derivatives (e.g. to link geometry to the kinematic of the flow along filaments), or for fields of higher dimensions (e.g. to study the saddle points of  high dimensions landscapes), following the tracks presented in \cite{pogoskel}; 
iii) extending the theory of critical sets to walls and manifolds of higher dimensions; 
iv) exploring the implication of the departure from a Gaussian JPDF on other statistics, such as the mean connectivity of the peaks of the cosmic web, following \cite{CGPP2010};
v) implementing second order PT, or the so-called renormalized perturbation theory \citep{Crocce} in order to extend the domain of applicability  of these asymptotic expansions, and possibly
build a generalized stable clustering model for our cumulants.
vi) extending the formalism to N-points statistics, e.g. to investigate the non-Gaussian correlation of peaks, or non-Gaussian count-in-cells.

\section*{Acknowledgments}
We thank S.~Colombi,  V.~Desjacques,  D.~Weinberg, F.~Bernardeau,
 S.~Prunet, and K.~Benabed for comments,  T.~Sousbie for providing us with his {\em \tt skeleton} code and D.~Munro for freely
distributing his Yorick programming language and OpenGL interface (available
at {\em\tt http://yorick.sourceforge.net/}). 
CG and CP thanks University of Alberta for hospitality in the spring
of 2010 and in the spring--summer of 2011.
Special thanks from L.~Vigroux and the Institut d'Astrophysique for funding some critical stages  of this work.
DP thanks the University of Oxford, the France-Canada Mobility fund, and the  CNRS (France) for
support through a ``poste rouge'' visiting position during Summer 2009 when
this investigation was initiated within  the framework of
the  Horizon  project, \texttt{www.projet-horizon.fr}. We warmly thank S. Prunet and S. Rouberol for running the \texttt{Horizon} cluster for us. 
CP acknowledges support from a Leverhulme visiting professorship at the Physics department of the University of Oxford,  and thanks Merton College, Oxford for a visiting
fellowship.

\bibliographystyle{mn2e}
\bibliography{DofZ}

\appendix

\section{Derivations of the Euler characteristic in 2 and 3D}
\label{sec:deriv}

\subsection{Euler characteristic in 2D}
\label{sec:deriveuler2D}

For a two-dimensional random field, the Euler characteristic $\chi(\nu)$
of the excursion set $x > \nu$
is given by:
\begin{equation}
\chi(\nu) = \frac{1}{{R_*^2}} \int_{-\infty}^\infty {\rm d} J_1 
\int_{\frac{\nu+\gamma J_1}{\sqrt{1-\gamma^2}}}^{\infty} {\rm d} \zeta
\int_0^\infty {\rm d} J_2 P_\mathrm{ext} (\zeta, J_1, J_2) I_2  , \label{eq:eulerformula}
\end{equation} where $P_\mathrm{ext}$ is the distribution function under the condition of zero gradient:
\begin{equation}  
P_\mathrm{ext}(\zeta, J_1, J_2) =  \int_0^{+\infty} {\rm d}q^2 P(\zeta, q^2, J_1, J_2) \delta_D (x_1) \delta_D (x_2).
\end{equation}
Using polar coordinates $
P_\mathrm{ext}(\zeta, J_1, J_2) =  \frac{1}{\pi} P(\zeta, q^2=0, J_1, J_2),
$
and
using the fact that $L_j (0) = 1$ for any $j$ we have:
\begin{equation}
P_\mathrm{ext}(\zeta, J_1, J_2) = \frac{1}{\pi} G_{\rm 2D}(\zeta, J_1, J_2)
\left[ 1 + 
\sum_{n=3}^\infty \sum_{i,j,k,l}^{i+2j+k+2 l=n} 
\frac{(-1)^{j+l}}{i!\;j!\; k!\; l!} 
\left\langle \zeta^i {q^2}^j {J_1}^k {J_2}^l \right\rangle_\mathrm{{\scriptscriptstyle GC}}
H_i\left(\zeta\right) H_k\left(J_1\right) L_l\left(J_2\right)
\right].
\label{eq:2DP_general_ext}
\end{equation}
Remarkably, the Euler characteristic can be evaluated completely for this general distribution
function.
For this we note that $I_2=\frac{1}{4}(J_1^2-J_2)=
\frac{1}{4}\left(H_2(J_1)+L_1(J_2)\right)$, thus, the calculations are reduced
to the integration of  products of Hermite and Laguerre polynomials.
As the first step, the integral over $J_2$ leaves only the terms
with $L_0(J_2)$ ($l=0$) or $L_1(J_2)$ ($l=1$) in the Gram-Charlier expansion:
\begin{multline}
\chi(\nu) = \frac{1}{4 \pi {R_*^2}}
\int_{-\infty}^\infty {\rm d} J_1 
\int_{\frac{\nu+\gamma J_1}{\sqrt{1-\gamma^2}}}^{\infty} {\rm d} \zeta
G (\zeta, J_1)  
\left[ \vphantom{\sum_{i,j,k}^{i+2 j+k=n}} H_2(J_1)  \right.
 + H_2(J_1) \sum_{n=3}^\infty \sum_{i,j,k}^{i+2 j+k=n} 
\frac{(-1)^{j}}{i!\;j!\; k!} 
\left\langle \zeta^i {q^2}^j {J_1}^k \right\rangle_\mathrm{{\scriptscriptstyle GC}}
H_i\left(\zeta\right) H_k\left(J_1\right) 
\\
+ \left. \sum_{n=3}^\infty \sum_{i,j,k}^{i+2 j+k+2=n} 
\frac{(-1)^{j+1}}{i!\;j!\; k!} 
\left\langle \zeta^i {q^2}^j {J_1}^k {J_2} \right\rangle_\mathrm{{\scriptscriptstyle GC}}
H_i\left(\zeta\right) H_k\left(J_1\right)
\right].
\end{multline}
Let us first establish the important relation between some coefficients of the expansion for an
isotropic field by evaluating $\chi(-\infty)$.  
Integrating over full range of $\zeta$ selects $i=0$ terms,
\begin{multline}
\chi(-\infty) = \frac{1}{4 \pi {R_*^2}}
\int_{-\infty}^\infty {\rm d} J_1 G (J_1)  
\left[ \vphantom{\sum_{i,j,k}^{i+2 j+k=n}} H_2(J_1)  \right.
+ H_2(J_1) \sum_{n=3}^\infty \sum_{j,k}^{2 j+k=n} 
\frac{(-1)^{j}}{j!\; k!} 
\left\langle {q^2}^j {J_1}^k \right\rangle_\mathrm{{\scriptscriptstyle GC}}
H_k\left(J_1\right) 
\\
+ \left. \sum_{n=3}^\infty \sum_{j,k}^{2 j+k+2=n} 
\frac{(-1)^{j+1}}{j!\; k!} 
\left\langle {q^2}^j {J_1}^k {J_2} \right\rangle_\mathrm{{\scriptscriptstyle GC}}
H_k\left(J_1\right)
\right] 
\end{multline}
and subsequent evaluation retains only $j=n/2-1$ in each even
order of the expansion 
\begin{equation}
\chi(-\infty) = \frac{1}{4\pi {R_*^2}}  
\sum_{j=1}^{\infty} \frac{(-1)^{j}}{j!} 
\left[
\left\langle {q^2}^j {J_1}^2 \right\rangle_\mathrm{{\scriptscriptstyle GC}}
- \left\langle {q^2}^j {J_2} \right\rangle_\mathrm{{\scriptscriptstyle GC}}
\right]
= \frac{1}{\pi {R_*^2}}  
\sum_{j=1}^{\infty} \frac{(-1)^{j}}{j!} 
\left\langle {q^2}^j {I_2} \right\rangle_\mathrm{{\scriptscriptstyle GC}} .
\label{eq:2De_infty}
\end{equation}
The limiting Euler number $\chi(-\infty)$ is determined by the
topology of the manifold which the sum of the cumulants in Eqs.~(\ref{eq:2De_infty}) must obey.
We note that in this derivation we have  obtained only the non-Gaussian contribution to $\chi(-\infty)$ 
which comes via hierarchy of the terms 
 $\left\langle {q^2}^j I_2 \right\rangle_\mathrm{{\scriptscriptstyle GC}}, ~ j\ge 1$.
 The Gaussian term of the same structure corresponds to $j=0$ and was set to
zero by our normalization $\langle I_2 \rangle = \langle J_1^2 \rangle - \langle J_2 \rangle = 0$.
It is an open question whether this is strictly consistent only with 
$\chi(-\infty)=0$ as, e.g., in an infinite Eucledian space, but requires more accurate treatment
on the 2D sphere where $\chi(-\infty)=2$. For now, we will leave the boundary
term $\chi(-\infty)$ included in our expressions but unspecified. This is 
also justified by the practical case of the boundary
topology  introduced by a mask on the data. In this case the moments of the field
remain the same as on unmasked manifold,
but $\chi(-\infty)$ measured in realizations will nevertherless reflect the topology of the mask,
being now disentangled from moment statistics.

Returning to the differential case, we use the following property of the
Hermite polynomials:
\begin{equation}
\int_a^\infty {\rm d} \zeta \; G(\zeta) H_i(\zeta) =  \left\{
\begin{array}{ll}\displaystyle
\frac{1}{2} \mathrm{Erfc}\left(\frac{a}{\sqrt{2}}\right)~, & i=0\,,\\
\displaystyle
\frac{1}{\sqrt{2\pi}} \exp\left(-\frac{a^2}{2} \right) H_{i-1}(a) ~, & i > 0\,.
\end{array}
\right.
\label{eq:Hermite_propI}
\end{equation}
Let us first treat the $i=0$ case, expanding an appearing product
of Hermite polynomials back into Hermite polynomials
$\displaystyle H_p(J_1) H_k(J_1) = \sum_{s=0}^{\mathrm{min}(p,k)} 
\frac{p! k!}{s!(p-s)!(k-s)!} H_{k+p-2s} 
\left(J_1\right) $, the contribution to the Euler characteristic reads
\begin{multline}
\lefteqn{\chi(\nu)_{i=0} = \frac{1}{8 \pi {R_*^2}}
\int_{-\infty}^\infty {\rm d} J_1 G(J_1)
\mathrm{Erfc} \left(\frac{\nu+\gamma J_1}{\sqrt{2(1-\gamma^2)}} \right)
\left[ \vphantom{\sum_{i,j,k}^{i+2 j+k=n}} H_2(J_1) + \right. }
\\
 \left. \sum_{n=3}^\infty \sum_{j,k}^{2 j+k=n} 
\frac{(-1)^{j}}{j!\; k!} 
\left\langle {q^2}^j {J_1}^k \right\rangle_\mathrm{{\scriptscriptstyle GC}}
\sum_{s=0}^{\mathrm{min}(2,k)} 
\frac{2!k!}{s!(2-s)!(k-s)!} H_{k+2-2s} \left(J_1\right) 
+ \sum_{n=3}^\infty \sum_{j,k}^{2 j+k+2=n} 
\frac{(-1)^{j+1}}{j!\; k!} 
\left\langle {q^2}^j {J_1}^k {J_2} \right\rangle_\mathrm{{\scriptscriptstyle GC}}
H_k\left(J_1\right)
\right]. \nonumber
\end{multline}
We now use 
\begin{equation}
\frac{1}{2} \int_{-\infty}^\infty {\rm d} J_1 \; G( J_1 )
\mathrm{Erfc} \left(\frac{\nu+\gamma J_1}{\sqrt{2(1-\gamma^2)}} \right)
H_k(J_1)
=  \left\{
\begin{array}{ll} \displaystyle
\frac{1}{2} \mathrm{Erfc}\left(\frac{\nu}{\sqrt{2}}\right)~, & k=0\,,\\
\displaystyle
\frac{1}{\sqrt{2\pi}} \exp\left(-\frac{\nu^2}{2} \right) 
(-1)^k \gamma^k H_{k-1}(\nu)
~, & k > 0\,,
\end{array}
\right.
\label{eq:Hermite_propII}
\end{equation}
to obtain
\begin{multline}
\chi(\nu)_{i=0} = \frac{1}{8 \pi {R_*^2}}
\mathrm{Erfc} \left(\frac{\nu}{\sqrt{2}} \right)
\sum_{j}^{\infty} \frac{(-1)^{j}}{j!} \left[ 
\left\langle {q^2}^j {J_1}^2 \right\rangle_\mathrm{{\scriptscriptstyle GC}}
- \left\langle {q^2}^j {J_2} \right\rangle_\mathrm{{\scriptscriptstyle GC}} 
\right]
 \\
+
\frac{1}{4 \pi \sqrt{2 \pi} R_*}
\exp\left(-\frac{\nu^2}{2}\right)
\left[ \vphantom{\sum_{i,j,k}^{i+2 j+k=n}} \gamma^2 H_1(\nu)  \right.
+ \sum_{n=3}^\infty \sum_{j,k}^{2 j+k=n} 
\frac{(-1)^{j+k}}{j!} 
\left\langle {q^2}^j {J_1}^k \right\rangle_\mathrm{{\scriptscriptstyle GC}}
\sum_{s=0}^{\mathrm{min}(2,k)} 
\frac{ 2! \gamma^{k+2-2s}}{s!(2-s)!(k-s)!} 
H_{k+1-2s} \left(\nu\right) 
 \\
 + \left. \sum_{n=3}^\infty \sum_{j=0,k=1}^{2 j+k+2=n} 
\frac{(-1)^{j+k+1} \gamma^k }{j!\; k!} 
\left\langle {q^2}^j {J_1}^k {J_2} \right\rangle_\mathrm{{\scriptscriptstyle GC}}
H_{k-1}\left(\nu\right)
\right],
\end{multline}
where in the latter summations only terms that lead to
non-negative index of Hermite polynomials 
should be included.
The remaining $i>0$ terms of the Euler number calculation are
\begin{multline}
\chi(\nu)_{i>0} = \frac{1}{4 \pi \sqrt{2 \pi} {R_*^2}}
\int_{-\infty}^\infty {\rm d} J_1 G(J_1)
\exp\left(-\frac{(\nu+\gamma J_1)^2}{2(1-\gamma^2)}\right) \times
\\
\left[ 
\sum_{n=3}^\infty \sum_{i=1,j,k}^{i+2 j+k=n} 
\frac{(-1)^{j}}{i!\; j!\; k!} 
\left\langle {\zeta^i q^2}^j {J_1}^k \right\rangle_\mathrm{{\scriptscriptstyle GC}}
H_{i-1}\left(-\frac{\nu+\gamma J_1}{\sqrt{1-\gamma^2}}\right)
\sum_{s}^{\mathrm{min}(2,k)} \frac{2!k!}{s!(2-s)!(k-s)!} H_{k+2-2s} \left(J_1\right) 
\right. \\
+ \left. \sum_{n=3}^\infty \sum_{i=1,j,k}^{i+2 j+k+2=n} 
\frac{(-1)^{j+1}}{i!\; j!\; k!} 
\left\langle \zeta^i {q^2}^j {J_1}^k {J_2} \right\rangle_\mathrm{{\scriptscriptstyle GC}}
H_{i-1}\left(-\frac{\nu+\gamma J_1}{\sqrt{1-\gamma^2}}\right)
H_k\left(J_1\right)
\right].
\end{multline}
We need the following convolution property of the Hermite polynomials
\begin{equation}
\frac{1}{\sqrt{2 \pi}}
\int_{-\infty}^\infty {\rm d} J_1 G(J_1)
\exp\left(-\frac{(\nu+\gamma J_1)^2}{2(1-\gamma^2)}\right)
H_{i}\left(-\frac{\nu+\gamma J_1}{\sqrt{1-\gamma^2}}\right)
H_k\left(J_1\right) 
= \frac{1}{\sqrt{2\pi}} \exp\left(-\frac{\nu^2}{2}\right) 
(-1)^k (1-\gamma^2)^{\frac{i+1}{2}} \gamma^k H_{i+k}(\nu)
\label{eq:Hermite_propIII}
\end{equation}
to obtain
\begin{multline}
\chi(\nu)_{i>0} = \frac{1}{4 \pi \sqrt{2 \pi} {R_*^2}}
\exp\left(-\frac{\nu^2}{2}\right)
\left[ 
\sum_{n=3}^\infty \sum_{i=1,j,k}^{i+2 j+k=n} 
\frac{(-1)^{j+k}}{i!\; j!} 
\left\langle {\zeta^i q^2}^j {J_1}^k \right\rangle_\mathrm{{\scriptscriptstyle GC}}
\left( 1-\gamma^2 \right)^{{i}/{2}}
\sum_{s=0}^{\mathrm{min}(2,k)} 
\frac{ 2! \gamma^{k+2-2s}}{s!(2-s)!(k-s)!} 
H_{i+k+1-2s} \left(\nu\right) 
\right. 
 \\
 + \left. \sum_{n=3}^\infty \sum_{i=1,j,k}^{i+2 j+k+2=n} 
\frac{(-1)^{j+k+1}}{i!\; j!\; k!} 
\left\langle \zeta^i {q^2}^j {J_1}^k {J_2} \right\rangle_\mathrm{{\scriptscriptstyle GC}}
\left(1-\gamma^2\right)^{{i}/{2}} \gamma^k
H_{i+k-1}\left(\nu\right)
\right].
\end{multline}
We see that it has the same form as the second part of the $\chi_{i=0}$
contribution 
which can be joined just by extending the range of $i$ to start
from zero. Thus, together with the result for $\chi(-\infty)$ 
we find the full Gram-Charlier expansion for the Euler characteristic, equation~(\ref{eq:genus_2D}).
\subsection{Euler characteristic in 3D}
\label{sec:deriveuler3D}
Following the 2D calculation, one just averages $-I_3=-\left( {J_1}^3 - 3 J_1 J_2 + 2 J_3 \right)/{27}$
over the full range of the second derivatives
\begin{equation}
\chi(\nu) = -  \frac{1}{{R_*^3}}
\int_{-\infty}^\infty  \!\!\!\! {\rm d} J_1 
{\int_{\frac{\nu+\gamma J_1}{\sqrt{1-\gamma^2}}}^\infty \!\!\! {\rm d} \zeta 
\int_0^\infty  \!\!\!\! {\rm d} J_2 \int_{-J_2^{3/2}}^{J_2^{3/2}} \!\! {\rm d} J_3
P_{\rm ext}(\zeta, J_1, J_2, J_3) } I_3.
\end{equation}
In 3D, we have
\begin{equation}P_{\rm ext}(\zeta, J_1, J_2, J_3)= \left( \frac{3}{2\pi} \right)^{3/2} P(\zeta,J_1,J_2,J_3). \end{equation}
Writing $I_3=\left(H_3(J_1)+ 
\frac{6}{5} H_1(J_1) L_1^{(3/2)}\left(\frac{5}{2} J_2\right) + 2 J_3 \right)/{27}  $,
one can perform all the integration explicitly using the orthogonality properties
of the polynomial expansion of $P_{\rm ext}$ given by equation (\ref{eq:3DP_general}).
First, integration over $J_3$ and $J_2$ gives the intermediate formula
\begin{multline}
\chi(\nu) = -\frac{1}{27 {R_*^3}} \left(\frac{3}{2\pi}\right)^{3/2}
\int_{-\infty}^\infty  \!\!\!\! {\rm d} J_1 
{\int_{\frac{\nu+\gamma J_1}{\sqrt{1-\gamma^2}}}^\infty \!\!\! {\rm d} \zeta 
G(\zeta, J_1) } 
\left[ \vphantom{ \sum_{n=3}^\infty} H_3\left(J_1\right) + \right. 
\sum_{n=3}^\infty \sum_{i,j,k}^{i+2 j+k=n} 
\frac{1}{i!\;j!\; k!} \left({\textstyle-\frac{3}{2}} \right)^j 
\left\langle \zeta^i {q^2}^j {J_1}^k \right\rangle_\mathrm{{\scriptscriptstyle GC}}
H_i\left(\zeta\right) H_k\left(J_1\right) H_3\left(J_1\right) 
\\
+  
\sum_{n=3}^\infty \sum_{i,j,k}^{i+2 j+k+2=n} 
\frac{-3 }{i!\;j!\; k! }  \left({\textstyle-\frac{3}{2}} \right)^j 
\left\langle \zeta^i {q^2}^j {J_1}^k {J_2} \right\rangle_\mathrm{{\scriptscriptstyle GC}}
H_i\left(\zeta\right) H_k\left(J_1\right) H_1\left(J_1\right)
+ \left.
\sum_{n=3}^\infty \sum_{i,j,k}^{i + 2j + k + 3=n} 
\frac{2}{i!\;j!\; k!}  \left({\textstyle-\frac{3}{2}} \right)^j  
\left\langle \zeta^i {q^2}^j {J_1}^k J_3 \right\rangle_\mathrm{{\scriptscriptstyle GC}}
H_i\left(\zeta\right) H_k\left(J_1\right)
\right] \nonumber
\end{multline}
where we have used $L_j^{(1/2)} (0)=(2j+1)!!/(2^j j!)$.
The Euler number at the threshold that encompass all the manifold
\begin{align}
\chi(-\infty) &= -\frac{1}{27 {R_*^3}} \left(\frac{3}{2\pi}\right)^{3/2}
\sum_{j=0}^{\infty} 
 \left({\textstyle-\frac{3}{2}} \right)^j 
\left[
\left\langle {q^2}^j {J_1}^3 \right\rangle_\mathrm{{\scriptscriptstyle GC}}
- 3 \left\langle {q^2}^j J_1 J_2 \right\rangle_\mathrm{{\scriptscriptstyle GC}}
+ 2 \left\langle {q^2}^j J_3 \right\rangle_\mathrm{{\scriptscriptstyle GC}}
\right]
\\ &= - \frac{1}{R_*} \left(\frac{3}{2\pi}\right)^{3/2}
\sum_{j=0}^{\infty} 
 \left(-\frac{3}{2} \right)^j
\left\langle {q^2}^j {I_3} \right\rangle_\mathrm{{\scriptscriptstyle GC}}.
\end{align}

As a function of the threshold, $\nu$, the integration is performed using
the properties of the Hermite polynomials, first to get
(given equations~\ref{eq:Hermite_propI}, \ref{eq:Hermite_propII}, \ref{eq:Hermite_propIII})
\begin{multline}
\chi(\nu)_{i=0} = -\frac{1}{27 {R_*^3}} \left(\frac{3}{2\pi}\right)^{3/2} \frac{1}{2}
\int_{-\infty}^\infty  \!\!\!\! {\rm d} J_1 
G(J_1) 
\mathrm{Erfc} \left(\frac{\nu+\gamma J_1}{\sqrt{2(1-\gamma^2)}} \right)
\left[ \vphantom{ \sum_{n=3}^\infty} H_3\left(J_1\right) + \right.
\sum_{n=3}^\infty \sum_{j,k}^{2 j+k=n} 
\frac{1}{(j!\; k!}  \left({\textstyle-\frac{3}{2}} \right)^j 
\left\langle {q^2}^j {J_1}^k \right\rangle_\mathrm{{\scriptscriptstyle GC}}
H_k\left(J_1\right) H_3\left(J_1\right) 
\\
+  
\sum_{n=3}^\infty \sum_{j,k}^{2 j+k+2=n} 
\frac{-3}{j!\; k! } \left({\textstyle-\frac{3}{2}} \right)^j 
\left\langle {q^2}^j {J_1}^k {J_2} \right\rangle_\mathrm{{\scriptscriptstyle GC}}
H_k\left(J_1\right) H_1\left(J_1\right)
+ \left.
\sum_{n=3}^\infty \sum_{j,k}^{2j + k + 3=n} 
\frac{2}{j!\; k!}  \left({\textstyle-\frac{3}{2}} \right)^j  
\left\langle {q^2}^j {J_1}^k J_3 \right\rangle_\mathrm{{\scriptscriptstyle GC}}
H_k\left(J_1\right)
\right]
\end{multline}
and
\begin{multline}
\chi(\nu)_{i>0} = -\frac{1}{27 {R_*^3}} \frac{3^{3/2}}{4\pi^2}
\int_{-\infty}^\infty  \!\!\!\! {\rm d} J_1 
G(J_1) 
\exp\left(-\frac{(\nu+\gamma J_1)^2}{2(1-\gamma^2)}\right)
\\
\left[
\sum_{n=3}^\infty \sum_{i=1,j,k}^{i+2 j+k=n} 
\frac{1}{i!\;j!\; k!}  \left({\textstyle-\frac{3}{2}} \right)^j 
\left\langle \zeta^i {q^2}^j {J_1}^k \right\rangle_\mathrm{{\scriptscriptstyle GC}}
H_{i-1}\left(-\frac{\nu+\gamma J_1}{\sqrt{1-\gamma^2}}\right)
H_k\left(J_1\right) H_3\left(J_1\right)
\right. 
\\   +
\sum_{n=3}^\infty \sum_{i=1,j,k}^{i+2 j+k+2=n} 
\frac{-3 }{i!\; j!\; k! }  \left({\textstyle-\frac{3}{2}} \right)^j 
\left\langle \zeta^i {q^2}^j {J_1}^k {J_2} \right\rangle_\mathrm{{\scriptscriptstyle GC}}
H_{i-1}\left(-\frac{\nu+\gamma J_1}{\sqrt{1-\gamma^2}}\right)
H_k\left(J_1\right) H_1\left(J_1\right)
\\
+ \left.
\sum_{n=3}^\infty \sum_{i=1,j,k}^{i+2j + k + 3=n} 
\frac{2 }{i!\; j!\; k!}  \left({\textstyle-\frac{3}{2}} \right)^j  
\left\langle \zeta^i {q^2}^j {J_1}^k J_3 \right\rangle_\mathrm{{\scriptscriptstyle GC}}
H_{i-1}\left(-\frac{\nu+\gamma J_1}{\sqrt{1-\gamma^2}}\right)
H_k\left(J_1\right)
\right]
\end{multline}
and, finally, equation~(\ref{eq:genus_3D}).

\section{Fits to extrema counts and skeleton in 3D}
\label{sec:fit}

\subsection{Differential extrema counts}
The Gaussian distribution of maxima is a classical result and fitting formulas have been proposed \emph{e.g.} in \cite{BBKS}. For the sake of consistency, we propose a formula using the same functional form as the next fits.
For a scale-invariant power spectrum of spectral index $n_\mathrm{s}$, the Gaussian distribution of maxima is accurately fitted by :
\begin{multline}
 \frac{\partial n_\mathrm{max}^{(0)}}{\partial\nu} = \frac{1}{10^3 \sqrt{2\pi}} e^{-\frac{\nu^2}{2}} \Bigg[
\left(\frac{97}{30}+\frac{n_\mathrm{s}}{143}-\frac{n_\mathrm{s}^2}{23}-\frac{n_\mathrm{s}^3}{85}\right)
+\left(\frac{182}{29}-\frac{4 n_\mathrm{s}}{25}-\frac{15 n_\mathrm{s}^2}{31}+\frac{n_\mathrm{s}^3}{78}\right) H_1(\nu)\\
+\left(\frac{368}{65}-\frac{2 n_\mathrm{s}}{19}-\frac{27 n_\mathrm{s}^2}{38}-\frac{n_\mathrm{s}^3}{23}\right) H_2(\nu)
+\left(\frac{97}{27}-\frac{2 n_\mathrm{s}}{23}-\frac{8 n_\mathrm{s}^2}{15}-\frac{n_\mathrm{s}^3}{22}\right) H_3(\nu)
+\left(\frac{61}{43}-\frac{n_\mathrm{s}}{44}-\frac{5 n_\mathrm{s}^2}{24}-\frac{n_\mathrm{s}^3}{50}\right) H_4(\nu)\\
+\left(\frac{29}{52}-\frac{n_\mathrm{s}}{32}-\frac{2 n_\mathrm{s}^2}{27}\right) H_5(\nu)
+\left(\frac{3}{31}-\frac{n_\mathrm{s}}{314}-\frac{n_\mathrm{s}^2}{85}\right) H_6(\nu)
+\left(\frac{1}{37}-\frac{n_\mathrm{s}}{383}-\frac{n_\mathrm{s}^2}{288}\right) H_7(\nu) \Bigg].
\end{multline}
The first order non Gaussian correction is fitted by:
\begin{multline}
  \frac{\partial n_\mathrm{max}^{(1)}}{\partial\nu} = \frac{\sigma}{10^3 \sqrt{2\pi}} e^{-\frac{\nu^2}{2}} \Bigg[
\left(\frac{38}{31}-\frac{15 n_\mathrm{s}^2}{34}-\frac{2 n_\mathrm{s}^3}{19}\right)
+\left(\frac{449}{64}-\frac{27 n_\mathrm{s}}{80}-\frac{84 n_\mathrm{s}^2}{67}-\frac{5 n_\mathrm{s}^3}{37}\right) H_1(\nu)\\
+\left(\frac{47}{4}-\frac{15 n_\mathrm{s}}{28}-\frac{35 n_\mathrm{s}^2}{38}+\frac{2 n_\mathrm{s}^3}{27}\right) H_2(\nu)
+\left(\frac{245}{16}-\frac{19 n_\mathrm{s}}{22}-\frac{9 n_\mathrm{s}^2}{7}+\frac{4 n_\mathrm{s}^3}{29}\right) H_3(\nu)
+\left(\frac{233}{28}-\frac{8 n_\mathrm{s}}{19}-\frac{17 n_\mathrm{s}^2}{36}+\frac{7 n_\mathrm{s}^3}{45}\right) H_4(\nu)\\
+\left(\frac{81}{17}-\frac{7 n_\mathrm{s}}{24}-\frac{25 n_\mathrm{s}^2}{46}+\frac{n_\mathrm{s}^3}{26}\right) H_5(\nu)
+\left(\frac{13}{14}-\frac{n_\mathrm{s}}{24}-\frac{4 n_\mathrm{s}^2}{43}+\frac{n_\mathrm{s}^3}{90}\right) H_6(\nu)
+\left(\frac{5}{17}-\frac{n_\mathrm{s}}{58}-\frac{n_\mathrm{s}^2}{19}-\frac{n_\mathrm{s}^3}{377}\right) H_7(\nu) \Bigg].
\end{multline}
For the saddle points, we have:
\begin{multline}
   \frac{\partial n_\mathrm{+--}^{(0)}}{\partial \nu} = \frac{1}{10^3 \sqrt{2\pi}} e^{-\frac{\nu^2}{2}} \Bigg[
\left( \frac{454}{47}+\frac{n_\mathrm{s}}{298}-\frac{n_\mathrm{s}^2}{25}-\frac{n_\mathrm{s}^3}{119} \right) 
+\left(\frac{153}{28}-\frac{4 n_\mathrm{s}}{27}-\frac{5 n_\mathrm{s}^2}{17}+\frac{n_\mathrm{s}^3}{21}\right) H_1(\nu)\\
+\left(\frac{3}{34}-\frac{n_\mathrm{s}}{31}+\frac{6 n_\mathrm{s}^2}{35}+\frac{n_\mathrm{s}^3}{22}\right) H_2(\nu)
+\left(-\frac{37}{92}-\frac{n_\mathrm{s}}{57}+\frac{10 n_\mathrm{s}^2}{39}+\frac{n_\mathrm{s}^3}{24}\right) H_3(\nu)
+\left(-\frac{1}{45}-\frac{n_\mathrm{s}}{95}+\frac{n_\mathrm{s}^2}{20}+\frac{n_\mathrm{s}^3}{112}\right) H_4(\nu)\\
+\left(-\frac{1}{42}-\frac{n_\mathrm{s}^2}{268}-\frac{n_\mathrm{s}^3}{308}\right) H_5(\nu)
+\left(-\frac{1}{54}+\frac{n_\mathrm{s}}{687}-\frac{n_\mathrm{s}^2}{371}-\frac{n_\mathrm{s}^3}{456}\right) H_6(\nu)
+\left(-\frac{1}{124}+\frac{n_\mathrm{s}^2}{528}\right) H_7(\nu) \Bigg]
\end{multline} and
\begin{multline}
   \frac{\partial n_\mathrm{+--}^{(1)}}{\partial \nu} = \frac{\sigma}{10^3 \sqrt{2\pi}} e^{-\frac{\nu^2}{2}} \Bigg[
\left(\frac{33}{28}-\frac{n_\mathrm{s}}{223}-\frac{5 n_\mathrm{s}^2}{14}-\frac{2 n_\mathrm{s}^3}{31}\right)
+\left(-\frac{30}{19}-\frac{25 n_\mathrm{s}}{88}+\frac{23 n_\mathrm{s}^2}{27}+\frac{9 n_\mathrm{s}^3}{32}\right) H_1(\nu)\\
+\left(\frac{31}{40}-\frac{4 n_\mathrm{s}}{9}+\frac{61 n_\mathrm{s}^2}{38}+\frac{31 n_\mathrm{s}^3}{61}\right) H_2(\nu)
+\left(\frac{43}{16}-\frac{n_\mathrm{s}}{9}+\frac{30 n_\mathrm{s}^2}{31}+\frac{6 n_\mathrm{s}^3}{37}\right) H_3(\nu)
+\left(-\frac{12}{23}-\frac{6 n_\mathrm{s}}{37}+\frac{32 n_\mathrm{s}^2}{51}+\frac{n_\mathrm{s}^3}{7}\right) H_4(\nu)\\
+\left(-\frac{6}{7}-\frac{n_\mathrm{s}}{32}+\frac{3 n_\mathrm{s}^2}{20}+\frac{n_\mathrm{s}^3}{52}\right) H_5(\nu)
+\left(-\frac{17}{32}-\frac{n_\mathrm{s}}{671}+\frac{4 n_\mathrm{s}^2}{47}+\frac{n_\mathrm{s}^3}{125}\right) H_6(\nu)
+\left(-\frac{2}{27}-\frac{n_\mathrm{s}}{246}+\frac{n_\mathrm{s}^2}{69}+\frac{n_\mathrm{s}^3}{446}\right) H_7(\nu) \Bigg].
\end{multline}
Replacing the field by its opposite (\emph{i.e.} replacing $\nu$ with $-\nu$ and the odd Gram-Charlier coefficients by their opposite) yields the expressions for minima and the second kind of saddle points:
\begin{align}
  \frac{\partial n_\mathrm{min}^{(0)}}{\partial \nu} (\nu)+\frac{\partial n_\mathrm{min}^{(1)}}{\partial \nu} (\nu)&=   \frac{\partial n_\mathrm{min}^{(0)}}{\partial \nu} (-\nu)-\frac{\partial n_\mathrm{max}^{(1)}}{\partial \nu} (-\nu),
&   \frac{\partial n_\mathrm{++-}^{(0)}}{\partial \nu} (\nu)+\frac{\partial n_\mathrm{++-}^{(1)}}{\partial \nu} (\nu)&=   \frac{\partial n_\mathrm{+--}^{(0)}}{\partial \nu} (-\nu)-\frac{\partial n_\mathrm{+--}^{(1)}}{\partial \nu} (-\nu).
\end{align}
The corresponding fits for the first order corrections are thus obtained by changing the signs of the even Hermite polynomials.

\subsection{Skeleton length}
For  $n_s\in [-2.5,0]$, the Gaussian prediction for the differential skeleton length is well fitted by
\begin{multline}
   \frac{\partial {L^\mathrm{skel}}^{(0)}}{\partial \nu} = \frac{1}{10^3 \sqrt{2\pi}} e^{-\frac{\nu^2}{2}} \Bigg[
\left(\frac{3041}{23}+\frac{4 n_\mathrm{s}}{39}-\frac{3 n_\mathrm{s}^2}{19}-\frac{n_\mathrm{s}^3}{307}\right)
+\left(\frac{1510}{11}-\frac{63 n_\mathrm{s}}{25}-\frac{226 n_\mathrm{s}^2}{23}+\frac{n_\mathrm{s}^3}{4}\right) H_1(\nu)\\
+\left(\frac{2401}{44}-\frac{23 n_\mathrm{s}}{50}-\frac{89 n_\mathrm{s}^2}{11}-\frac{32 n_\mathrm{s}^3}{41}\right) H_2(\nu)
+\left(\frac{297}{38}+\frac{25 n_\mathrm{s}}{63}-\frac{64 n_\mathrm{s}^2}{33}-\frac{40 n_\mathrm{s}^3}{93}\right) H_3(\nu)
\Bigg],
\end{multline}
while the first order correction is fitted by
\begin{multline}
  \frac{\partial {L^\mathrm{skel}}^{(1)}}{\partial \nu} = \frac{\sigma}{10^3 \sqrt{2\pi}} e^{-\frac{\nu^2}{2}} \Bigg[
\left( -\frac{2450}{33}-\frac{719 n_\mathrm{s}}{41}+\frac{2 n_\mathrm{s}^2}{19}+\frac{67 n_\mathrm{s}^3}{24}\right)
+\left(-\frac{2143}{57}-\frac{1332 n_\mathrm{s}}{19}+\frac{910 n_\mathrm{s}^2}{9}+\frac{2679 n_\mathrm{s}^3}{64}\right) H_1(\nu)\\
+\left(\frac{4479}{38}-\frac{1952 n_\mathrm{s}}{23}+\frac{5912 n_\mathrm{s}^2}{37}+\frac{1909 n_\mathrm{s}^3}{31}\right) H_2(\nu)
+\left(\frac{5239}{32}-\frac{14842 n_\mathrm{s}}{291}+\frac{3788 n_\mathrm{s}^2}{35}+\frac{715 n_\mathrm{s}^3}{18}\right) H_3(\nu)\\
+\left(\frac{4926}{59}-\frac{388 n_\mathrm{s}}{23}+\frac{2979 n_\mathrm{s}^2}{76}+\frac{178 n_\mathrm{s}^3}{13}\right) H_4(\nu)
+\left(\frac{834}{35}-\frac{118 n_\mathrm{s}}{35}+\frac{240 n_\mathrm{s}^2}{29}+\frac{74 n_\mathrm{s}^3}{27}\right) H_5(\nu)\\
+\left(\frac{26}{9}-\frac{7 n_\mathrm{s}}{24}+\frac{21 n_\mathrm{s}^2}{26}+\frac{19 n_\mathrm{s}^3}{77}\right) H_6(\nu)
\Bigg].
\end{multline}

\section{Computing the skeleton length in the Hessian eigenframe}
\label{sec:skelhess}

The calculations made in Section~\ref{sec:skel2D} used the assumption that the properties of the components of the gradients are the same in the Hessian eigenframe as in any frame. This assumption translates into
%
$ \langle {\tilde x_1}^2 \rangle = \langle {x_1}^2 \rangle = {1}/{2}$, $\langle {\tilde x_2}^2 \rangle = \langle {x_2}^2 \rangle = {1}/{2}$,
and
$ \langle \zeta {\tilde x_2}^2 \rangle = \langle \zeta {x_2}^2 \rangle$, $  \langle \tilde {x_2}^2 J_1 \rangle = \langle {x_2}^2 J_1 \rangle$.
These equalities are true in the Gaussian case: the first and second derivatives are independent so that the statistical properties of the gradient do not depend on the Hessian eigendirections. However it is no longer true when non-Gaussianities give rise to correlations between the gradient and the Hessian.
To avoid this assumption, one can measure the values of $ \langle \zeta {\tilde x_2}^2 \rangle$ and $\langle \tilde {x_2}^2 J_1 \rangle$. To do so the Hessian needs to be diagonalized at each point, which makes the measurements a bit more difficult. Furthermore the cumulants measured in the Hessian eigenframe seem difficult to predict via the perturbation theory described in the next section.
\\
The assumption on the variances $ \langle {\tilde x_1}^2 \rangle$ and $ \langle {\tilde x_2}^2 \rangle$ is more subtle. These terms are not explicitly present in the expansion. But they are actually implicitly assumed when the Gaussian kernel is made to built the Gram-Charlier expansion. If these variances are taken in a fixed frame, the isotropy and the normalization of the gradient $\langle q^2 \rangle =1$ fix these variances to $1/2$. But if the Hessian eigenframe is used, these variables are still Gaussian but their variance can change. The non-Gaussianities then affect the Gaussian kernel and the expression of the expansion. However this problem can be easily solved by using these variances to normalize the variables. We can use the variable
\( Q = {1}/{\sqrt{2\langle {\tilde x_2}^2 \rangle}} \tilde x_2 \) instead of $x_2$ to make sure that the variance is fixed and that the expansion using this new variable will not change.
To summarize, if the properties of the gradient are not assumed to be the same in the Hessian eigenframe as in a fixed frame, one needs to diagonalize the Hessian, compute the components of the gradient in the eigenframe, normalize them to a fixed variance, measure the cumulants $\langle \zeta Q^2 \rangle$ and $\langle Q^2 J_1 \rangle$ and use the previous analytical predictions with the new normalized variable $Q$ instead of the original $x_2$. Figure~\ref{fig:sklnongausshess} shows the results of such a procedure. The difference with the previous results is almost negligible: the statistical properties of the components of the gradient are almost the same in the Hessian eigenframe as in a fixed frame and considering the apparition of a difference with non-Gaussianities is just a higher-order correction.
\begin{figure}
  \includegraphics[width=0.49\textwidth]{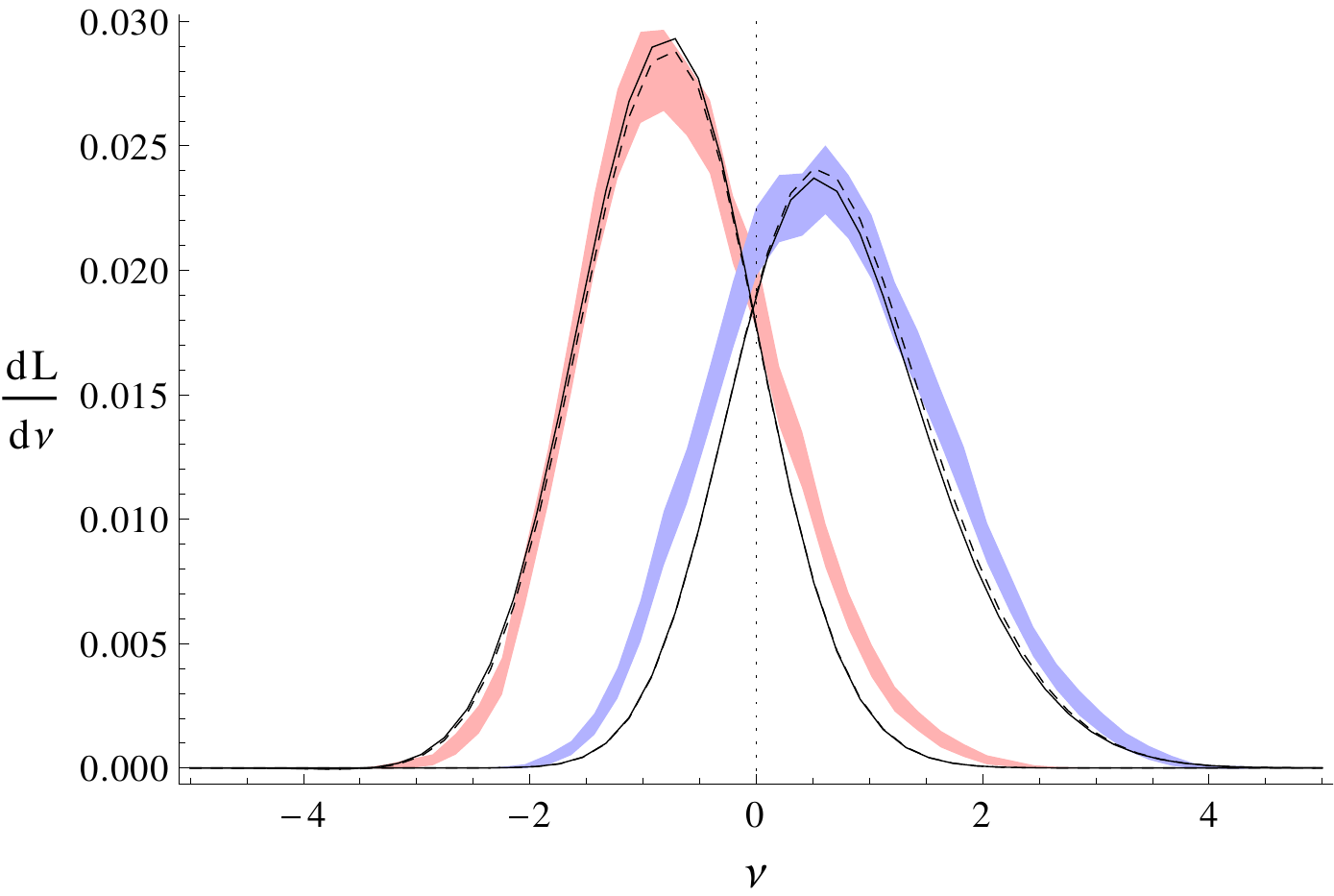}
\caption{Prediction for the 2D skeleton length using the components of the gradient in any frame (\textit{dashed line}) or in the Hessian eigenframe (\textit{solid line}). A small improvement is seen, especially near the maximum of the skeleton curve. However the correction remains very small, and its implementation also requires  diagonalizing the Hessian and measuring two cumulants that will not be predicted by  perturbation theory.}
\label{fig:sklnongausshess}
\end{figure}

\section{Connecting the Edgeworth to the Gram-Charlier expansion}
\label{sec:Edgeworth}
In this section, we give a formal derivation of the Edgeworth series \citep{Blinnikov}. For convenience, we restrict it to the case of a monovariate distribution, but it can be generalized directly \citep{Kendall}. Understanding the link between the Gram-Charlier and the Edgeworth expansions will be the key to a physically motivated resummation of the expansion.
For a given PDF $P(x)$, we define the moment generating function
\begin{equation} M(t)=\int_{-\infty}^{+\infty} P(x) e^{tx} {\rm d}x = \sum_{p=0}^\infty \frac{\left\langle  x^p \right\rangle}{p!} t^p\,, \end{equation}
and the cumulant generating function
\begin{equation} C(t)= \sum_{p=2}^\infty \frac{\left\langle  x^p \right\rangle_c}{p!} t^p .\end{equation}
By definition of the cumulants, we have
\begin{equation} M(t)=\exp \left( C(t) \right).\end{equation}
Now, let's take the Fourier transform of the PDF:
\begin{equation} \tilde P(t)=\frac{1}{\sqrt{2\pi}} \int_{-\infty}^{+\infty} P(x) e^{itx} {\rm d}x. \end{equation}
We have
\begin{equation} \tilde P(t)=M(it)=\exp(C(it)). \end{equation}
We can get the PDF by taking the inverse Fourier transform:
\begin{equation} P(x) = \frac{1}{\sqrt{2\pi}} \int_{-\infty}^{+\infty} e^{-itx} \exp \left( \sum_{p=2}^\infty \frac{\left\langle  x^p \right\rangle_c}{p!} i^p t^p \right) {\rm d}t.\end{equation}
In order to have an expansion around the Gaussian case, we extract the $p=2$ term which corresponds to the purely Gaussian contribution:
\begin{equation} P(x) = \frac{1}{\sqrt{2\pi}} \int_{-\infty}^{+\infty} e^{-itx} e^{-\frac{\sigma^2}{2} t^2} \exp\left( \sum_{p=3}^\infty \frac{\left\langle  x^p \right\rangle_c}{p!} i^p t^p \right) {\rm d}t \end{equation} where $\sigma^2 \equiv \langle x^2 \rangle$.
If we expand the last exponential, we will thus have the Gaussian term
\begin{equation} \frac{1}{\sqrt{2\pi}} \int_{-\infty}^{+\infty} e^{-itx} e^{-\frac{\sigma^2}{2} t^2} {\rm d}t = \frac{1}{\sqrt{2\pi \sigma^2}} e^{-\frac{x^2}{2\sigma^2}}\,, \end{equation}
and the other terms will involve
\begin{equation} \frac{1}{\sqrt{2\pi}} \int_{-\infty}^{+\infty} e^{-itx} e^{-\frac{\sigma^2}{2} t^2} i^p t^p {\rm d}t = (-1)^p \frac{{\rm d}^p}{{\rm d}x^p} \left(\frac{1}{\sqrt{2\pi \sigma^2}} e^{-\frac{x^2}{2\sigma^2}}\right). \end{equation}
The expansion can therefore be formally written as
\begin{equation} P(x)= \exp \left( \sum_{p=3}^\infty \frac{\left\langle  x^p \right\rangle_c}{p!} (-1)^p \frac{{\rm d}^p}{{\rm d}x^p} \right) \frac{1}{\sqrt{2\pi \sigma^2}} e^{-\frac{x^2}{2\sigma^2}}. \end{equation}
Let us  expand the exponential in Taylor series:
\begin{equation} P(x) = \left(1+ \left( \sum_{p=3}^\infty \frac{\left\langle  x^p \right\rangle_c}{p!} (-1)^p \frac{{\rm d}^p}{{\rm d}x^p} \right) + \frac{1}{2} \left( \sum_{p=3}^\infty \frac{\left\langle  x^p \right\rangle_c}{p!} (-1)^p \frac{{\rm d}^p}{{\rm d}x^p} \right)^2 + \frac{1}{6} \left( \sum_{p=3}^\infty \frac{\left\langle  x^p \right\rangle_c}{p!} (-1)^p \frac{{\rm d}^p}{{\rm d}x^p} \right)^3 + .. \right) G(x) \,,
\label{eq:C11}
\end{equation}
where $G(x)=\exp({-{x^2}/{2\sigma^2}})/{\sqrt{2\pi \sigma^2}} $ is the Gaussian PDF. Developing powers of the operator allows us to rewrite equation~(\ref{eq:C11}) as 
\begin{multline}
P(x) = \left(1+ \left( \sum_{p=3}^\infty \frac{\left\langle  x^p \right\rangle_c}{p!} (-1)^p \frac{{\rm d}^p}{{\rm d}x^p} \right) + \frac{1}{2} \left( \sum_{p_1=3}^\infty \sum_{p_2=3}^\infty \frac{\left\langle  x^{p_1} \right\rangle_c \left\langle  x^{p_2} \right\rangle_c}{p_1! p_2!} (-1)^{p_1+p_2} \frac{{\rm d}^{p_1+p_2}}{{\rm d}x^{p_1+p_2}} \right) + \right. \\ \left. \frac{1}{6} \left( \sum_{p_1=3}^\infty \sum_{p_2=3}^\infty \sum_{p_3=3}^\infty \frac{\left\langle  x^{p_1} \right\rangle_c\left\langle  x^{p_2} \right\rangle_c\left\langle  x^{p_3} \right\rangle_c}{p_1! p_2! p_3!} (-1)^{p_1+p_2+p_3} \frac{{\rm d}^{p_1+p_2+p_3}}{{\rm d}x^{p_1+p_2+p_3}} \right) + \dots \right) G(x) .
\end{multline}
By definition of the Hermite polynomials:
$(-1)^k {{\rm d}^k}G/{{\rm d}x^k}=  H_k({x}/{\sigma}) G(x)/\sigma^k $
it follows that
\begin{multline}
P(x) = \left(1+ \left( \sum_{p=3}^\infty \frac{\left\langle  x^p \right\rangle_c \sigma^{-p}}{p!} H_p(\frac{x}{\sigma}) \right) + \frac{1}{2} \left( \sum_{p_1=3}^\infty \sum_{p_2=3}^\infty \frac{\left\langle  x^{p_1} \right\rangle_c \left\langle  x^{p_2} \right\rangle_c \sigma^{-p_1-p_2}}{p_1! p_2!} H_{p_1+p_2}(\frac{x}{\sigma}) \right) +\right. \\ \left. \frac{1}{6} \left( \sum_{p_1=3}^\infty \sum_{p_2=3}^\infty \sum_{p_3=3}^\infty \frac{\left\langle  x^{p_1} \right\rangle_c\left\langle  x^{p_2} \right\rangle_c\left\langle  x^{p_3} \right\rangle_c \sigma^{-p_1-p_2-p_3}}{p_1! p_2! p_3!} H_{p_1+p_2+p_3}(\frac{x}{\sigma}) \right) + .. \right) G(x)  \,.
\end{multline}
Joining the terms having the same Hermite polynomial would give the Gram-Charlier expansion. The Gram-Charlier coefficient $\langle x^n \rangle_\mathrm{{\scriptscriptstyle GC}}$ can therefore be expressed in terms of cumulants:
\begin{equation}
 \langle x^n \rangle_\mathrm{{\scriptscriptstyle GC}} = \langle x^n \rangle_c + \frac{n!}{2!} \sum_{p_1, p_2=3}^{p_1+p_2=n} \frac{\left\langle  x^{p_1} \right\rangle_c\left\langle  x^{p_2} \right\rangle_c}{p_1! p_2!} +
\frac{n!}{3!} \sum_{p_1, p_2, p_3=3}^{p_1+p_2+p_3=n} \frac{\left\langle  x^{p_1} \right\rangle_c \left\langle  x^{p_2} \right\rangle_c \left\langle  x^{p_3} \right\rangle_c}{p_1! p_2! p_3!} + \dots
\label{eq:GCtocum}
\end{equation} where the notation $\displaystyle \sum_{p_1, p_2=3}^{p_1+p_2=n}$ means a summation on $p_1$ and $p_2$ from 3 to $\infty$ under the condition $p_1+p_2=n$.
In the context of gravitational instability, we have $\langle x^n \rangle \propto \sigma^{2n-2}$. The expansion can be reordered according to the order of the cumulants, introducing the coefficients $S_p={\left\langle x^p \right\rangle_c}/{\sigma^{2p-2}}$:
\begin{multline}
P(x) = \left(1+ \left( \sum_{p=3}^\infty \frac{S_p \sigma^{p-2}}{p!} H_p(\frac{x}{\sigma}) \right) + \frac{1}{2} \left( \sum_{p_1=3}^\infty \sum_{p_2=3}^\infty \frac{S_{p_1} S_{p_2} \sigma^{p_1+p_2-4}}{p_1! p_2!} H_{p_1+p_2}(\frac{x}{\sigma}) \right) + \right. \\ \left. \frac{1}{6} \left( \sum_{p_1=3}^\infty \sum_{p_2=3}^\infty \sum_{p_3=3}^\infty \frac{S_{p_1} S_{p_2} S_{p_3} \sigma^{p_1+p_2+p_3-6}}{p_1! p_2! p_3!} H_{p_1+p_2+p_3}(\frac{x}{\sigma}) \right) + .. \right) G(x).
\end{multline}
The first terms are:
\begin{multline}
P(x) = \left(1+ \left( \frac{S_3 \sigma}{3!} H_3(\frac{x}{\sigma})+\frac{S_4 \sigma^2}{4!} H_4(\frac{x}{\sigma})+\frac{S_5 \sigma^3}{5!} H_5(\frac{x}{\sigma}) + ...\right) +\right. \\ \left. \frac{1}{2} \left( \frac{S_3 S_3 \sigma^2}{3!3!} H_6(\frac{x}{\sigma}) + \frac{S_3 S_4 \sigma^3}{3!4!} H_7(\frac{x}{\sigma})+ \frac{S_4 S_3 \sigma^3}{4!3!} H_7(\frac{x}{\sigma}) + ..\right) + \frac{1}{6} \left( \frac{S_3 S_3 S_3 \sigma^3}{3! 3! 3!} H_9(\frac{x}{\sigma}) + ... \right) + .. \right) G(x) .
\end{multline}
Ordering the terms w.r.t their order in $\sigma$ gives the Edgeworth expansion:
\begin{multline}
P(x) = \left(1+ \sigma \frac{S_3}{6} H_3(\frac{x}{\sigma})+ \sigma^2 \left( \frac{S_4}{24} H_4(\frac{x}{\sigma}) + \frac{{S_3}^2 \sigma^2}{72} H_6(\frac{x}{\sigma}) \right)  +\right. 
 \left. \sigma^3 \left(\frac{S_5}{120} H_5(\frac{x}{\sigma}) + \frac{S_3 S_4}{144} H_7(\frac{x}{\sigma})+ \frac{{S_3}^3 }{1296} H_9(\frac{x}{\sigma}) \right) + .. \right) G(x), \nonumber
\end{multline}
which corresponds to the expression found in the literature \citep{Bernardeau}.
Equation (\ref{eq:GCtocum}) allows us to determine the orders to which a given coefficient contributes to the summation. Indeed, a coefficient with $n$ fields can be decomposed as a sum of:
\begin{itemize}
 \item a linear (in cumulants) term, which is of order $\sigma^{n-2}$, 
 \item if $n>3$, a quadratic term, of order $\sigma^{p_1-2}+ \sigma^{p_2-2}=\sigma^{n-4} $,
 \item if $n>6$, a cubic term, of order $\sigma^{p_1-2}+ \sigma^{p_2-2}+\sigma^{p_3-2}=\sigma^{n-6} $,
 \item \dots
 \item if $n>3k$, a product of $k$ cumulants, of order $\sigma^{n-2k-2}$.
\end{itemize}
Thus the orders in $\sigma$ can be obtained as follow:
\begin{itemize}
 \item the $\sigma$ order comes from the $n=3$ terms,
 \item the $\sigma^2$ order comes from the $n=4$ term and part of the $n=6$ one,
 \item the $\sigma^3$ order comes from the $n=5$ term and part of the $n=7$ and $n=9$ ones,
 \item the $\sigma^4$ order comes from part of the $n=6$, $n=8$, $n=10$ and $n=12$ terms,
 \item \dots
\end{itemize}
In the next section, we will see how to compute explicitly the required elements of the expansion up to some order in $\sigma$.
\section{Critical sets  in the Edgeworth PT expansion}
\label{sec:sigma2}
\subsection{Truncating the Gram-Charlier expansion to order $\sigma^2$ }

\begin{table}
\begin{center}
\begin{tabular}{rl}
\begin{tabular}[t]{|c|c|}
\hline
Coefficient & $\sigma^2$\\ \hline
$\langle \zeta^6 \rangle_\mathrm{{\scriptscriptstyle GC}}$ & $10 \langle \zeta^3 \rangle^2$ \\ \hline
$\langle \zeta^5 J_1 \rangle_\mathrm{{\scriptscriptstyle GC}}$ & $10 \langle \zeta^3 \rangle \langle \zeta^2 J_1 \rangle$ \\ \hline
$\langle \zeta^4 {J_1}^2 \rangle_\mathrm{{\scriptscriptstyle GC}}$ & $6 \langle \zeta^2 J_1 \rangle^2+4 \langle \zeta^3\rangle \langle \zeta {J_1}^2\rangle$ \\ \hline
$\langle \zeta^3 {J_1}^2 \rangle_\mathrm{{\scriptscriptstyle GC}}$ & $9 \langle \zeta^2 J_1 \rangle \langle \zeta {J_1}^2 \rangle +\langle \zeta^3\rangle \langle {J_1}^3\rangle$ \\ \hline
$\langle \zeta^2 {J_1}^4 \rangle_\mathrm{{\scriptscriptstyle GC}}$ & $6 \langle \zeta {J_1}^2 \rangle^2+4 \langle {J_1}^3\rangle \langle \zeta^2 J_1 \rangle$ \\ \hline
$\langle \zeta {J_1}^5 \rangle_\mathrm{{\scriptscriptstyle GC}}$ & $10 \langle {J_1}^3 \rangle \langle \zeta {J_1}^2 \rangle$ \\ \hline
$\langle {J_1}^6 \rangle_\mathrm{{\scriptscriptstyle GC}}$ & $10 \langle {J_1}^3 \rangle^2$ \\ \hline
$\langle \zeta^4 q^2 \rangle_\mathrm{{\scriptscriptstyle GC}}$ & $4 \langle \zeta^3 \rangle \langle \zeta q^2 \rangle$ \\ \hline
$\langle \zeta^3 q^2 J_1 \rangle_\mathrm{{\scriptscriptstyle GC}}$ & $ \langle \zeta^3 \rangle \langle q^2 J_1 \rangle + 3 \langle \zeta q^2 \rangle \langle \zeta^2 J_1 \rangle$ \\ \hline
$\langle \zeta^2 q^2 {J_1}^2 \rangle_\mathrm{{\scriptscriptstyle GC}}$ & $ 2 \langle \zeta q^2 \rangle \langle \zeta {J_1}^2 \rangle + 2 \langle q^2 J_1 \rangle \langle \zeta^2 J_1 \rangle$ \\ \hline
$\langle \zeta q^2 {J_1}^3 \rangle_\mathrm{{\scriptscriptstyle GC}}$ & $ \langle {J_1}^3 \rangle \langle \zeta q^2 \rangle + 3 \langle q^2 J_1 \rangle \langle \zeta {J_1}^2 \rangle$ \\ \hline
$\langle q^2 {J_1}^4 \rangle_\mathrm{{\scriptscriptstyle GC}}$ & $4 \langle {J_1}^3 \rangle \langle q^2 J_1 \rangle$ \\ \hline

$\langle q^6 \rangle_\mathrm{{\scriptscriptstyle GC}}$ & $0$ \\ \hline
 \end{tabular}
&
\begin{tabular}[t]{|c|c|}
\hline
Coefficient & $\sigma^2$\\ \hline
$\langle \zeta^4 J_2 \rangle_\mathrm{{\scriptscriptstyle GC}}$ & $4 \langle \zeta^3 \rangle \langle \zeta J_2 \rangle$ \\ \hline
$\langle \zeta^3 J_1 J_2 \rangle_\mathrm{{\scriptscriptstyle GC}}$ & $ \langle \zeta^3 \rangle \langle J_1 J_2\rangle + 3 \langle \zeta J_2 \rangle \langle \zeta^2 J_1 \rangle$ \\ \hline
$\langle \zeta^2 {J_1}^2 J_2 \rangle_\mathrm{{\scriptscriptstyle GC}}$ & $ 2 \langle \zeta J_2 \rangle \langle \zeta {J_1}^2 \rangle + 2 \langle J_1 J_2 \rangle \langle \zeta^2 J_1 \rangle$ \\ \hline
$\langle \zeta {J_1}^3 J_2 \rangle_\mathrm{{\scriptscriptstyle GC}}$ & $ \langle {J_1}^3 \rangle \langle \zeta J_2 \rangle + 3 \langle J_1 J_2 \rangle \langle \zeta {J_1}^2 \rangle$ \\ \hline
$\langle {J_1}^4 J_2 \rangle_\mathrm{{\scriptscriptstyle GC}}$ & $4 \langle {J_1}^3 \rangle \langle J_1 J_2 \rangle$ \\ \hline

$\langle \zeta^2 q^2 J_2 \rangle_\mathrm{{\scriptscriptstyle GC}}$ & $2 \langle \zeta q^2 \rangle \langle \zeta J_2 \rangle$ \\ \hline
$\langle \zeta q^2 J_1 J_2 \rangle_\mathrm{{\scriptscriptstyle GC}}$ & $\langle \zeta q^2 \rangle \langle J_1 J_2 \rangle+\langle q^2 J_1 \rangle \langle \zeta J_2 \rangle$ \\ \hline
$\langle q^2 {J_1}^2 J_2 \rangle_\mathrm{{\scriptscriptstyle GC}}$ & $2 \langle q^2 J_1 \rangle \langle J_1 J_2 \rangle$ \\ \hline

$\langle \zeta^3 J_3  \rangle_\mathrm{{\scriptscriptstyle GC}}$ & $\langle \zeta^3 \rangle \langle J_3 \rangle$ \\ \hline
$\langle \zeta^2 J_1 J_3  \rangle_\mathrm{{\scriptscriptstyle GC}}$ & $\langle \zeta^2 J_1 \rangle \langle J_3 \rangle$ \\ \hline
$\langle \zeta {J_1}^2 J_3  \rangle_\mathrm{{\scriptscriptstyle GC}}$ & $\langle \zeta {J_1}^2 \rangle \langle J_3 \rangle$ \\ \hline
$\langle {J_1}^3 J_3  \rangle_\mathrm{{\scriptscriptstyle GC}}$ & $\langle {J_1}^3 \rangle \langle J_3 \rangle$ \\ \hline

$\langle \zeta q^2 J_3  \rangle_\mathrm{{\scriptscriptstyle GC}}$ & $\langle \zeta q^2 \rangle \langle J_3 \rangle$ \\ \hline
$\langle q^2 J_1 J_3  \rangle_\mathrm{{\scriptscriptstyle GC}}$ & $\langle q^2 J_1 \rangle \langle J_3 \rangle$ \\ \hline

 \end{tabular}\\
\\
\begin{tabular}[t]{|c|c|}
\multicolumn{2}{c}{2D}\\
\hline
Coefficient & $\sigma^2$\\ \hline
$\langle \zeta^2 q^4 \rangle_\mathrm{{\scriptscriptstyle GC}}$ & $ 4 \langle \zeta q^2 \rangle^2$ \\ \hline
$\langle \zeta q^4 J_1 \rangle_\mathrm{{\scriptscriptstyle GC}}$ & $ 4 \langle \zeta q^2 \rangle \langle q^2 J_1 \rangle $ \\ \hline
$\langle q^4 {J_1}^2 \rangle_\mathrm{{\scriptscriptstyle GC}}$ & $ 4 \langle q^2 J_1\rangle^2$ \\ \hline
\end{tabular}
&
\begin{tabular}[t]{|c|c|}
\multicolumn{2}{c}{3D}\\
\hline
Coefficient & $\sigma^2$\\ \hline
$\langle \zeta^2 q^4 \rangle_\mathrm{{\scriptscriptstyle GC}}$ & $ \frac{10}{3} \langle \zeta q^2 \rangle^2$ \\ \hline
$\langle \zeta q^4 J_1 \rangle_\mathrm{{\scriptscriptstyle GC}}$ & $ \frac{10}{3} \langle \zeta q^2 \rangle \langle q^2 J_1 \rangle $ \\ \hline
$\langle q^4 J_1 \rangle_\mathrm{{\scriptscriptstyle GC}}$ & $ \frac{10}{3} \langle q^2 J_1\rangle^2$ \\ \hline
$\langle q^4 J_2 \rangle_\mathrm{{\scriptscriptstyle GC}}$ & $\frac{5}{6} \langle q^2 J_1 \rangle^2$ \\ \hline
\end{tabular}
\end{tabular}
\end{center}
\caption{$\sigma^2$ order terms from the $n=6$ Gram Charlier order. Only terms contributing to the Euler characteristic are shown. Terms involving $J_3$ do not exist in 2D, and  those involving $q^4$ differ in 2 and 3 dimensions.}
\label{tab:6th}
\end{table}
According to Appendix \ref{sec:Edgeworth}, the $\sigma^2$ term arises from the 4th order cumulants and from part of the 6th order Gram-Charlier coefficients. Indeed, the 6th order coefficients contain quadratic $S_3^2$-like terms. For example, Equation \ref{eq:GCtocum} gives $\langle x^6 \rangle_\mathrm{{\scriptscriptstyle GC}} = \langle x^6 \rangle_c + 10 \langle x^3 \rangle_c ^2$.
In general, these contributions can also be computed by expressing the Gram-Charlier coefficients in terms of moments using the expressions of the orthogonal polynomials (see equations (\ref{eq:GCtomoments2D}), (\ref{eq:GCtomoments3D1}) and (\ref{eq:GCtomoments3D2})), and then expressing moments in terms of cumulants. Perturbation theory ensures us that a term involving a cumulant with $p$ fields will be of order $\sigma^{p-2}$\footnote{Or $\sigma^{2p-2}$ if not normalized.}.
For example, the 6th order Hermite polynomial is :
$   H_6(\zeta)=\zeta^6-15\zeta^4+45\zeta^2-15$.
So:
\[ \langle \zeta^6 \rangle_\mathrm{{\scriptscriptstyle GC}}\equiv \langle H_6(\zeta) \rangle  = \langle \zeta^6 \rangle -15 \langle \zeta^4 \rangle + 30\,, \] where we have used the fact that $\langle \zeta^2 \rangle = 1$.
We can then use the decomposition of moments into cumulants:
\[  \langle \zeta^4 \rangle = \langle \zeta^4 \rangle_c + 3  \langle \zeta^2 \rangle^2, \quad
  \langle \zeta^6 \rangle = \langle \zeta^6 \rangle_c + 15  \langle \zeta^4 \rangle_c \langle \zeta^2 \rangle  + 10  \langle \zeta^3 \rangle^2 + 15\langle \zeta^2 \rangle^3 .\]
In the end,
\[ \langle \zeta^6 \rangle_\mathrm{{\scriptscriptstyle GC}} = \langle \zeta^6 \rangle_c + 10 \langle \zeta^3 \rangle^2. \]
The first term is of order $\sigma^4$ and the second one is of order $\sigma^2$.
Table \ref{tab:6th} give all $\sigma^2$ terms coming from the 6th order Gram-Charlier coefficients used in the computation of the Euler characteristic.
The expansion to 9th order is available online (\texttt{http://www.iap.fr/users/pichon/Gram}).
\subsection{Minkowski functionals}
\label{sec:MinkovFunc}
\subsubsection{2D length of isocontours}
In order to illustrate how to use Table \ref{tab:6th}, let us consider the reordering of the length of the isocontours in 2 dimensions presented in Section~\ref{sec:minkov2D}. To obtain an expression up to $\sigma^2$ order, one has to compute the $n=3$, 4 and 6 term from the general expression~(\ref{eq:defCont2Dfinal}):
\begin{multline}
 {\cal L}(\nu)=\frac{1}{2\sqrt{2}} e^{-\frac{\nu^2}{2}} \Big( 1 + \frac{1}{6} \langle x^3 \rangle H_3(\nu) + \frac{1}{2} \langle xq^2 \rangle H_1(\nu) + \frac{1}{24} \langle x^4 \rangle_\mathrm{c} H_4(\nu) + \frac{1}{4} \langle x^2 q^2 \rangle_\mathrm{c} H_2(\nu) - \frac{1}{16} \langle q^4 \rangle_\mathrm{c} H_0(\nu) \\+ \frac{1}{720} \langle x^6 \rangle_\mathrm{{\scriptscriptstyle GC}} H_6(\nu) + \frac{1}{48} \langle x^4 q^2 \rangle_\mathrm{{\scriptscriptstyle GC}} H_4(\nu) - \frac{1}{128} \langle x^2 q^4 \rangle_\mathrm{{\scriptscriptstyle GC}} H_2(\nu) +\frac{1}{96} \langle q^6 \rangle_\mathrm{{\scriptscriptstyle GC}} H_0(\nu) \Big)
\,.
\end{multline}
Using the results from table~\ref{tab:6th}\footnote{Since no second derivative is involved in this simple example, $x$ can be used instead of $\zeta$.}, this expression can be truncated at the $\sigma^2$ order:
\begin{multline}
 {\cal L}(\nu)=\frac{1}{2\sqrt{2}} e^{-\frac{\nu^2}{2}} \Big( 1 + \frac{1}{6} \langle x^3 \rangle H_3(\nu) + \frac{1}{2} \langle xq^2 \rangle H_1(\nu) + \frac{1}{24} \langle x^4 \rangle_\mathrm{c} H_4(\nu) + \frac{1}{4} \langle x^2 q^2 \rangle_\mathrm{c} H_2(\nu) - \frac{1}{16} \langle q^4 \rangle_\mathrm{c} H_0(\nu) \\+ \frac{1}{72} \langle x^3 \rangle^2 H_6(\nu) + \frac{1}{12} \langle x^3 \rangle \langle x q^2 \rangle H_4(\nu) - \frac{1}{32} \langle x q^2 \rangle^2 H_2(\nu) \Big)\,.
\end{multline}
\subsubsection{3D surface of the isocontours}
Following the same route, equation~(\ref{eq:defCont3Dfinal}) in Section~\ref{sec:mink3D} becomes to the required order
\begin{multline}
 {\cal A}(\nu)=\frac{2}{\sqrt{3}\pi} e^{-\frac{\nu^2}{2}} \Bigg( 1 + \frac{3}{2} \langle x q^2 \rangle H_1(\nu) + \frac{1}{2} \langle x^3 \rangle H_3(\nu)
- \frac{9}{40} \langle q^4 \rangle_\mathrm{c} H_0(\nu) + \frac{3}{4} \langle x^2 q^2 \rangle_\mathrm{c} H_2(\nu) + \frac{1}{8} \langle x^4 \rangle_\mathrm{c} H_4(\nu)
\\+\frac{27}{560} \langle q^6 \rangle_\mathrm{{\scriptscriptstyle GC}} H_0(\nu)- \frac{9}{80} \langle x^2 q^4 \rangle_\mathrm{{\scriptscriptstyle GC}} H_2(\nu) + \frac{1}{16} \langle x^4 q^2 \rangle_\mathrm{{\scriptscriptstyle GC}} H_4(\nu) + \frac{1}{240} \langle x^6 \rangle_\mathrm{{\scriptscriptstyle GC}} H_8(\nu) \Bigg). \label{eq:tmp1}
\end{multline}
Using the expression of the $\sigma^2$ contribution of the $n=6$ terms, we get the $\sigma^2$ expression,
and its truncation to $\sigma^2$ order reads:
\begin{multline}
 {\cal A}(\nu)=\frac{2}{\sqrt{3}\pi} e^{-\frac{\nu^2}{2}} \Bigg( 1 + \frac{3}{2} \langle x q^2 \rangle H_1(\nu) + \frac{1}{2} \langle x^3 \rangle H_3(\nu)
- \frac{9}{40} \langle q^4 \rangle_\mathrm{c} H_0(\nu) + \frac{3}{4} \langle x^2 q^2 \rangle_\mathrm{c} H_2(\nu) + \frac{1}{8} \langle x^4 \rangle_\mathrm{c} H_4(\nu)
\\- \frac{3}{8} \langle x q^2 \rangle^2 H_2(\nu) + \frac{1}{4} \langle x q^2 \rangle \langle x^3 \rangle H_4(\nu) + \frac{1}{24} \langle x^3 \rangle^2 H_8(\nu) \Bigg)\,.
\end{multline}
All other terms in equation~(\ref{eq:tmp1}) contribute to orders in $\sigma$ higher than 2.
\subsection{2D Euler characteristic}
It is interesting to look at the first correction to the Euler characteristic since it has been predicted in other papers using a somewhat different approach \citep{Matsubara0, Matsubara}.
The $\sigma$ order term is given by the $n=3$ term of the Gram-Charlier expansion.
In 2 dimensions, equation~(\ref{eq:genus_2D}) yields
\begin{multline}
 \chi(\nu) =\frac{1}{4\pi \sqrt{2\pi} {R_*^2}} e^{-\frac{\nu^2}{2}} \Bigg[ \gamma^2 H_1 (\nu) + \left( 2 \gamma \left\langle q^2 J_1 \right\rangle + \sqrt{1-\gamma^2} \left\langle \zeta {J_1}^2 \right\rangle - \gamma \left\langle {J_1}^3 \right\rangle - \sqrt{1-\gamma^2} \left\langle \zeta J_2 \right\rangle +\gamma \left\langle J_1 J_2 \right\rangle \right) H_0(\nu) \\
+ \left(-\gamma^2 \sqrt{1- \gamma^2} \left\langle \zeta q^2 \right\rangle + \gamma^3 \left\langle q^2 J_1 \right\rangle -\gamma (1-\gamma^2) \left\langle \zeta^2 J_1 \right\rangle + 2\gamma^2 \sqrt{1-\gamma^2} \left\langle \zeta {J_1}^2 \right\rangle - \gamma^3 \left\langle {J_1}^3 \right\rangle \right) H_2(\nu) \\
+ \left( \frac{\gamma^2}{6} (1-\gamma^2)^{3/2} \left\langle \zeta^3 \right\rangle + \frac{\gamma^3}{2} (1-\gamma^2) \left\langle \zeta^2 J_1 \right\rangle + \frac{\gamma^4}{2} \sqrt{1-\gamma^2} \left\langle \zeta {J_1}^2 \right\rangle -\frac{\gamma^5}{6} \left\langle {J_1}^3 \right\rangle \right) H_4(\nu) \Bigg].
\end{multline}
This result can be simplified using the variable $x$ and $I_2$:
\begin{equation}
  \chi(\nu) =\frac{1}{4\pi \sqrt{2\pi} {R_*^2}} e^{-\frac{\nu^2}{2}} \Bigg[ \gamma^2 H_1 (\nu) + \left( 2 \gamma  \left\langle q^2 I_1\right\rangle +4 \left\langle x  I_2\right\rangle  \right) H_0(\nu)-\left( \gamma ^2 \left\langle x q^2 \right\rangle + \gamma  \left\langle x ^2 I_1\right\rangle \right) H_2(\nu)+\frac{\gamma^2}{6} \left\langle x ^3\right\rangle  H_4(\nu) \Bigg].
\end{equation}
The expression given by \cite{Matsubara} for the two-dimensional genus is
\begin{equation}
 G_2(\nu) = \frac{1}{(2\pi)^{3/2}} \left(\frac{\sigma_1}{\sqrt{2} \sigma} \right)^2 e^{-\frac{\nu^2}{2}} \Bigg[ H_1(\nu) + \left(\frac{S^{(0)}}{6} H_4(\nu) + \frac{2 S^{(1)}}{3} H_2(\nu) +  \frac{S^{(2)}}{3} \right) \sigma  \Bigg]\,,
\end{equation}
where
\begin{align}
 S^{(0)}&=\frac{\left\langle \rho^3 \right\rangle}{{\sigma}^4}, & S^{(1)} &= -\frac{3}{4} \frac{ \left\langle \rho^2 \nabla^2 \rho \right\rangle }{{\sigma}^2 {\sigma_1}^2}, & S^{(2)} &= - \frac{3d}{2(d-1)} \frac{ \left\langle (\nabla \rho)^2 \nabla^2 \rho \right\rangle }{{\sigma_1}^4}.
\end{align} We use here the unnormalized field, $\rho$, since \citeauthor{Matsubara} has a different normalization convention for the derivatives.
With our normalization and for $d=2$, these quantities can be expressed in terms of the rotational invariants:
\begin{align}
S^{(0)}&=\frac{\left\langle x^3 \right\rangle}{\sigma}, & S^{(1)} &= -\frac{3}{4} \frac{ \left\langle x^2 I_1 \right\rangle }{\gamma \sigma}, & S^{(2)} &= - \frac{3}{\gamma} \left\langle q^2 I_1 \right\rangle.
\end{align}
Factorizing $\gamma^2$ from our expression for the Euler characteristic gives the same prefactor since $\gamma/R_*=\sigma_1 / \sigma$. The $H_0(\nu$) term is obviously the same:
\begin{equation} \sigma \frac{S^{(0)}}{6} = \frac{1}{6} \left\langle x^3 \right\rangle. \end{equation}
Using the isotropy relations given in \cite{Matsubara}, one can check the identities:
\begin{align} \left\langle x q^2 \right\rangle &= -\frac{1}{2\gamma} \left\langle x^2 I_1 \right \rangle, & \left\langle x I_2 \right\rangle &= -\frac{3}{4} \gamma \left\langle q^2 I_1 \right\rangle, \label{eq:isotropy}\end{align} which are needed to prove that the coefficients of  $H_2(\nu)$ and $H_4(\nu)$  agree.
Similarly one can express the Euler characteristic up to $\sigma^2$ order. The result is more compact when expressed in terms of $x$ and $I_2$:
\begin{multline}
  \chi(\nu) =\frac{1}{4\pi \sqrt{2\pi} {R_*^2}} e^{-\frac{\nu^2}{2}} \Bigg( \gamma^2 H_1 (\nu) + \left( 2 \gamma  \left\langle q^2 I_1\right\rangle +4 \left\langle x  I_2\right\rangle  \right) H_0(\nu)-\left( \gamma ^2 \left\langle x q^2 \right\rangle + \gamma  \left\langle x ^2 I_1\right\rangle \right) H_2(\nu)+\frac{\gamma^2}{6} \left\langle x ^3\right\rangle  H_4(\nu)
\\+ \left( \frac{\gamma^2}{2} \langle q^4 \rangle_\mathrm{c} + 2\gamma \langle x q^2 J_1 \rangle_\mathrm{c} + 2\langle x^2 I_2 \rangle_\mathrm{c}  \right) H_1(\nu) - \left( \frac{\gamma^2}{2} \langle x^2 q^2 \rangle_\mathrm{c}  + \frac{\gamma}{3} \langle x^3 J_1 \rangle_\mathrm{c} \right) H_3(\nu) + \frac{\gamma^2}{24} \langle x^4 \rangle_\mathrm{c} H_5(\nu)
\\- \left( 4 \gamma \langle x q^2 \rangle \langle q^2 J_1 \rangle + \langle q^2 J_1 \rangle \langle x^2 J_1 \rangle + 4\langle x q^2 \rangle \langle x I_2 \rangle \right) H_1(\nu) \\+ \left(\gamma^2 \langle x q^2 \rangle^2 + \frac{\gamma}{3} \langle x^3 \rangle \langle q^2 J_1 \rangle + \gamma \langle x q^2 \rangle \langle x^2 J_1 \rangle + \frac{1}{4} \langle x^2 J_1 \rangle^2 + \frac{2}{3} \langle x^3 \rangle \langle x I_2 \rangle \right) H_3(\nu) \\- \left( \frac{\gamma}{6} \langle x q^2 \rangle \langle x^3 \rangle + \frac{\gamma}{6} \langle x^3 \rangle \langle x^2 J_1 \rangle \right) H_5(\nu) + \frac{\gamma^2}{72} \langle x^3 \rangle^2 H_7(\nu)
\Bigg).
\end{multline}
This result can be simplified using the isotropy relations~(\ref{eq:isotropy}):
\begin{multline}
  \chi(\nu) =\frac{1}{4\pi \sqrt{2\pi} {R_*^2}} e^{-\frac{\nu^2}{2}} \Bigg( \gamma^2 H_1 (\nu) - \gamma  \left\langle q^2 I_1\right\rangle  H_0(\nu)-\frac{1}{2} \gamma  \left\langle x ^2 I_1\right\rangle H_2(\nu)+\frac{\gamma^2}{6} \left\langle x ^3\right\rangle  H_4(\nu)
\\+ \left( \frac{\gamma^2}{2} \langle q^4 \rangle_\mathrm{c} + 2\gamma \langle x q^2 J_1 \rangle_\mathrm{c} + 2\langle x^2 I_2 \rangle_\mathrm{c}  \right) H_1(\nu) - \left( \frac{\gamma^2}{2} \langle x^2 q^2 \rangle_\mathrm{c}  + \frac{\gamma}{3} \langle x^3 J_1 \rangle_\mathrm{c} \right) H_3(\nu) + \frac{\gamma^2}{24} \langle x^4 \rangle_\mathrm{c} H_5(\nu)
\\- \frac{1}{2} \langle q^2 J_1 \rangle \langle x^2 J_1 \rangle H_1(\nu) - \frac{\gamma}{6} \langle x^3 \rangle \langle q^2 J_1 \rangle H_3(\nu) - \frac{\gamma}{12} \langle x^3 \rangle \langle x^2 J_1 \rangle H_5(\nu) + \frac{\gamma^2}{72} \langle x^3 \rangle^2 H_7(\nu) \Bigg).
\end{multline}
This expression agrees with \cite{Matsubara2010}  given the following identities:
\begin{align}
 \langle x^2 q^2 \rangle_c &= -\frac{1}{3\gamma} \langle x^3 J_1 \rangle_\mathrm{c}\,, & \langle q^4 \rangle_\mathrm{c} + \frac{3}{\gamma} \langle x \,q^2 J_1 \rangle_\mathrm{c} + \frac{2}{\gamma^2} \langle x^2 I_2 \rangle_\mathrm{c} &=0\,.
\end{align}
These relations are consistent with the measurements.
The compactness of the expansion in invariants allows us to compute the $\sigma^3$ correction:
\begin{multline}
  \chi^{(3)}(\nu) = \left( \frac{3}{2} \gamma \langle q^4 \rangle_\mathrm{c} \langle q^2 J_1 \rangle - \gamma \langle q^4 J_1 \rangle_\mathrm{c} + \frac{3}{\gamma} \langle q^2 J_1 \rangle^2 \langle x^2 J_1 \rangle - \frac{4}{\gamma} \langle q^2 I_2 \rangle \langle x^2 J_1 \rangle + \frac{3}{2} \gamma \langle q^4 \rangle_\mathrm{c} \langle q^4 J_1 \rangle + 4 \langle q^2 J_1 \rangle \langle x q^2 J_1 \rangle_\mathrm{c} - 4 \langle x q^2 I_2 \rangle_\mathrm{c} \right) H_0(\nu)
\\+ \left( \frac{1}{3} \langle x^3 \rangle \langle q^2 J_1 \rangle^2 - \frac{1}{3} \langle q^2 J_1 \rangle \langle x^3 J_1 \rangle_\mathrm{c} - \frac{\gamma}{2} \langle q^2 J_1 \rangle \langle x^2 q^2 \rangle_\mathrm{c} + \frac{\gamma}{4} \langle q^4 \rangle_\mathrm{c} \langle x^2 J_1 \rangle + \gamma \langle x^2 q^2 J_1 \rangle_\mathrm{c} - \frac{1}{4\gamma} \langle q^2 J_1 \rangle\langle x^2 J_1 \rangle^2 + \langle x^2 J_1 \rangle \langle x q^2 J_1 \rangle_\mathrm{c} \right.\\\left. + \frac{1}{\gamma} \langle x^2 I_2 \rangle_\mathrm{c} \langle x^2 J_1 \rangle - \frac{2}{3} \langle q^2 I_2 \rangle_\mathrm{c} \langle x^2 \rangle + \frac{\gamma^2}{2} \langle x q^4 \rangle_\mathrm{c} - \frac{3}{4} \gamma^2 \langle q^4 \rangle_\mathrm{c} \langle x q^4 \rangle_\mathrm{c} + \frac{2}{3} \langle x^3 I_2 \rangle_\mathrm{c} \right) H_2(\nu)+\nonumber
\end{multline}
\begin{multline}
+ \left( \frac{1}{3} \gamma \langle x^3 \rangle \langle x q^2 J_1 \rangle_\mathrm{c}  - \frac{1}{24} \gamma \langle x^4 \rangle_\mathrm{c} \langle q^2 J_1 \rangle - \frac{1}{12} \langle x^3 \rangle \langle q^2 J_1 \rangle \langle x^2 J_1 \rangle - \frac{1}{12} \gamma \langle x^4 J_1 \rangle_\mathrm{c} + \frac{1}{12} \gamma^2 \langle q^4 \rangle_\mathrm{c} \langle x^3 \rangle - \frac{1}{6} \gamma^2 \langle x^3 q^2 \rangle_\mathrm{c} + \frac{1}{3} \langle x^3 \rangle \langle x^2 I_2 \rangle_\mathrm{c} \right) H_4(\nu)
\\+ \left(\frac{1}{120} \gamma^2 \langle x^5 \rangle_\mathrm{c} - \frac{1}{72} \gamma \langle x^3 \rangle^2 \langle q^2 J_1 \rangle - \frac{1}{18} \gamma \langle x^3 \rangle \langle x^3 J_1 \rangle_\mathrm{c} - \frac{1}{48} \gamma \langle x^4 \rangle_\mathrm{c} \langle x^2 J_1 \rangle - \frac{1}{12} \gamma^2 \langle x^3 \rangle \langle x^2 q^2 \rangle_\mathrm{c}  \right) H_6(\nu)
\\+ \left( \frac{1}{144} \gamma^2 \langle x^3 \rangle \langle x^4 \rangle_\mathrm{c} - \frac{1}{144} \gamma \langle x^3 \rangle^2 \langle x^2 J_1 \rangle \right) H_8(\nu)
+ \frac{1}{1296} \gamma^2 \langle x^2 \rangle^3 H_{10} (\nu)\,.\label{eq:defchi3}
\end{multline}

\subsection{3D Euler characteristic}

In 3D, the first-order term from equation~(\ref{eq:genus_3D}) is
\begin{multline}
 \chi(\nu) = \frac{1}{27 {R_*^3} } \left( \frac{3}{2\pi} \right)^{3/2} \frac{1}{\sqrt{2\pi}} e^{-\frac{\nu^2}{2}} \Bigg[ \gamma^3 H_2(\nu) \\
+ \left( \frac{9}{2} \gamma^2 \left\langle q^2 J_1 \right\rangle + 3 \gamma \sqrt{1-\gamma^2} \left\langle \zeta {J_1}^2 \right\rangle - 3\gamma^2 \left\langle {J_1}^3 \right\rangle -3 \gamma \sqrt{1-\gamma^2} \left\langle \zeta J_2 \right\rangle + 3 \gamma^2 \left\langle J_1 J_2 \right\rangle \right) H_1(\nu)\\
+ \left( -\frac{3}{2} \gamma^3 \sqrt{1-\gamma^2} \left\langle \zeta q^2 \right\rangle + \frac{3}{2} \gamma^4 \left\langle q^2 J_1 \right\rangle -\frac{3}{2}\gamma^2 (1-\gamma^2) \left\langle \zeta^2 J_1 \right\rangle +3\gamma^3 \sqrt{1-\gamma^2} \left\langle \zeta {J_1}^2 \right\rangle -\frac{3}{2} \gamma^4 \left\langle {J_1}^3 \right\rangle \right) H_3(\nu)\\
+ \left( \frac{1}{6} \gamma^3 (1-\gamma^2)^{3/2} \left\langle \zeta^3 \right\rangle - \frac{1}{2} \gamma^4 (1-\gamma^2) \left\langle \zeta^2 J_1 \right \rangle + \frac{1}{2} \gamma^5 \sqrt{1-\gamma^2} \left\langle \zeta {J_1}^2 \right\rangle - \frac{1}{6} \gamma^6 \left\langle {J_1}^3 \right\rangle \right) H_5(\nu) \Bigg].
\end{multline}
As in 2D, it can be simplified by using the variables $x$, $I_2$ and $I_3$:
\begin{multline}
  \chi(\nu) = \frac{1}{27 {R_*^3} } \left( \frac{3}{2\pi} \right)^{3/2} \frac{1}{\sqrt{2\pi}} e^{-\frac{\nu^2}{2}} \Bigg[ \gamma^3 H_2(\nu)
+  \left( \frac{9}{2} \gamma^2 \left\langle q^2 I_1 \right\rangle + 9 \gamma \left\langle x I_2 \right\rangle \right) H_1(\nu) \\
- \left( \frac{3}{2} \gamma^3 \left\langle x q^2 \right\rangle + \frac{3}{2} \gamma^2 \left\langle x^2 I_1 \right\rangle \right) H_3(\nu)
+ \frac{1}{6} \gamma^3 \left\langle x^3 \right\rangle H_5(\nu) \Bigg].
\end{multline}

\citeauthor{Matsubara}'s expression for the three dimensional genus is
\begin{equation}
 G_3 (\nu) = \frac{1}{(2\pi)^2} \left( \frac{\sigma_1}{\sqrt{3} \sigma} \right)^3 e^{-\frac{\nu^2}{2}} \Bigg[ H_2(\nu) + \left( \frac{S^{(0)}}{6} H_5(\nu) + S^{(1)} H_3(\nu) + S^{(2)} H_1(\nu) \right) \sigma\Bigg]\,,
\end{equation} with
\begin{align}
 S^{(0)} &= \frac{1}{\sigma} \left\langle x^3 \right\rangle, & S^{(1)} &= -\frac{3}{4} \frac{ \left\langle x^2 I_1 \right\rangle }{\gamma \sigma}, & S^{(2)} &= - \frac{9}{4} \frac{ \left\langle q^2 I_1 \right\rangle }{\gamma \sigma}.
\end{align}
Factorizing the expression for $\chi(\nu)$ with $\gamma^3$ leads to the same prefactor except for the sign. This sign difference comes from the fact that $\chi(\nu)$ is the Euler characteristic of the excursion set while $G_3(\nu)$ is the genus of the isocontour and thus $\chi(\nu)=-G_3(\nu)$. Using the isotropy identities given in equation~(\ref{eq:isotropy}), it is then easy to show that each term coincides.
Finally, the 3D Euler characteristic up to $\sigma^2$ order expressed in terms of $x$, $I_2$ and $I_3$ is:
\begin{multline}
 \chi(\nu) = \frac{1}{27 {R_*^3} } \left( \frac{3}{2\pi} \right)^{3/2} \frac{1}{\sqrt{2\pi}} e^{-\frac{\nu^2}{2}} \Bigg[ \gamma^3 H_2(\nu) 
 + \left( \frac{9}{2} \gamma^2 \langle q^2 J_1 \rangle + 9 \gamma \langle x I_2 \rangle \right) H_1(\nu) - \left( \frac{3}{2} \gamma^3 \langle x q^2 \rangle + \frac{3}{2} \gamma^2 \langle x^2 J_1 \rangle \right) H_3(\nu) \\+ \frac{1}{6} \gamma^3 \langle x^3 \rangle H_5(\nu)
 - \left( \frac{27}{2} \gamma \langle q^2 I_2 \rangle_\mathrm{c} + 27 \langle x I_3 \rangle_\mathrm{c} \right) H_0(\nu) + \left( \frac{9}{8} \gamma^3 \langle q^4 \rangle_c + \frac{9}{2} \gamma^2 \langle x q^2 J_1 \rangle_\mathrm{c} + \frac{9}{2} \gamma \langle x^2 I_2 \rangle_\mathrm{c} \right) H_2(\nu) \\- \left( \frac{3}{4} \gamma^3 \langle x^2 q^2 \rangle_\mathrm{c} + \frac{1}{2} \gamma^2 \langle x^2 J_1 \rangle_\mathrm{c} \right) H_4(\nu) + \frac{1}{24} \gamma^3 \langle x^4 \rangle_\mathrm{c} H_6(\nu)
 + \left( \frac{135}{16} \gamma \langle q^2 J_1 \rangle^2 + \frac{27}{2} \langle q^2 J_1 \rangle \langle x I_2 \rangle + \frac{81}{2} \langle x q^2 \rangle \langle I_3 \rangle \right) H_0(\nu) \\- \left( \frac{45}{4} \gamma^2 \langle x q^2 \rangle \langle q^2 J_1 \rangle + \frac{9}{2} \gamma \langle q^2 J_1 \rangle \langle x^2 J_1 \rangle + \frac{27}{2} \gamma \langle x q^2 \rangle \langle x I_2 \rangle + \frac{9}{2} \langle x^2 J_1 \rangle \langle x I_2 \rangle + \frac{9}{2} \langle x^3 I_3 \rangle \right) H_2(\nu) \\+ \left(\frac{15}{8} \gamma^3 \langle x q^2 \rangle^2 + \frac{3}{4} \gamma^2 \langle x^3 \rangle \langle q^2 J_1 \rangle + \frac{9}{4} \gamma^2 \langle x q^2 \rangle \langle x^2 J_1 \rangle + \frac{3}{4} \gamma \langle x^2 J_1 \rangle^2 + \frac{3}{2} \gamma \langle x^3 \rangle \langle x I_2 \rangle \right) H_4(\nu) \\- \left( \frac{1}{4} \gamma^3 \langle x q^2 \rangle \langle x^3 \rangle + \frac{1}{4} \gamma^2 \langle x^3 \rangle \langle x^2 J_1 \rangle \right) H_6(\nu) + \frac{1}{72} \gamma^3 \langle x^3 \rangle^2 H_8(\nu)
\Bigg]\,.
\end{multline}

\section{Geometric cumulants from perturbation theory}
\label{sec:PTappendix}

In this section, we will compute the 3rd order cumulants involved in the non-Gaussian expansion in the case where non-Gaussianities arise from the non-linear gravitational evolution.

\subsection{Derivation for an EdS universe }
\label{PTderiv}

For a three dimensional field, the  cumulants involving the field and its derivatives can be written as
\begin{equation} \left\langle \left( \partial_1^{\alpha_1} \partial_2^{\alpha_2} \partial_3^{\alpha_3} \delta \right) ( \partial_1^{\beta_1} \partial_2^{\beta_2} \partial_3^{\beta_3} \delta ) \left( \partial_1^{\gamma_1} \partial_2^{\gamma_2} \partial_3^{\gamma_3} \delta \right) \right\rangle 
\label{eq:defcum}
\end{equation} where $\partial_i^n $ is the $n$-th derivative with respect to the $i$-th coordinate.
Using gravitational  perturbation theory \citep{Bernardeau}, we expand the field as $\delta=\delta^{(1)} + \delta^{(2)} + \dots$. Inserting this expansion in  equation~(\ref{eq:defcum}) gives the lowest non-vanishing order contribution to the cumulant:
\begin{multline} \left\langle ( \partial_1^{\alpha_1} \partial_2^{\alpha_2} \partial_3^{\alpha_3} \delta ) ( \partial_1^{\beta_1} \partial_2^{\beta_2} \partial_3^{\beta_3} \delta ) ( \partial_1^{\gamma_1} \partial_2^{\gamma_2} \partial_3^{\gamma_3} \delta ) \right\rangle = 
 \left\langle ( \partial_1^{\alpha_1} \partial_2^{\alpha_2} \partial_3^{\alpha_3} \delta^{(1)} ) ( \partial_1^{\beta_1} \partial_2^{\beta_2} \partial_3^{\beta_3} \delta^{(1)} ) ( \partial_1^{\gamma_1} \partial_2^{\gamma_2} \partial_3^{\gamma_3} \delta^{(2)} ) \right\rangle \\ +
 \left\langle ( \partial_1^{\alpha_1} \partial_2^{\alpha_2} \partial_3^{\alpha_3} \delta^{(1)} ) ( \partial_1^{\beta_1} \partial_2^{\beta_2} \partial_3^{\beta_3} \delta^{(2)} ) ( \partial_1^{\gamma_1} \partial_2^{\gamma_2} \partial_3^{\gamma_3} \delta^{(1)} ) \right\rangle +
 \left\langle ( \partial_1^{\alpha_1} \partial_2^{\alpha_2} \partial_3^{\alpha_3} \delta^{(2)} ) ( \partial_1^{\beta_1} \partial_2^{\beta_2} \partial_3^{\beta_3} \delta^{(1)} ) ( \partial_1^{\gamma_1} \partial_2^{\gamma_2} \partial_3^{\gamma_3} \delta^{(1)} ) \right\rangle. \label{eq:defIIs}
\end{multline}
Let us compute the first term in equation~(\ref{eq:defIIs}), ${\cal I}\equiv \langle ( \partial_1^{\alpha_1} \partial_2^{\alpha_2} \partial_3^{\alpha_3} \delta^{(1)} )  ( \partial_1^{\beta_1} \partial_2^{\beta_2} \partial_3^{\beta_3} \delta^{(1)} ) ( \partial_1^{\gamma_1} \partial_2^{\gamma_2} \partial_3^{\gamma_3} \delta^{(2)} ) \rangle $, which is best expressed in Fourier space as (with $\mathbf{k}^{[j]}$ the $j$th component of vector $\mathbf k$)
\begin{multline}
{\cal I}= 
 \int {\rm d}^3 \mathbf{k_1} {\rm d}^3 \mathbf{k_2} {\rm d}^3 \mathbf{k_3} e^{i (\mathbf{k_1}+\mathbf{k_2}+\mathbf{k_3})\cdot\mathbf{x}}
 \left\langle \left( i\mathbf{k}_1^{[1]} \right)^{\alpha_1} \left( i\mathbf{k}_1^{[2]} \right)^{\alpha_2} \left( i\mathbf{k}_1^{[3]} \right)^{\alpha_3} \delta^{(1)}(\mathbf{k_1}) \right. \\
 \left.  \left( i\mathbf{k}_2^{[1]} \right)^{\beta_1} \left( i\mathbf{k}_2^{[2]} \right)^{\beta_2} \left( i\mathbf{k}_2^{[3]} \right)^{\beta_3} \delta^{(1)}(\mathbf{k_2})  
  \left( i \mathbf{k}^{[1]}_3 \right)^{\gamma_1} \left( i\mathbf{k}_3^{[2]} \right)^{\gamma_2} \left( i \mathbf{k}^{[3]}_3 \right)^{\gamma_3} \delta^{(2)}(\mathbf{k_3}) \right\rangle .
  \label{eq:IFour}
\end{multline} 
In equation~(\ref{eq:IFour}), we assumed zero smoothing of the field for now.
Using the notations of  \cite{Bernardeau}, given the Euler and continuity equation for an EdS universe, we can express $\delta^{(2)}$ in terms of $\delta^{(1)}$ as :
\begin{equation}
 \delta^{(2)} (\mathbf{k}) = \int {\rm d}^3 \mathbf{k_1} {\rm d}^3 \mathbf{k_2} \delta^D(\mathbf{k}-\mathbf{k_1}-\mathbf{k_2} )F_2 (\mathbf{k_1},\mathbf{k_2}) \delta^{(1)}(\mathbf{k_1}) \delta^{(1)}(\mathbf{k_2}) \,,
\quad {\rm with}
\quad
 F_2(\mathbf{k_1},\mathbf{k_2}) = \frac{5}{7} + \frac{\mathbf{k_1}\cdot \mathbf{k_2}}{{k_1}^2} + \frac{2}{7} \frac{\left( \mathbf{k_1}\cdot \mathbf{k_2}\right)^2}{{k_1}^2 {k_2}^2}.
 \label{eq:I2}
\end{equation}
Thus equation~(\ref{eq:I2}) becomes:
\begin{multline}
{\cal I}=\
 \int {\rm d}^3 \mathbf{k_1} {\rm d}^3 \mathbf{k_2} {\rm d}^3 \mathbf{k_4} {\rm d}^3 \mathbf{k_5} e^{i (\mathbf{k_1}+\mathbf{k_2}+\mathbf{k_4}+\mathbf{k_5})\cdot\mathbf{x}}
\left( i\mathbf{k}_1^{[1]} \right)^{\alpha_1} \left( i\mathbf{k}_1^{[2]} \right)^{\alpha_2} \left( i\mathbf{k}_1^{[3]} \right)^{\alpha_3}
\left( i\mathbf{k}_2^{[1]} \right)^{\beta_1} \left( i\mathbf{k}_2^{[2]} \right)^{\beta_2} \left( i\mathbf{k}_2^{[3]} \right)^{\beta_3} \\
\left( i \left(\mathbf{k}^{[1]}_4+\mathbf{k}^{[1]}_5 \right)  \right)^{\gamma_1} \left( i \left((\mathbf{k}^{[2]}_4+\mathbf{k}^{[2]}_5 \right)  \right)^{\gamma_2}
 \left( i \left(\mathbf{k}^{[3]}_4+\mathbf{k}^{[3]}_5\right)  \right)^{\gamma_3} F_2 (\mathbf{k_4},\mathbf{k_5})
\left\langle \delta^{(1)} (\mathbf{k_1}) \delta^{(1)} (\mathbf{k_2}) \delta^{(1)} (\mathbf{k_4}) \delta^{(1)} (\mathbf{k_5}) \right\rangle .
 \end{multline}
To simplify, we can combine the derivatives and $F_2$ into a  generalized shape factor:
\begin{multline}
  \mathcal{F}_{\alpha,\beta,\gamma} (\mathbf{k_1},\mathbf{k_2},\mathbf{k_4},\mathbf{k_5})=
\left( i\mathbf{k}_1^{[1]} \right)^{\alpha_1} \left( i\mathbf{k}_1^{[2]} \right)^{\alpha_2} \left( i\mathbf{k}_1^{[3]} \right)^{\alpha_3}
\left( i\mathbf{k}_2^{[1]} \right)^{\beta_1} \left( i\mathbf{k}_2^{[2]} \right)^{\beta_2} \left( i\mathbf{k}_2^{[3]} \right)^{\beta_3}\\
\left( i \left(\mathbf{k}^{[1]}_4+\mathbf{k}^{[1]}_5 \right)  \right)^{\gamma_1} \left( i \left((\mathbf{k}^{[2]}_4+\mathbf{k}^{[2]}_5 \right)  \right)^{\gamma_2}
 \left( i \left(\mathbf{k}^{[3]}_4+\mathbf{k}^{[3]}_5\right)  \right)^{\gamma_3} F_2 (\mathbf{k_4},\mathbf{k_5}), \label{eq:defF2gen}
 \end{multline} which gives
\begin{equation}
{\cal I} =
 \int {\rm d}^3 \mathbf{k_1} {\rm d}^3 \mathbf{k_2} {\rm d}^3 \mathbf{k_4} {\rm d}^3 \mathbf{k_5} e^{i (\mathbf{k_1}+\mathbf{k_2}+\mathbf{k_4}+\mathbf{k_5})\cdot\mathbf{x}}
\mathcal{F}_{\alpha,\beta,\gamma} (\mathbf{k_1},\mathbf{k_2},\mathbf{k_4},\mathbf{k_5})
\left\langle \delta^{(1)} (\mathbf{k_1}) \delta^{(1)} (\mathbf{k_2}) \delta^{(1)} (\mathbf{k_4}) \delta^{(1)} (\mathbf{k_5}) \right\rangle .
 \end{equation}
The integrated moment is then expanded using Wick's theorem:
\begin{multline}
   \left\langle \delta^{(1)} (\mathbf{k_1}) \delta^{(1)} (\mathbf{k_2}) \delta^{(1)} (\mathbf{k_4}) \delta^{(1)} (\mathbf{k_5}) \right\rangle  =
\left\langle \delta^{(1)} (\mathbf{k_1}) \delta^{(1)} (\mathbf{k_2}) \right\rangle \left\langle \delta^{(1)} (\mathbf{k_4}) \delta^{(1)} (\mathbf{k_5}) \right\rangle + \\
 \left\langle \delta^{(1)} (\mathbf{k_1}) \delta^{(1)} (\mathbf{k_4}) \right\rangle \left\langle \delta^{(1)} (\mathbf{k_2}) \delta^{(1)} (\mathbf{k_5}) \right\rangle +
 \left\langle \delta^{(1)} (\mathbf{k_1}) \delta^{(1)} (\mathbf{k_5}) \right\rangle \left\langle \delta^{(1)} (\mathbf{k_2}) \delta^{(1)} (\mathbf{k_4}) \right\rangle.
\end{multline}
The first term leads to $F_2(\mathbf{k_4},-\mathbf{k_4})$ which vanishes, while the second and third terms are equivalent:
\begin{equation}
 {\cal I}  = 
2 \int {\rm d}^3 \mathbf{k_1}  {\rm d}^3 \mathbf{k_2} {\rm d}^3 \mathbf{k_4} {\rm d}^3 \mathbf{k_5} 
\mathcal{F}_{\alpha,\beta,\gamma} (\mathbf{k_1},\mathbf{k_2},\mathbf{k_4},\mathbf{k_5}) \delta^D(\mathbf{k_1} +\mathbf{k_4}) P(k_1) \delta^D(\mathbf{k_2} +\mathbf{k_5}) P(k_2).
 \end{equation}
Integrating over $\mathbf{k_4}$ and $\mathbf{k_5}$ gives:
\begin{equation}
{\cal I}  =
2 \int {\rm d}^3 \mathbf{k_1} {\rm d}^3 \mathbf{k_2}
\mathcal{F}_{\alpha,\beta,\gamma} (\mathbf{k_1},\mathbf{k_2}) P(k_1) P(k_2)\,,
   \end{equation} 
   where we use the notation $\mathcal{F}_{\alpha,\beta,\gamma} (\mathbf{k_1},\mathbf{k_2})=\mathcal{F}_{\alpha,\beta,\gamma} (\mathbf{k_1},\mathbf{k_2},-\mathbf{k_1},-\mathbf{k_2})$ (see equation~(\ref{eq:defcalG}) in the main text).
Finally, combining the three terms in equation~(\ref{eq:defIIs}) gives us the cumulant:
\begin{equation}
\left\langle \left( \partial_1^{\alpha_1} \partial_2^{\alpha_2} \partial_3^{\alpha_3} \delta \right) ( \partial_1^{\beta_1} \partial_2^{\beta_2} \partial_3^{\beta_3} \delta ) \left( \partial_1^{\gamma_1} \partial_2^{\gamma_2} \partial_3^{\gamma_3} \delta \right) \right\rangle=2 \int {\rm d}^3 \mathbf{k_1} {\rm d}^3 \mathbf{k_2}
\left( \mathcal{F}_{\alpha,\beta,\gamma} (\mathbf{k_1},\mathbf{k_2})+ \mathcal{F}_{\beta,\gamma,\alpha} (\mathbf{k_1},\mathbf{k_2})+ \mathcal{F}_{\gamma,\alpha,\beta} (\mathbf{k_1},\mathbf{k_2})  \right) P(k_1) P(k_2). \nonumber 
 \end{equation}
Let us now take into account the smoothing of the field over a scale, $R$; the cumulants become:
\begin{equation}
2 \int {\rm d}^3 \mathbf{k_1} {\rm d}^3 \mathbf{k_2} 
\left[ \mathcal{F}_{\alpha,\beta,\gamma} (\mathbf{k_1},\mathbf{k_2})+ \mathcal{F}_{\beta,\gamma,\alpha} (\mathbf{k_1},\mathbf{k_2})+ \mathcal{F}_{\gamma,\alpha,\beta} (\mathbf{k_1},\mathbf{k_2})  \right] P(k_1) P(k_2) W(k_1 R)W(k_2 R)  W(|\mathbf{k_1}+\mathbf{k_2}| R) \,. \label{eq:finalcum}
 \end{equation}
Equation~(\ref{eq:finalcum}) is the general expression for the cumulants used in the main text.
The appearance of the dependence on the relative orientation of the wave
vectors $\vk_{1}$ and $\vk_{2}$ in the filter factor $W(|\vk_{1}+\vk_{2}|R)$
is the source of  most  of the complexity of the theory and, in some sense,
the essence of  perturbation theory. It reflects the fact that
the nonlinear field, smoothed at radius R, is not determined solely
by the average quantities at this radius, but also but what happens
at shorter scales.
For Gaussian filtering, the filter function can be expanded in Legendre series with respect
to $\vk_{1}\cdot\vk_{2}$ \citep{Lokas}:
\begin{equation}
W(|\vk_{1}+\vk_{2}|R)=\exp\left(-\frac{1}{2}(k_{1}^{2}+k_{2}^{2})R^{2}\right)\sum_{\ell=0}^{\infty}(-1)^{\ell}(2\ell+1)P_{\ell}\left(\frac{\vk_{1}\cdot\vk_{2}}{k_1 k_2}\right)I_{\ell+1/2}(k_{1}k_{2}R^{2})\sqrt{\frac{\pi}{2k_{1}k_{2}R^{2}}}.
\end{equation}
Then,
$\mathcal{F}_{\alpha,\beta,\gamma}$ can be decomposed on the basis of the Legendre polynomials and the integration over the angles then just requires the orthogonality relation of the Legendre polynomials:
\begin{equation}
 \int_{-1}^1 {\rm d}x P_m(x) P_n(x)= \frac{2}{2m+1} \delta_{m,n}. 
\end{equation} It also means that in practice, one does not need Legendre polynomials of order higher than the degree of the integrated term (here, 2 from $F_2$ plus the number of derivatives) and can truncate the expansion.
The result is an integral of Bessel functions, which can be expressed using hypergeometric functions. The results for all the independent third order cumulants is given in Appendix~\ref{sec:PTresults}.

\subsection{Gaussian filtered scale invariant geometric cumulants }
\label{sec:PTresults}
Using the above defined procedure, the cumulants can be computed for a Gaussian filter. For a scale-invariant power-spectrum of index ${ n}$ (called $n_s$ in the main text), the results can be analytically expressed using the hypergeometric function $_2 F_1 (a,b,c,x)$.
For example, the result for the skewness is already known \citep{Lokas}:
\begin{equation}
\frac{1}{\sigma} \langle x^3 \rangle = 3\; {_2 F_1} \left(\frac{3+{ n}}{2},\frac{3+n}{2},\frac{3}{2},\frac{1}{4} \right) - \frac{1}{7} (8+7n) {\;_2 F_1} \left(\frac{3+n}{2},\frac{3+n}{2},\frac{5}{2},\frac{1}{4} \right)\,.
\end{equation}
Proceeding the same way with the cumulants involving derivatives of the field yields:
\begin{equation}
\frac{1}{\sigma} \langle x {x_1}^2 \rangle = \frac{4(48+62n+21n^2)}{21n^2} {\;_2 F_1} \left(\frac{3+n}{2},\frac{3+n}{2},\frac{3}{2},\frac{1}{4} \right) - \frac{6(3+n)(8+7n)}{21n^2} {\;_2 F_1} \left(\frac{3+n}{2},\frac{5+n}{2},\frac{3}{2},\frac{1}{4} \right),
\end{equation}
\begin{multline} \frac{1}{\sigma} \langle x {x_{11}}^2 \rangle = -\frac{4(30720+51456n+37092n^2+13828n^3+2579n^4+189n^5)}{35n^2 (2+n)^2 (4+n)^2 (5+n)^2} \Bigg( (4+2n) {\;_2 F_1} \left(\frac{3+n}{2},\frac{3+n}{2},-\frac{1}{2},\frac{1}{4} \right) \\ +3 {\;_2 F_1} \left(\frac{3+n}{2},\frac{5+n}{2},-\frac{1}{2},\frac{1}{4} \right)\Bigg) + \frac{3(3840+5016n+2748n^2+693n^3+63n^4)}{35n^2 (2+n)(4+n)(5+n)(6+n)} \\\Bigg(3 {\;_2 F_1} \left(\frac{3+n}{2},\frac{7+n}{2},-\frac{1}{2},\frac{1}{4} \right) + n {\;_2 F_1} \left(\frac{3+n}{2},\frac{7+n}{2},\frac{1}{2},\frac{1}{4} \right) \Bigg),
\end{multline}
\begin{multline}
 \frac{1}{\sigma} \langle x x_{11} x_{22} \rangle = \frac{ 4( -46080 - 37984n -2848 n^2 +5718 n^3 +1949 n^4 +189 n^5)}{105 n^2 (2+n)^2 (4+n)^2 (5+n)^2 } \Bigg( (4+2n)  {\;_2 F_1} \left(\frac{3+n}{2},\frac{3+n}{2},-\frac{1}{2},\frac{1}{4} \right) \\ +3  {\;_2 F_1} \left(\frac{3+n}{2},\frac{5+n}{2},-\frac{1}{2},\frac{1}{4} \right) \Bigg) + \frac{ 3(-5760-2624n+608n^2+476 n^3+63n^4 )}{105 n^2 (2+n)(4+n)(5+n)(6+n)} \\\Bigg( 3 {\;_2 F_1} \left(\frac{3+n}{2},\frac{7+n}{2},-\frac{1}{2},\frac{1}{4} \right) + n  {\;_2 F_1} \left(\frac{3+n}{2},\frac{7+n}{2},\frac{1}{2},\frac{1}{4} \right) \Bigg),
\end{multline}
\begin{multline}
  \frac{1}{\sigma} \langle x {x_{12}}^2 \rangle = -\frac{4(9120+96176n+57062n^2+17883n^3+2894n^4+189n^5)}{105n^2(2+n)^2 (4+n)^2 (5+n)^2} \Bigg( (4+n) {\;_2 F_1} \left(\frac{3+n}{2},\frac{3+n}{2},-\frac{1}{2},\frac{1}{4} \right) \\+3{\;_2 F_1} \left(\frac{3+n}{2},\frac{5+n}{2},-\frac{1}{2},\frac{1}{4} \right) \Bigg) + \frac{3(8640+8836n+3818n^2+791n^3+63n^4)}{105n^2 (2+n)(4+n)(5+n)(6+n)} \\\Bigg( 3{\;_2 F_1} \left(\frac{3+n}{2},\frac{7+n}{2},-\frac{1}{2},\frac{1}{4} \right) +n {\;_2 F_1} \left(\frac{3+n}{2},\frac{7+n}{2},\frac{1}{2},\frac{1}{4} \right) \Bigg).
\end{multline}
The last ones also depend on the spectral parameter
$ \gamma=\sqrt{{(n+3)}/{(n+5)}}$: 
\begin{multline}
\frac{1}{\sigma} \langle x_1 x_2 x_{12} \rangle = - \frac{\gamma}{42n^2(2+n)} \Bigg( (448+260n+42n^2)  {\;_2 F_1} \left(\frac{3+n}{2},\frac{5+n}{2},\frac{3}{2},\frac{1}{4} \right) \\-(96+50n+7n^2)   {\;_2 F_1} \left(\frac{5+n}{2},\frac{5+n}{2},\frac{3}{2},\frac{1}{4} \right) \Bigg),
\end{multline}
\begin{multline}
\frac{1}{\sigma} \langle x_{12} x_{13} x_{23} \rangle = - \frac{8 \gamma}{245n^2(2+n)(4+n)} \Bigg( (7680+6384n+2262n^2+395n^3+28n^4) {\;_2 F_1} \left(\frac{3+n}{2},\frac{7+n}{2},\frac{3}{2},\frac{1}{4} \right) \\- (4320+3438n+1180n^2+201n^3+14n^4) {\;_2 F_1} \left(\frac{5+n}{2},\frac{7+n}{2},\frac{3}{2},\frac{1}{4} \right) \Bigg),
\end{multline}
\begin{multline}
\frac{1}{\sigma} \langle x_{11} {x_{12}}^2 \rangle = - \frac{4 \gamma}{735n^2(2+n)(4+n)} \Bigg( 4(25600+21280n+8002n^2+1557n^3+126n^4)  {\;_2 F_1} \left(\frac{3+n}{2},\frac{7+n}{2},\frac{3}{2},\frac{1}{4} \right) \\-3 (19200+15280n+5582n^2+1059n^3+54n^4)  {\;_2 F_1} \left(\frac{5+n}{2},\frac{7+n}{2},\frac{3}{2},\frac{1}{4} \right) \Bigg),
\end{multline}
\begin{multline}
\frac{1}{\sigma} \langle x_{11} {x_{23}}^2 \rangle = - \frac{4 \gamma}{735n^2(2+n)(4+n)} \Bigg( 8(1280+1064n+608n^2+186n^3+21n^4) {\;_2 F_1} \left(\frac{3+n}{2},\frac{7+n}{2},\frac{3}{2},\frac{1}{4} \right) \\-3 (1920+1528n+862n^2+255n^3+28n^4)  {\;_2 F_1} \left(\frac{5+n}{2},\frac{7+n}{2},\frac{3}{2},\frac{1}{4} \right) \Bigg),
\end{multline}
\begin{multline}
\frac{1}{\sigma} \langle x_{11} x_{22} x_{33} \rangle = - \frac{4 \gamma}{245n^2(2+n)(4+n)} \Bigg( 4(-5120-4256n-1046n^2-23n^314n^4) {\;_2 F_1} \left(\frac{3+n}{2},\frac{7+n}{2},\frac{3}{2},\frac{1}{4} \right) \\+ (11520+9168n+2134n^2+39n^3-28n^4)  {\;_2 F_1} \left(\frac{5+n}{2},\frac{7+n}{2},\frac{3}{2},\frac{1}{4} \right) \Bigg),
\end{multline}
\begin{multline}
\frac{1}{\sigma} \langle {x_{11}}^2 x_{22} \rangle = - \frac{4 \gamma}{735n^2(2+n)(4+n)} \Bigg( 4(-1024-8512n-706n^2+675n^3+126n^4) {\;_2 F_1} \left(\frac{3+n}{2},\frac{7+n}{2},\frac{3}{2},\frac{1}{4} \right) \\-3 (7680+6112n+410n^2-471n^3-84n^4)  {\;_2 F_1} \left(\frac{5+n}{2},\frac{7+n}{2},\frac{3}{2},\frac{1}{4} \right) \Bigg) ,
\end{multline}
\begin{multline}
\frac{1}{\sigma} \langle {x_1}^2 x_{22} \rangle = \frac{\gamma}{21n^2(2+n)} \Bigg( (448+260n+42n^2)  {\;_2 F_1} \left(\frac{3+n}{2},\frac{5+n}{2},\frac{3}{2},\frac{1}{4} \right) \\- 3(96+50n+7n^2)  {\;_2 F_1} \left(\frac{5+n}{2},\frac{5+n}{2},\frac{3}{2},\frac{1}{4} \right) \Bigg),
\end{multline}
\begin{multline}
\frac{1}{\sigma} \langle {x_{11}}^3 \rangle = - \frac{12 \gamma}{245n^2(2+n)(4+n)} \Bigg( 4(1024+8512n+3478n^2+767n^3+70n^4) {\;_2 F_1} \left(\frac{3+n}{2},\frac{7+n}{2},\frac{3}{2},\frac{1}{4} \right) \\- (23040+18336n+7306n^2+1569n^3+140n^4)  {\;_2 F_1} \left(\frac{5+n}{2},\frac{7+n}{2},\frac{3}{2},\frac{1}{4} \right) \Bigg) ,
\end{multline}
\begin{equation}
\frac{1}{\sigma} \langle x^2 x_{11} \rangle = -\frac{\gamma}{21n^2} \left( 2(48+62n+21n^2)  {\;_2 F_1} \left(\frac{3+n}{2},\frac{3+n}{2},\frac{3}{2},\frac{1}{4} \right) - 3(3+n)(8+7n)  {\;_2 F_1} \left(\frac{3+n}{2},\frac{5+n}{2},\frac{3}{2},\frac{1}{4} \right) \right).
\end{equation}
All other cumulants are either identical to one of these because of isotropy, or null (mostly because of the parity of the derivatives). These formulae are thus sufficient to build any combination (e.g. the cumulants of the rotational invariants). In Section~\ref{sec:measurements},  Tables~\ref{tab:S3} and \ref{tab:cumulants} give numerical values and show they agree with measurements.

\begin{table}
\begin{center}
\begin{tabular}{|l|cc|cc|cc|}
\hline
 & \multicolumn{2}{c|}{$n_\mathrm{s}=0$} &
\multicolumn{2}{c|}{$n_\mathrm{s}=-1$} &
\multicolumn{2}{c|}{$n_\mathrm{s}=-2$}\\ \hline
 & prediction & measurement & prediction & measurement & prediction &
measurement\\ \hline
$\langle x^3\rangle / \sigma$ & 3.144 & $3.08\pm 0.09$ & 3.468 & $3.53 \pm
0.15$& 4.022 & $4.3\pm0.4$\\ \hline
$\langle x {x_1}^2\rangle / \sigma$ & 0.699 & $0.68 \pm 0.01$& 0.771 & $0.795
\pm 0.032$ & 0.894 & $1.0 \pm 0.1$\\ \hline
$\langle x {x_{11}}^2\rangle / \sigma$ & 0.567 & $0.554\pm0.009$ & 0.612 &
$0.636 \pm 0.026$ & 0.661 & $0.8 \pm 0.1$\\ \hline
$\langle x x_{11} x_{22} \rangle / \sigma$ & 0.361 & $0.348 \pm 0.006$ & 0.340 &
$0.361 \pm 0.018$ & 0.311& $0.44\pm0.06$\\ \hline
$\langle x {x_{12}}^2\rangle / \sigma$ & 0.103 & $0.102 \pm 0.002$& 0.136 &
$0.137 \pm 0.005$ & 0.175& $0.20\pm0.02$\\ \hline
$\langle x_1 x_2 x_{12}\rangle / \sigma$ &0.111& $0.107 \pm 0.002$ & 0.096 &
$0.106 \pm 0.007$ &0.0784& $0.13 \pm 0.02$\\ \hline
$\langle x_{12} x_{13} x_{23} \rangle / \sigma$ & 0.0124 & $0.0122 \pm 0.0003$&
0.007 & $0.010 \pm 0.01$ & 0.00254 & $0.009\pm 0.003$\\ \hline
$\langle x_{11} {x_{12}}^2\rangle / \sigma$ & -0.0068 & $-0.0061 \pm 0.0006$ &
-0.016 & $-0.014 \pm 0.001$ & -0.0237& $-0.025 \pm 0.003$\\ \hline
$\langle x_{11} {x_{23}}^2\rangle / \sigma$ & -0.0316 & $-0.0306 \pm 0.0006$ &
-0.031 & $-0.034 \pm 0.002$ &-0.0288& $-0.046 \pm 0.007$\\ \hline
$\langle x_{11} x_{22} x_{33}\rangle / \sigma$ & -0.120 & $-0.116 \pm 0.002$ &
-0.108 & $-0.121 \pm 0.008$ & -0.0914& $-0.15 \pm 0.03$\\ \hline
$\langle{x_{11}}^2 x_{22} \rangle / \sigma$ & -0.183 & $-0.177 \pm 0.004$ &
-0.170 & $-0.188 \pm 0.012$ & -0.149& $-0.24 \pm 0.04$\\ \hline
$\langle{x_1}^2 x_{22}\rangle / \sigma$ & -0.222 & $-0.214 \pm 0.005$& -0.193 &
$-0.213 \pm 0.014$ &-0.157& $-0.25\pm 0.04$\\ \hline
$\langle{x_{11}}^3\rangle / \sigma$ & -0.210 & $-0.203 \pm 0.004$ & -0.235 &
$-0.244 \pm 0.011$ &-0.244& $-0.35 \pm 0.05$\\ \hline
$\langle x^2 x_{11}\rangle / \sigma$ &-1.08& $-1.05 \pm 0.02$& -1.090 & $-1.097
\pm 0.055$ & -1.03& $-1.3 \pm 0.1$\\ \hline
\end{tabular}
\end{center}
\caption{Predicted and measured (Sec.~\ref{sec:measurements}.)  third order 3D
coordinate cumulants. According to perturbation theory, the dimensionless third
order cumulants are proportional to $ \sigma$.}
\label{tab:S3}
\end{table}

\section{Euler characteristic algorithm}
\label{sec:euleralgorithm}

The very fast, linear in the number of pixels, {\sc Euler2D}
algorithm that we use to measure the Euler characteristic in 2D is based on the Gauss-Bonnet theorem.
According to this theorem, the Euler characteristic of a region can be obtained by integrating
the curvature over the boundary surface. We however note that explicit integration of the curvature is not needed,
since the any continuous deformation of the regions conserve the topology. Hence, on a grid 
the curvature can be thought of as concentrated in the corners of the cells at the boundary between the cells above
and below the threshold and just counted by addition. What is needed is to assign the appropriate
curvature weight for each boundary grid vertex, depending on which cells that form it are below and which are above
the threshold.  The sum of the weights over all the vertices will give us the Euler characteristic
of the excursion set. Our implementation of the algorithm is complemented by the cluster analysis which
allow to count the Euler characteristics of the individual isolated regions.
 
To determine the weights, we use a bootstrap method: we consider elementary configurations
for which the Euler characteristic is known in order to determine the number to associate
with a new geometrical configuration.

For a 2D regular grid, the geometric configurations can be classified in six categories,
represented in Figure~\ref{fig:euleralgorithm1}.
\begin{figure}
 \begin{center}
\subfigure[$\chi=1$]{\includegraphics[height=0.125\textheight]{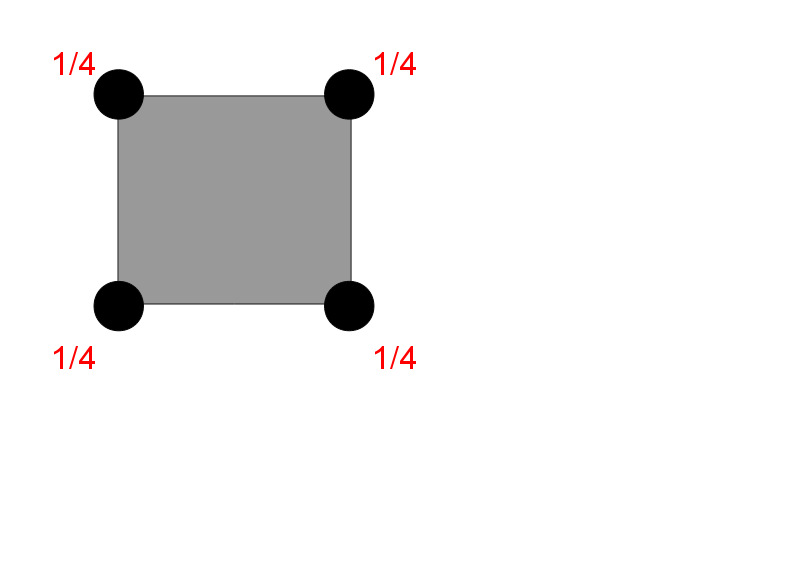}\label{fig:euleralgorithmA}}
\subfigure[$\chi=1$]{\includegraphics[height=0.125\textheight]{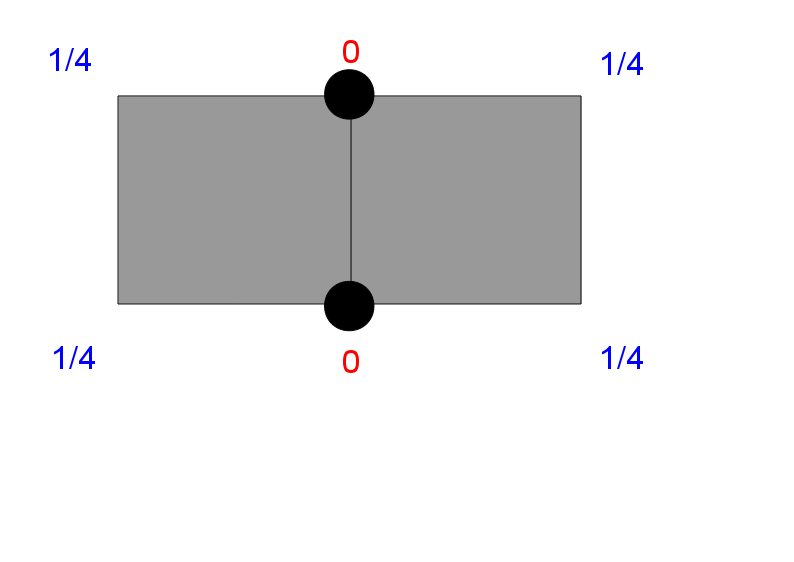}\label{fig:euleralgorithmB}}
\subfigure[$\chi=1$]{\includegraphics[height=0.125\textheight]{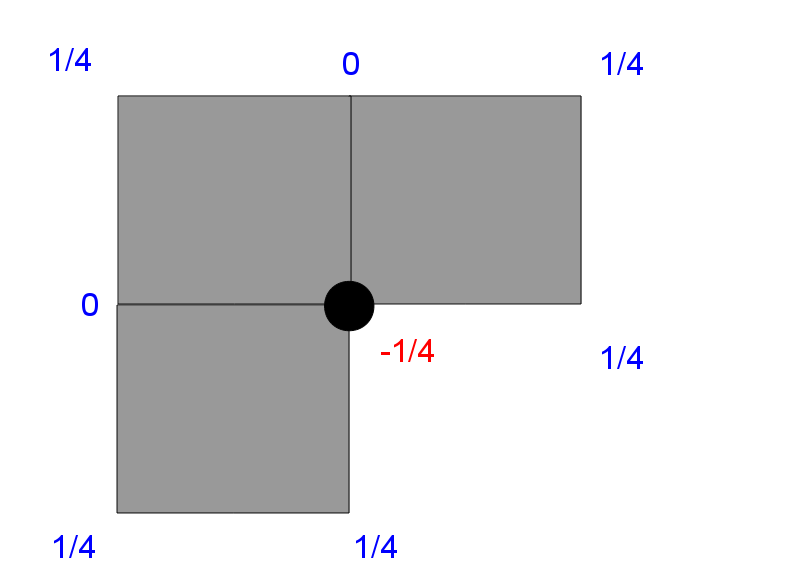}\label{fig:euleralgorithmC}}
\subfigure[$\chi=1$ or $\chi=2$]{\includegraphics[height=0.125\textheight]{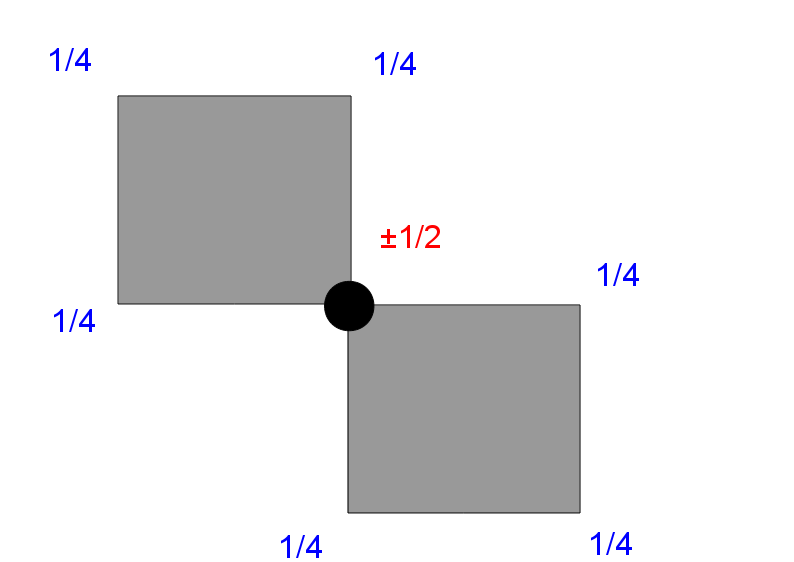}\label{fig:euleralgorithmD}}
\end{center}
\caption{Simple configurations used to determine the number of the geometrical configurations.
 The Euler characteristic of the example is written below the drawings. The red numbers are the new weights
that are determined at the current step of the bootstrap, while the blue ones are the previously determined ones.}
\end{figure}
The number associated with the vertex in the 
first configuration (when all four surrounding pixels are not part of a region) is trivial:
it is zero, since adding some empty pixels around a region does not change its topology.
Similarly a vertex surrounded by four pixels has also the weight of zero since vertices
interior to the excursion set does not change its Euler characteristic. 
The second configuration can be studied by looking at the case of a region made of
only one isolated pixel (Figure~G\ref{fig:euleralgorithmA}): 
its Euler characteristic is unity and all the four corners of the pixel are equivalent
and should thus carry the weight $1/4$. These vertices are of a type when three neighbours are outside the
region and one is inside. Recording this value, we can know look at the next category of pixel.
 Figure~G\ref{fig:euleralgorithmB} shows that if the region is made of two adjacent pixels,
the Euler characteristic is still unity and the intermediate corners of pixels must contribute the weight of zero since
four corner vertices are of the type considered before and carry the weight of $1/4$.
We can then consider a L-shaped region made of three pixels (Figure~G\ref{fig:euleralgorithmC})
to find that a vertex surrounded by three pixels in a region has a weight of $-1/4$.
The last remaining configuration is more problematic: when two pixels touch only by a corner 
(Figure~G\ref{fig:euleralgorithmD}),
 one has to decide whether they are part of the same region or not. 
Depending on this choice, the weight is $1/2$ or $-1/2$. 
In case we do track the topology of individual regions one of the decisions has to be made, which will bias
the statistical result (but will be exact for a given definition of a connected region). 
If we are interested in statistical analysis of the total Euler characteristic of the complete excursion set
we note that, statistically in 2D, the two pixels touching  by just the vertex are as likely
to be connected as they are not.  Thus, we make a statistical choice of assigning the weight zero to this
last vertex type.

The final assignment of curvature weights is given in the Figure~\ref{fig:euleralgorithm1}. One can easily repeat
the procedure in 3D, or generalize it onto more complicated grids.
It can be noticed that for the simple case of a 2D regular grid, the result is simply the number
 of turns the boundary contour rotates at a given corner \citep{Weinberg1988}. 
Furthermore, with our statistical prescription, the weights associated with  different geometrical
 configurations only depends on the number of surrounding pixels in the region, not on their configuration,
 which makes coding the algorithm very simple.

\begin{figure}
\begin{center}
\begin{tabular}{cccccc}
\includegraphics[width=0.15\textwidth]{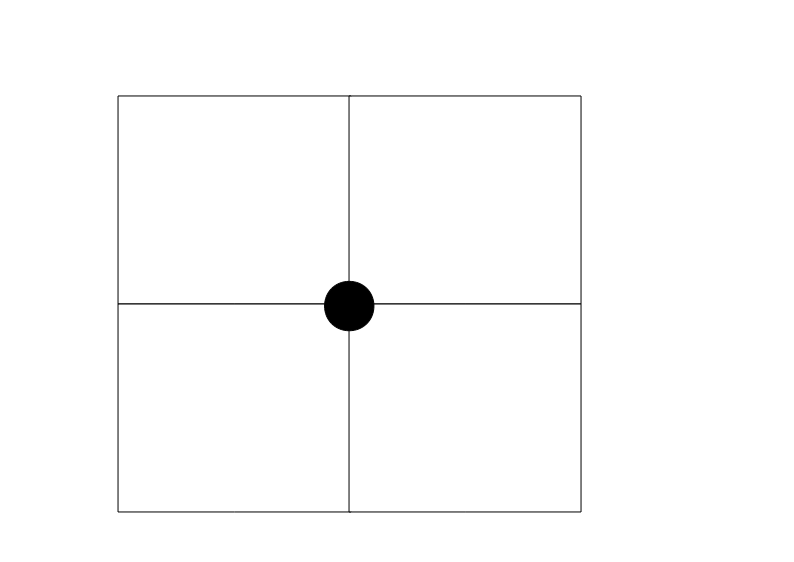} & \includegraphics[width=0.15\textwidth]{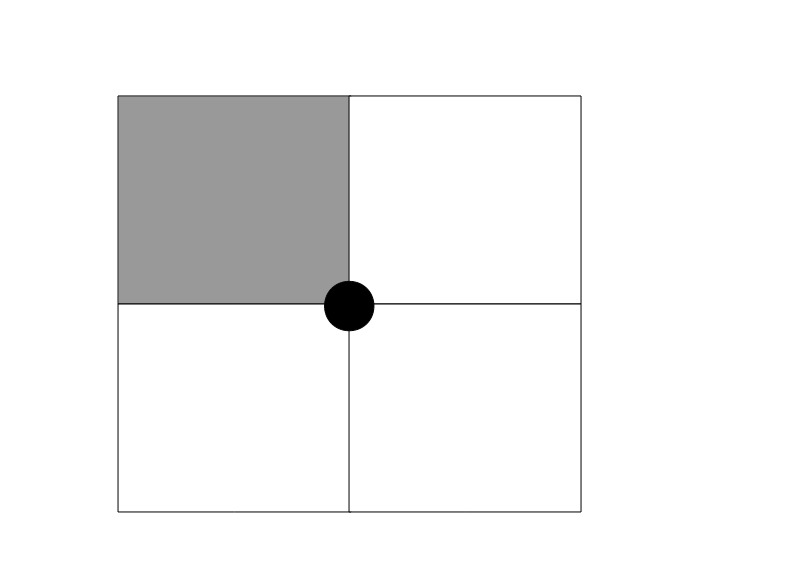} & \includegraphics[width=0.15\textwidth]{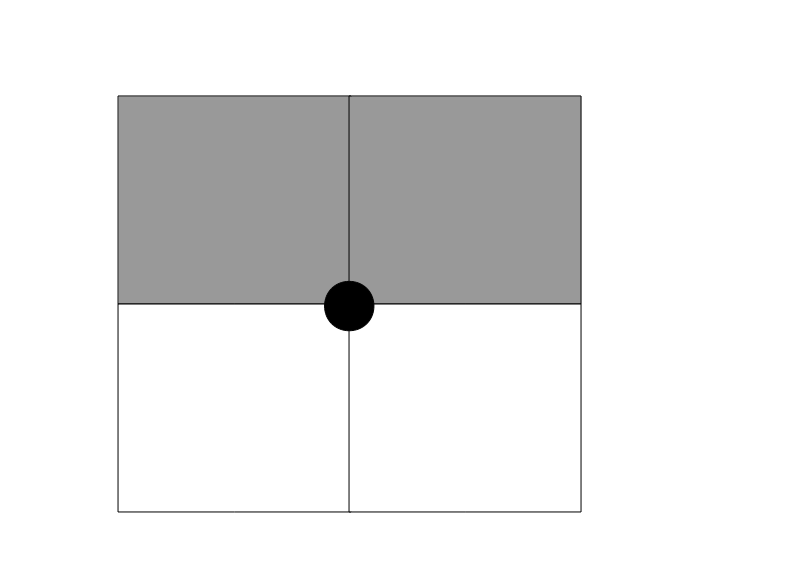} & 
\includegraphics[width=0.15\textwidth]{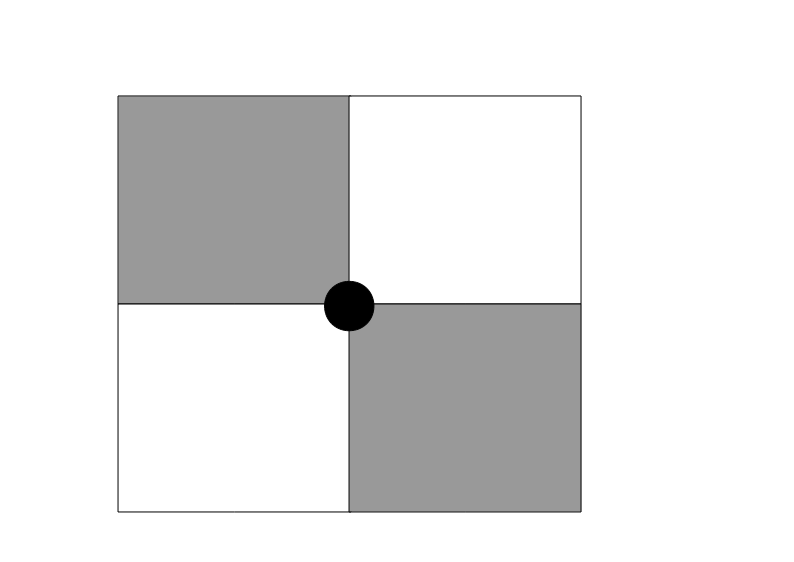} & \includegraphics[width=0.15\textwidth]{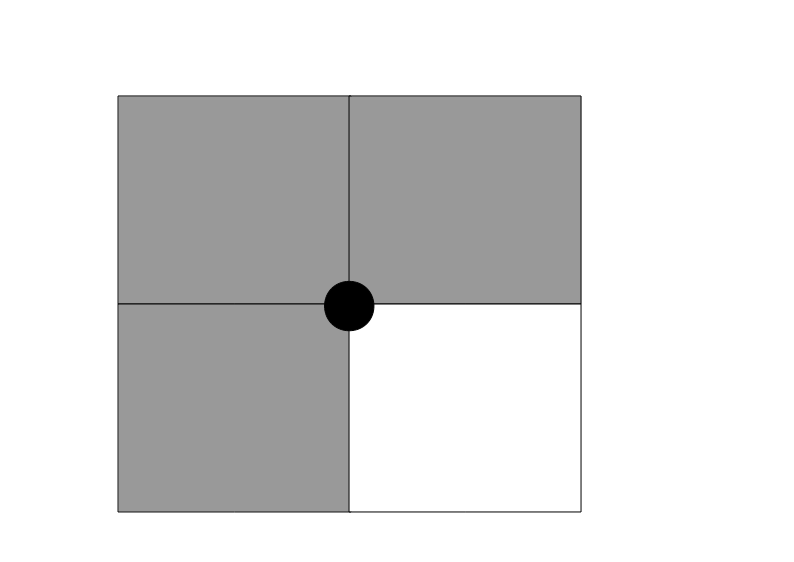} & \includegraphics[width=0.15\textwidth]{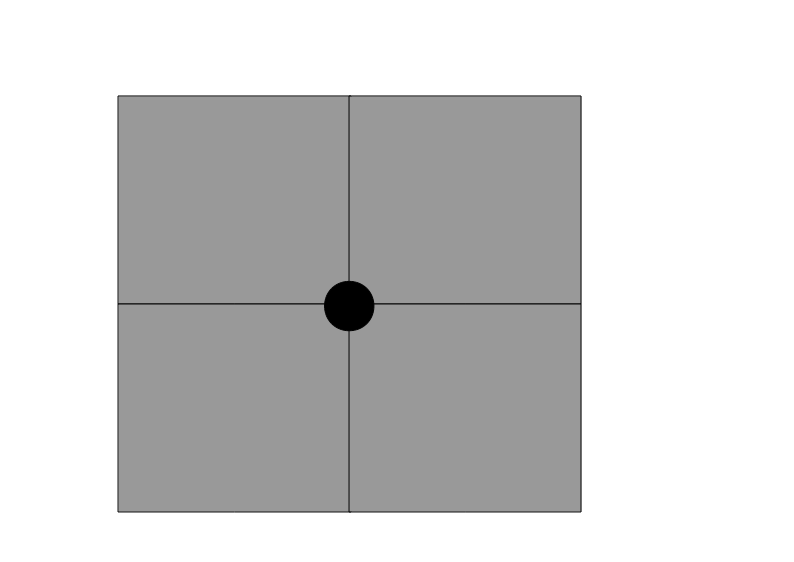} \\
0 & $\frac{1}{4}$ & 0 & 0 & $-\frac{1}{4}$ & 0 
\end{tabular}
\end{center}
\caption{Examples of the 6 geometrical configurations that can appear with a 2D regular grid.
 The grey pixels are part of the region while the white ones are outside of it. 
The numbers indicate the curvature weight with ``statistical'' choice for connectivity in
the case of touching vertices (the fourth plot).
}
\label{fig:euleralgorithm1}
\end{figure}

In complete implementation the genus algorithm is complimented by the cluster analysis, so that the Euler 
characteristic of each individual isolated cluster can be tracked.
This approach has also been implemented for 3D grids and more complicated HEALPix pixelization
in {\sc Euler3D} and {\sc EulerHealpix} families of codes. The codes are available from the authors.
\end{document}